\documentclass[prd,aps,10pt,twocolumn,nofootinbib,showpacs,notitlepage]{revtex4-2}

\usepackage{amsmath}
\usepackage{amssymb}
\usepackage{graphicx}

\newcommand{\be}{\begin{equation}}
\newcommand{\ee}{\end{equation}}
\newcommand{\planck}{\textsc{Planck}{}}
\newcommand{\dd}{\mathrm{d}}

\begin{document}
\title{Pre-inflationary scalar perturbations on closed universes in loop quantum cosmology}
\author{Lucas M. V. Montese}
\email{lucasmvmo@gmail.com}
\author{N. Yokomizo}
\email{yokomizo@fisica.ufmg.br}
\affiliation{Departamento de F\'isica - ICEx, Universidade Federal de Minas Gerais, CP 702, 30161-970, Belo Horizonte, MG, Brazil}

\begin{abstract}
The inflationary and pre-inflationary evolution of scalar modes of cosmological perturbations in a closed universe is analyzed for a loop quantum cosmology model with an inflationary regime consistent with the constraints on inflation derived from observations of the cosmic microwave background. The effective background dynamics includes a bounce induced by quantum gravity effects. Initial conditions for the perturbations are set before the bounce, and the perturbations are numerically evolved until the end of inflation, allowing the determination of the primordial power spectrum for comoving curvature perturbations. The power spectrum includes corrections due to the presence of spatial curvature and excitations produced before the onset of slow-roll inflation. The deviations from the usual power-law spectrum can be significant for observable modes even for spatial curvatures much smaller than the observational limit obtained without considering the pre-inflationary evolution of the perturbations.
\end{abstract}
\date{\today}
\maketitle


\section{Introduction}
\label{sec:intro}

A main prediction of inflation is a power-law spectrum for cosmological perturbations at the end of inflation \cite{Dodelson:2020,Weinberg:2008}, setting initial conditions for the linear perturbations of the standard $\Lambda$CDM model. The evolution of these primordial perturbations can explain the origin of the anisotropies observed in the cosmic microwave background (CMB) and the formation of the large-scale structure of the universe. The inflationary predictions have been largely confirmed by the observations, but anomalies were detected in the CMB power spectra at large angular scales \cite{Schwarz:2015cma}. Although the statistical significance of the anomalies is not decisive, they might be the result of neglected physical effects that could be included in more refined models. Mechanisms leading to modifications in the usual inflationary predictions have been explored in several works. Corrections to the primordial power spectra at the scale of the anomalies can be produced by a non-negligible spatial curvature at the onset of inflation \cite{Bonga:2016iuf,Bonga:2016cje,Handley:2019anl,Renevey:2020zdj,Renevey:2022xzh,Hergt:2022fxk,Kiefer:2021iko}, a relatively short duration of inflation \cite{Contaldi:2003zv,Boyanovsky:2006pm,Ramirez:2011kk,Handley:2014bqa,Cicoli:2014bja}, and quantum gravity effects in the inflationary and pre-inflationary regime \cite{Ashtekar:2015dja,Agullo:2023rev,Chataignier:2023rkq,Martin:2000xs}.


Effects of spatial curvature can be studied in a simple extension of the standard model with an extra parameter $\Omega_k$ describing the spatial curvature. Negative, positive and vanishing values of $\Omega_k$ correspond, respectively, to closed, open and flat universes. The analysis of the \planck{} Collaboration, combining their observations of the CMB and external baryon acoustic oscillations (BAO) measurements \cite{Planck:2018vyg}, gives an estimate of $\Omega_k = 0.001\pm 0.002$, consistent with a flat universe. It has been argued, though, that the CMB and BAO measurements are statistically inconsistent datasets, and should not be combined for parameter estimation \cite{Handley:2019tkm}. Without resorting to external data, the \planck{} data alone leads to an interval $-0.095 < \Omega_k < -0.007$ at the $99\%$ confidence level, favoring a closed universe \cite{Planck:2018vyg}. The spatial curvature in a closed universe can also explain an anomalous lensing amplitude and removes internal tensions in the estimation of cosmological parameters at distinct angular scales with CMB data \cite{DiValentino:2019qzk}. These results led to a renewed interest in the analysis of cosmological models with positive spatial curvature (i.e., for $\Omega_k < 0$). Estimates of $\Omega_k$ from alternative datasets are discussed in \cite{Vagnozzi:2020rcz,Vagnozzi:2020dfn,Dhawan:2021mel}.


Inflationary models in closed universes have been analyzed in several works \cite{Bonga:2016iuf,Bonga:2016cje,Handley:2019anl,Renevey:2020zdj,Renevey:2022xzh,Hergt:2022fxk,Kiefer:2021iko}. The presence of spatial curvature during inflation affects the predictions for the primordial power spectra in two distinct ways. First, although negligible at the end of inflation, the spatial curvature can be considerable at the early phases of inflation, since during slow-roll the energy density of the inflaton is approximately constant, while that associated with the curvature falls as $a^{-2}$, where $a$ is the scale factor. Spatial curvature effects are thus enhanced towards the past and can significantly change the dynamics of the background when the largest observable modes cross the Hubble horizon. Secondly, the equations describing the evolution of the perturbations acquire additional terms dependent on the spatial curvature that affect the evolution of the normal modes for wavelengths close to the radius of the universe. These effects combine to produce corrections to the power spectra at the end of inflation. For the scalar modes, the corrections include a suppression of power at large scales accompanied by oscillations in the spectrum, and lead to a better fit to the temperature anisotropies of the CMB \cite{Handley:2019anl}.


A common assumption underlying inflationary models is that the cosmological perturbations are in a vacuum state at the onset of inflation. In a flat universe, the perturbations are set in a Bunch-Davies vacuum state in the far past \cite{Dodelson:2020,Weinberg:2008}, while in closed universes an instantaneous vacuum is considered at the onset of inflation \cite{Bonga:2016iuf,Bonga:2016cje,Handley:2019anl,Hergt:2022fxk}. Excitations may have been produced in previous eras, however, which are neglected in this procedure, but can be computed in a pre-inflationary extension of an inflationary model. This can lead to potentially observable consequences. In \cite{Renevey:2020zdj,Renevey:2022xzh}, for instance, an extrapolation towards the past of inflation on a closed universe was analyzed. For special sets of initial conditions, a classical bounce dominated by spatial curvature can occur. Considering this case, and setting the perturbations in a vacuum state at an initial time in the contracting phase, it was found that excitations are produced at the bounce, which lead to modifications in the power spectra at the end of inflation. These can affect the largest observable modes, for a certain range of spatial curvatures.

The curvature bounce for inflation in a closed universe is not generic. In general, the extension of inflation towards the past produces an initial singularity, similarly as in flat universes, where inflationary models based on general relativity always meet a Big Bang singularity \cite{Borde:1996pt}. A description of the pre-inflationary regime in a closed universe, when a classical bounce does not take place, must then involve Planck scale physics, as the energy density diverges at the singularity. Moreover, as stressed in \cite{Martin:2000xs}, if the universe has been expanding for a sufficiently long time, observable modes of the cosmological perturbations had physical wavelengths smaller than the Planck length in the far past. At such scales, quantum fluctuations of the geometry might become relevant for the dynamics of the modes. This leads to the so-called trans-Planckian problem in cosmology: observable modes emerge from scales whose description requires a theory of quantum gravity, which  becomes indispensable for the analysis of the pre-inflationary regime and its eventual phenomenological implications.

A framework allowing the description of quantized perturbations on a quantum spacetime is required to address the trans-Planckian issue. In this work, we are interested in an approach to the problem proposed in the context of loop quantum cosmology (LQC) \cite{Ashtekar:2015dja,Agullo:2023rev}. A formalism for quantized fields on quantum cosmological spacetimes was developed in \cite{Ashtekar:2009mb}, and subsequently applied for the construction of a pre-inflationary extension of single-field inflation in a flat universe in \cite{Agullo:2012sh,Agullo:2013ai}. In this model, the background geometry is described by a wavefunction $\Psi(a,\phi_0)$ assigning quantum amplitudes to the scale factor and homogeneous background scalar field $\phi_0$, and background quantities in the equations of the perturbations become operators. Neglecting backreaction, it turns out that the evolution of the perturbations on the quantum background is equivalent to that on a classical background with a dressed metric $\tilde{g}$ including quantum corrections. For sharply peaked states of the background, the dressed metric has an effective dynamics dictated by a well-known modified Friedmann equation of LQC. In the effective dynamics, the Big Bang singularity is replaced by a quantum bounce.

Phenomenological implications of the model for the CMB were analyzed in \cite{Agullo:2013ai,Agullo:2015tca}. Setting the perturbations in a vacuum state before the bounce, excitations are produced as they cross the bounce, leading to modifications in the primordial power spectra. If the number of $e$-folds between the bounce and the onset of the observable phase of inflation is not too large ($\sim 12$ $e$-folds), such modifications can affect the largest observable modes. A quadratic inflaton potential was originally considered in the references \cite{Agullo:2012sh,Agullo:2013ai,Agullo:2015tca}. A Starobinsky potential was later analyzed in \cite{Bonga:2015kaa,Bonga:2015xna}. Non-Gaussianities were studied for both potentials in \cite{Agullo:2018bs}. In all these works, a flat universe was considered.

This set of results shows that the primordial power spectra can carry imprints, at the scale of the largest observable modes, both of a non-negligible spatial curvature near the onset of inflation and of excitations produced at a quantum bounce. How these effects combine in a model where both are present has not been analyzed so far, however. Our main goal in the present work is to fill this gap. We consider an LQC pre-inflationary extension of inflation in a closed universe modelled after the dressed metric approach employed for the case of a flat universe in \cite{Ashtekar:2015dja,Agullo:2023rev}. We consider the case of single-field inflation with a Starobinsky potential and focus on scalar modes of the cosmological perturbations. The effective dynamics of the background is described by the modified Friedmann equation for a closed universe obtained in \cite{Ashtekar:2006es}. We derive equations for the perturbations in a gauge analogous to the spatially flat gauge employed in flat universes. Setting the perturbations in a vacuum state in the pre-bounce era, we numerically evolve the perturbations until the end of inflation, and determine the primordial power spectrum for comoving curvature perturbations at the end of inflation.

This work is organized as follows. In Sec.~\ref{sec:hamiltonian-formalism}, we develop the classical Hamiltonian formalism for the background and scalar modes of linear cosmological perturbations in a spatially spherical gauge in a closed universe. In Sec.~\ref{sec:lqc-effective-dynamics}, we review general aspects of the effective dynamics and of the dressed metric approach of LQC, and introduce our model. The evolution of the background is then analyzed in Sec.~\ref{sec:effective-background}, and that of the perturbations in Sec.~\ref{sec:evolution-perturbations}, where we determine the primordial power spectrum. We consider backgrounds consistent with observational constraints on inflation, and vary the spatial curvature within such bounds. We conclude in Sec.~\ref{ref:conclusion} with a summary and discussion of our results.

\section{Hamiltonian formalism}
\label{sec:hamiltonian-formalism}

\subsection{ADM formalism}

The action for the gravitational field $g_{\mu \nu}$ interacting with a scalar field $\phi$ is given by
\begin{equation}
S = \int d^4x \sqrt{-g} \left[ \frac{1}{2\kappa} R - \frac{1}{2} g^{\mu \nu} \partial_\mu \phi \partial_\nu \phi - V(\phi) \right] \, ,
\label{eq:EH-action}
\end{equation}
where $\kappa=8\pi G$. The extremization of the action under variations of the metric tensor leads to the Einstein equation:
\[
R_{\mu \nu} - \frac{1}{2} R g_{\mu \nu} = \kappa T_{\mu \nu} \, .
\]
The energy-momentum tensor of the scalar field reads:
\be
T_{\mu \nu} = (\nabla_\mu \phi) (\nabla_\nu \phi) - \frac{1}{2} g_{\mu \nu} \nabla^\rho \phi \nabla_\rho \phi - g_{\mu \nu} V \, .
\label{eq:energy-momentum}
\ee
Varying the action with respect to the scalar field, we obtain the Klein-Gordon equation:
\be
\Box \phi - \frac{dV}{d\phi} = 0 \, .
\label{eq:kg}
\ee

The metric is expressed in terms of ADM variables as
\begin{equation}
ds^2 = - (N^2 - N^i N_i) dt^2 + 2 N_i dt dx^i + \gamma_{ij} dx^i dx^j \, ,
\label{eq:adm-metric}
\end{equation}
where $N$ is the lapse function, $N^i$ is the shift vector and $\gamma_{ij}$ is the spatial metric on spatial slices $\Sigma$ of constant time. In matrix representation, the metric tensor $g_{\mu \nu}$ and its inverse $g^{\mu \nu}$ are given by:
\begin{align}
g_{\mu \nu} = \begin{pmatrix}
- N^2 + N^i N_i	&	N_i	\\
N_i			&	\gamma_{ij}
\end{pmatrix} , 
\nonumber \\
g^{\mu \nu} = \begin{pmatrix}
- \frac{1}{N^2}		&	\frac{N_i}{N^2}	\\
\frac{N_i}{N^2}		&	\gamma^{ij} -\frac{N^i N_i}{N^2}
\end{pmatrix} .
\label{eq:metric-matrix}
\end{align}
The determinant of the metric is $g= -N^2 \gamma$.

Let $D_i$ be the three-dimensional covariant derivative compatible with the spatial metric, $D_i \gamma_{jk} = 0$. The extrinsic curvature of the spatial leaves is given by
\be
K_{ij} = \frac{1}{2N} (-\dot{\gamma}_{ij} + D_j N_i + D_i N_j) \, .
\label{eq:K-adm}
\ee
The Ricci scalar $R$ of $g_{\mu \nu}$ can be expressed in terms of the Ricci scalar $\mathcal{R}$ of the intrinsic metric $\gamma_{ij}$ and the extrinsic curvature $K$ as
\be
R =\mathcal{R} + K^{ij} K_{ij} - K^2 \, ,
\label{eq:R-adm}
\ee
where $K=\gamma^{ij} K_{ij}$ is the trace of the extrinsic curvature.

The formulas \eqref{eq:K-adm} and \eqref{eq:R-adm} allow the action \eqref{eq:EH-action} to be written in terms of the ADM variables. We take $\gamma_{ij}$ and $\phi$ as configuration variables, and $N,N_i$ will be recognized as Lagrange multipliers. The conjugate momenta are:
\begin{align}
\tilde{\pi}^{ij} &= \frac{\delta S}{\delta \dot{\gamma}_{ij}} = \frac{\sqrt{\gamma}}{2 \kappa} (K \gamma^{ij} - K^{ij})  \, , 
\label{eq:p-gamma-def} \\
\tilde{\pi}_\phi &= \frac{\delta S}{\delta \dot{\phi}}  = - \sqrt{-g} \, \partial^0 \phi = \frac{\sqrt{\gamma}}{N} ( \dot{\phi} - N^i \partial_i \phi ) \, . 
\label{eq:p-phi-def}
\end{align}
The configuration variables $\gamma_{ij}$ and $\phi$ are tensor fields, while the conjugate momenta $\tilde{\pi}^{ij}$ and $\tilde{\pi}_\phi$ are tensor densities, as indicated by the tildes. The nonzero Poisson brackets among these canonical variables are:
\begin{align}
\{\gamma_{ij}(\vec x),\tilde{\pi}^{kl}(\vec y)\} &= \delta_{(i}^k \delta_{j)}^l \delta(\vec x - \vec y) \, , 
\label{eq:gamma-brackets}
\\
\{\phi(\vec x), \tilde{\pi}_\phi(\vec y)\} &= \delta(\vec x - \vec y) \, .
\label{eq:phi-brackets}
\end{align}

Applying a Legendre transformation, we obtain the ADM Hamiltonian:
\be
H = \int \dd^3 x (N \mathcal{H} + N^i \mathcal{H}_i) \, ,
\label{eq:adm-hamiltonian}
\ee
where the scalar (or Hamiltonian) constraint $\mathcal{H}$ and the diffeomorphism constraint $\mathcal{H}_i$ read:
\begin{align}
\mathcal{H} &= \frac{2 \kappa}{\sqrt{\gamma}} \left( \tilde{\pi}^{ij} \tilde{\pi}_{ij} - \frac{\tilde{\pi}^2}{2} \right)- \frac{\sqrt{\gamma}}{2 \kappa} \mathcal{R} + \frac{\tilde{\pi}_\phi^2}{2 \sqrt{\gamma}} \nonumber \\
	& \qquad + \sqrt{\gamma} V + \frac{\sqrt{\gamma}}{2} (\partial_i \phi) (\partial^i \phi) \, , 
\label{eq:constraint-scalar}
\\
\mathcal{H}_i &= -2 D_j (\tilde{\pi}^{jk} \gamma_{ki}) + \tilde{\pi}_\phi \partial_i \phi \, ,
\label{eq:constraint-diff}
\end{align}
and spatial indices are lowered and raised with the spatial metric $\gamma_{ij}$ and its inverse.

\subsection{Mode expansion of linear perturbations}

Let us consider a spacetime that can be well approximated by a perturbed closed FLRW universe. The canonical variables can then be written in the form:
\begin{align}
\gamma_{ij} = \mathring{\gamma}_{ij} + \delta \gamma_{ij} \, &, \qquad 
\tilde{\pi}^{ij} = \mathring{\tilde{\pi}}^{ij} + \delta \tilde{\pi}^{ij} \, , \nonumber 
\\
\phi = \phi_0 + \delta \phi \, &, \qquad \tilde{\pi}_\phi= \tilde{\pi}_{\phi_0} + \delta \tilde{\pi}_\phi \, ,
\label{eq:first-order-expansion-vars}
\end{align}
where the circle over a quantity indicates a background, unperturbed quantity. For scalar quantities, we use a subscript $0$ to represent a background quantity. The background spatial geometry $\mathring{\gamma}_{ij}$ describes a $3$-sphere with a time-dependent radius $a(t)$. The unperturbed scalar field depends only on time, $\phi_0=\phi_0(t)$, due to homogeneity. The lapse function and shift vector are expressed in terms of a background contribution and a perturbation,
\be
N = N_0 + \delta N \, , 
\quad
N^i = \delta N^i \, ,
\label{eq:multipliers-pert}
\ee
where the background contribution to the shift vector must vanish due to isotropy, and the background contribution to the lapse function is a function only of time, $N_0=N_0(t)$, due to homogeneity.

The background geometry is a closed FLRW universe with line element:
\begin{equation}
d\mathring{s}^2 = - N_0^2(t) dt^2 + a(t)^2 \Omega_{ij}(x) dx^i dx^j \, ,
\label{eq:background-metric}
\end{equation}
where $\Omega_{ij}$ is the metric of the unit $3$-sphere and $a(t)$ is the scale factor. The unperturbed spatial metric reads:
\[
\mathring{\gamma}_{ij} = a(t)^2 \Omega_{ij} \, .
\]
We denote the three-dimensional covariant derivative compatible with $\mathring{\gamma}_{ij}$ by $\mathring{D}_k$, and the covariant derivative compatible with the comoving metric $\Omega_{ij}$ by $\bar{D}_k$, and put $\bar{D}^i = \Omega^{ij}\bar{D}_j$, where $\Omega^{ij}$ is the inverse of $\Omega_{ij}$.

The extrinsic curvature and the scalars of curvature of the background geometry read:
\[
\mathring{\mathcal{R}} = \frac{6}{a^2} \,, \quad \mathring{K}_{ij} = - \frac{1}{a} \frac{\dot{a}}{N_0} \mathring{\gamma}_{ij} \, ,\quad \mathring{R} = \frac{6}{a^2} \left( 1 - \frac{\dot{a}^2}{N_0^2} \right) \, .
\]
From Eq.~\eqref{eq:p-gamma-def} and the formula for $\mathring{K}_{ij}$, we obtain for the zeroth-order momentum conjugated to the spatial metric:
\be
\mathring{\tilde{\pi}}^{ij} = - \sqrt{\Omega} \frac{1}{\kappa} \frac{\dot{a}}{N_0} \Omega^{ij} \, .
\label{eq:zeroth-pi-tilde}
\ee
where $\Omega$ is the determinant of $\Omega_{ij}$. The zeroth-order conjugate field momentum is obtained from Eq.~\eqref{eq:p-phi-def}:
\be
\tilde{\pi}_{\phi_0} = \sqrt{\Omega} a^3 \frac{\dot{\phi}_0}{N_0} \, .
\label{eq:zeroth-pi-phi}
\ee

The perturbed metric and scalar fields can be expanded in terms of hyperspherical harmonics on the $3$-sphere $S^3$, as we will now discuss. Hyperspherical harmonics on $S^3$ are reviewed in the Appendix \ref{sec:harmonics}, to which we refer for their definition and basic properties.

\subsubsection{Mode expansion of the scalar field}

A scalar function on the $3$-sphere can be expanded in terms of scalar hyperspherical harmonics $Q_{nlm}$. The scalar field and its conjugate momentum can be represented at any time as
\begin{align}
\phi &= \sum_{nlm} f_{nlm} Q_{nlm} \, , \\
\tilde{\pi}_\phi &= \sqrt{\Omega} \sum_{nlm} \pi^f_{nlm} Q_{nlm} \, ,
\end{align}
with expansion coefficients given by:
\begin{align}
f_{nlm}(t) &= \int \dd \Omega \, Q_{nlm}(\vec x) \phi(t,\vec x) \, , \\
\pi^f_{nlm}(t) &= \int \dd^3x \, Q_{nlm}(\vec x) \tilde{\pi}_\phi(t,\vec x) \, .
\end{align}
The field momentum $\tilde{\pi}_\phi$ is a spatial density, which requires the inclusion of the factor of $\sqrt{\Omega}$, the volume element of the unit $3$-sphere, in its expansion. We obtain for the Poisson brackets of the mode amplitudes:
\[
\{f_{nlm},\pi^f_{n'l'm'}\} = \delta_{nn'} \delta_{ll'} \delta_{mm'} \, .
\]
The subscripts $nlm$ in the amplitudes $f$ and $\pi^f$ will often be omitted to lighten the notation.

The background scalar field $\phi_0$ corresponds to the mode $n=1$,
\be
\phi_0 = f_{100} Q_{100} = \frac{1}{V_0} \int \dd \Omega \, \phi \, ,
\label{eq:background-field-def}
\ee
where we used the explicit form of the homogeneous mode $Q_{100}=1/\sqrt{V_0}$. It describes the average value of the field over the three-sphere. Similarly, for the conjugate momentum, the homogeneous contribution is given by the average:
\[
\tilde{\pi}_{\phi_0}= \pi^f_{100} Q_{100} \sqrt{\Omega} = \frac{\sqrt{\Omega}}{V_0} \int \dd^3x \, \tilde{\pi}_\phi \, .
\]

The quantities $\phi_0$ and $\tilde{\pi}_{\phi_0}$ were defined as a scalar function and a scalar density over the $3$-sphere, respectively. The function $\phi_0$ is independent of the position, however, as is the scalar function
\[
\pi_{\phi_0} := \int \dd^3x \, \tilde{\pi}_\phi
\]
related to the momentum density of the field through
\[
\mathring{\tilde{\pi}}_{\phi_0} = \frac{\sqrt{\Omega}}{V_0} \pi_{\phi_0} \, .
\]
Interpreting $\phi_0$ as the function of time defined by Eq.~\eqref{eq:background-field-def}, we find that
\[
\{\phi_0, \pi_{\phi_0}\} = \frac{1}{V_0} \int \dd \Omega \, \dd^3x' \, \{\phi(\vec x), \tilde{\pi}_\phi(\vec x')\}  = 1\, .
\]

The linear perturbations of the scalar field and its momentum are described by the sum of the remaining contributions with $n\neq1$ in their mode expansions,
\begin{align}
\delta \phi &= \sum_{n=2}^\infty \sum_{lm} f_{nlm} Q_{nlm} \, , \nonumber \\
\delta \tilde{\pi}_\phi &= \sum_{n=2}^\infty \sum_{lm} \pi^f_{nlm} Q_{nlm} \sqrt{\Omega} \, .
 \label{eq:field-perturbations-mode-expansion}
\end{align}
It follows that
\[
\{\delta \phi(\vec x), \delta \tilde{\pi}_\phi(\vec y)\} = \delta(\vec x - \vec y) - \frac{\sqrt{\Omega}}{V_0} \, .
\]

\subsubsection{Mode expansion of the spatial metric}

Let us now consider the decomposition of the metric. Any symmetric tensor field on the $3$-sphere can be expanded in terms of tensor hyperspherical harmonics which, as in the case of a flat space, can be decomposed into scalar, vector and tensor modes in an SVT decomposition. There are two scalar, two vector and two tensor modes. Following \cite{Halliwell:1984eu}, we choose for the scalar modes:
\begin{align}
(A^1_{nlm})_{ij} &= \frac{1}{3} \Omega_{ij} Q_{nlm} \, , \quad n =1,2,3,\dots \, ,\nonumber \\
(A^2_{nlm})_{ij} &= \left( \frac{1}{n^2-1} \bar{D}_i \bar{D}_j + \frac{1}{3} \Omega_{ij} \right) Q_{nlm} \, , \; n =3,4,\dots \, ,
\label{eq:metric-scalar-modes}
\end{align}
and let $(A^\alpha_{nlm})_{ij}$, for $\alpha=3,\dots,6$, represent the vector ($\alpha=3,4$) and tensor ($\alpha=5,6$) modes. The explicit form of the vector and tensor modes is described in \cite{Halliwell:1984eu} and will not be relevant for our purposes. The modes $A^\alpha_{nlm}$ form an orthogonal basis for the space of symmetric tensors on the $3$-sphere. The indices $nlm$ are the curved analogue of the components of the momentum $\vec k$ in a Fourier expansion on a flat space.

Introducing normalization factors $N_\alpha$ through
\[
\int d\Omega (A^\alpha_{nlm})_{ij} (A^{\alpha'}_{n'l'm'})^{ij} = N_\alpha \, \delta_{\alpha \alpha'}\delta_{n n'} \delta_{m m'} \delta_{n n'} \,, 
\]
where we lower and raise indices in the harmonics with the metric $\Omega_{ij}$ and its inverse, we have:
\begin{align}
N_1&=\frac{1}{3} \, , & &N_2 = \frac{2}{3} \frac{n^2-4}{n^2-1} \, , \nonumber \\
 N_3&= N_4= 2(n^2-4) \, , & &N_5=N_6 = 1\, .
\end{align}
The resolution of the identity in the space of symmetric tensors for this choice of orthogonal basis reads:
\begin{equation}
\delta_i^{(r} \delta_j^{s)} \delta(x-y)= \sum_{nlm} \sum_\alpha \frac{1}{N_\alpha} (A^\alpha_{nlm})_{ij}(x) (A^\alpha_{nlm})^{rs}(y) \, .
\label{eq:tensor-A-orthonormality}
\end{equation}

The spatial metric and its conjugate momentum can be represented in terms of an SVT decomposition as
\begin{align}
\gamma_{ij} &= \sum_{nlm} \sum_\alpha \gamma^\alpha_{nlm} (A^\alpha_{nlm})_{ij} \, , 
\label{eq:gamma-svt}
\\
\tilde{\pi}^{rs} &= \sqrt{\Omega} \sum_{nlm} \sum_\alpha \frac{1}{N_\alpha} \pi^\alpha_{nlm} (A^\alpha_{nlm})^{rs} \, ,
\label{eq:pi-gamma-svt}
\end{align}
with expansion coefficients
\begin{align}
\gamma^\alpha_{nlm}(t) &= \frac{1}{N_\alpha} \int \dd \Omega \, (A^\alpha_{nlm})^{ij}(\vec x) \gamma_{ij}(t,\vec x) \, ,
\label{eq:gamma-modes} \\
\pi^\alpha_{nlm}(t) &= \int \dd^3 x \, (A^\alpha_{nlm})_{ij}(\vec x) \tilde{\pi}^{ij}(t,\vec x)  \, .
\label{eq:pi-modes}
\end{align}
The Poisson brackets of the expansion coefficients are obtained from Eqs.~\eqref{eq:gamma-brackets}, \eqref{eq:gamma-modes} and \eqref{eq:pi-modes}, which lead to:
\begin{equation}
\{ \gamma^\alpha_{nlm} , \pi^{\alpha '}_{n'l'm'}\} = \delta^{\alpha \alpha '}  \delta_{nn'} \delta_{m m'} \delta_{l l'} \, .
\label{eq:expansion-coeff-bracket}
\end{equation}
As for the field perturbations, the subscripts $nlm$ in the amplitudes $\gamma^\alpha$ and $\pi^\alpha$ will be omitted when possible.

The mode $(100)$ is proportional to $\Omega_{ij}$ and as such describes the background metric. The background metric and associated momentum are then expressed in terms of hyperspherical harmonics as:
\begin{align*}
\mathring{\gamma}_{ij} &= \gamma^1_{100} (A^1_{100} )_{ij}  \, , \\
\mathring{\tilde{\pi}}^{ij} &= \sqrt{\Omega} \frac{1}{N_1} \pi^1_{100} (A^1_{100} )_{ij} \, .
\end{align*}
From Eq.~\eqref{eq:expansion-coeff-bracket},
\[
\{ \mathring{\gamma}_{ij}(\vec x),\mathring{\tilde{\pi}}^{rs}(\vec y) \} = \frac{\sqrt{\Omega}}{V_0} \frac{1}{3} \Omega_{ij}(\vec x) \Omega^{rs}(\vec y) \, .
\]
As for the scalar field, the perturbations of the metric and its conjugate momentum are the sums of all contributions with $n\neq1$ in their mode expansions:
\begin{align}
\delta \gamma_{ij}(\vec x) &= \sum_{n \neq 1}^\infty \sum_{lm\alpha} \gamma^\alpha_{nlm} (A^\alpha_{nlm})_{ij}(\vec x) \nonumber \\
\delta \tilde{\pi}^{rs}(\vec y) &= \sqrt{\Omega} \sum_{n \neq 1}^\infty \sum_{lm\alpha}  \frac{1}{N_\alpha} \pi^\alpha_{nlm} (A^\alpha_{nlm})^{rs}(\vec y) \, .
\label{eq:metric-perturbations-mode-expansion}
\end{align}
The nonzero Poisson brackets are:
\begin{align*}
\{ \delta \gamma_{ij}(\vec x), \delta \tilde{\pi}^{rs}(\vec y) \} =  \delta_i^{(r} \delta_j^{s)} \delta(\vec x-\vec y) - \frac{\Omega_{ij}(\vec x) \Omega^{rs}(\vec y) }{3\sqrt{V_0}} \, .
\end{align*}

In this work, we are interested in the dynamics of the scalar modes. Accordingly, we will only keep terms with $\alpha=1,2$ in the SVT decompositions \eqref{eq:gamma-svt} and \eqref{eq:pi-gamma-svt}. Inserting the explicit form of the scalar modes in these decompositions, we write the scalar perturbations of the spatial metric and its conjugate momentum in the form:
\begin{widetext}
\begin{align}
\gamma_{ij} &= \sum_{n=1}^\infty \sum_{lm} \left[ \gamma^1_{nlm} \frac{1}{3} \Omega_{ij} Q_{nlm}\right] + \sum_{n=3}^\infty \sum_{lm} \left[ \gamma^2_{nlm} \left( \frac{1}{n^2-1} \bar{D}_i \bar{D}_j + \frac{1}{3} \Omega_{ij} \right) Q_{nlm} \right]\, , 
\label{eq:metric-scalar-pert}
\\
\tilde{\pi}^{rs} & = \sqrt{\Omega} \left\{ \sum_{n=1}^\infty \sum_{lm} \left[ \frac{\pi^1_{nlm}}{N_1} \frac{1}{3} \Omega^{rs} Q_{nlm}\right] + \sum_{n=3}^\infty \sum_{lm} \left[ \frac{\pi^2_{nlm}}{N_2} \left( \frac{1}{n^2-1} \bar{D}^r \bar{D}^s + \frac{1}{3} \Omega^{rs} \right) Q_{nlm} \right] \right\}\, .
\label{eq:pi-metric-scalar-pert}
\end{align}
\end{widetext}

\subsubsection{Mode expansion of the lapse function and shift vector}

The lapse function and shift vector are expanded in terms of hyperspherical harmonics in the form:
\begin{align}
N &= N_0 + \sum_{n \neq 1} \sum_{lm} g_{nlm} Q_{nlm}\, , \\
N^i &= \delta N^i = \sum_{nlm} k_{nlm} \bar{D}^i Q_{nlm} \, ,
\end{align}
where the homogeneous part of the lapse function is
\[
N_0 = g_{100} Q_{100} = \frac{1}{V_0} \int \dd\Omega \, N \, ,
\]
and the expansion of the shift vector only includes scalar modes.

\subsection{Background Hamiltonian}
\label{sec:zeroth-adm}

The background is completely specified by two dynamical variables, the scale factor $a(t)$ and the mean field $\phi_0(t)$, and the unperturbed Lagrange multiplier $N_0(t)$. The unperturbed action obtained from \eqref{eq:EH-action} is given in terms of these quantities by:
\begin{align}
S_0 &= \int \dd t \, V_0 N_0 a^3 \left[ \frac{3}{\kappa a^2} \left( 1 - \frac{\dot{a}^2}{N_0^2} \right) + \frac{1}{2} \frac{\dot{\phi_0}^2}{N_0^2} - V(\phi_0) \right] \nonumber \\
	& = \int \dd t \, L_0 \, .
\label{eq:0th-action}
\end{align}
The conjugate momenta are:
\begin{align}
\pi_a &= \frac{\partial L_0}{\partial \dot{a}} = - 6 \frac{V_0}{\kappa N_0} a \dot{a} \, , 
\label{eq:pi-a}
\\
\pi_{\phi_0} &= \frac{\partial L_0}{\partial \dot{\phi}_0} = \frac{V_0}{N_0} a^3 \dot{\phi}_0 \, .
\label{eq:pi-phi0}
\end{align}
The background momenta are related to the zeroth-order momentum densities given in Eqs.~\eqref{eq:zeroth-pi-tilde} and \eqref{eq:zeroth-pi-phi} through:
\begin{align}
\mathring{\tilde{\pi}}^{ij} &= \frac{\sqrt{\Omega}}{V_0} \frac{\Omega^{ij}}{6a} \pi_a \, ,  \nonumber \\
\mathring{\tilde{\pi}}_\phi &=  \frac{\sqrt{\Omega}}{V_0} \pi_{\phi_0} \, .
\label{eq:momentum-densities-bg}
\end{align}

Inserting the formulas \eqref{eq:momentum-densities-bg} for the momentum densities in the general expression \eqref{eq:constraint-scalar} for the scalar constraint, we obtain the background Hamiltonian in terms of the background degrees of freedom $a,\phi_0$ and their conjugate momenta:
\begin{align}
H^{(0)} &= \int \dd^3 x \, N_0 \mathcal{H}^{(0)} \nonumber \\
	&= N_0 V_0 \left[ - \frac{\kappa \pi_a^2}{12 a V_0^2} + \frac{\pi_{\phi_0}^2}{2 a^3 V_0^2} + a^3 V - \frac{3a}{\kappa} \right] \, .
\label{eq:hamiltonian-constraint-zeroth}
\end{align}
The diffeomorphism constraint vanishes identically at zeroth-order, due to homogeneity. The lapse function $N_0(t)$ can be chosen at will. If one chooses $N_0(t)=1$, then the time coordinate $t$ corresponds to the proper time of comoving observers. Expressing the momenta in terms of time derivatives (Eqs.~\eqref{eq:pi-a} and \eqref{eq:pi-phi0}), we verify that the zeroth-order scalar constraint $\mathcal{H}^{(0)} \approx 0$ is simply the Friedmann equation,
\begin{equation}
\frac{1}{a^2}\left( \frac{\dot{a}}{N_0} \right)^2 = \frac{\kappa}{3} \rho - \frac{1}{a^2} \, .
\label{eq:friedmann-h0}
\end{equation}

The Hamilton equations for the scale factor read:
\begin{align*}
\dot{a} &= \{a,H^{(0)}\}  = - \frac{N_0}{V_0} \frac{\kappa}{6a} \pi_a \, , \\
\dot{\pi}_a &= \{\pi_a,H^{(0)}\} \\
& \Rightarrow \frac{1}{a} \frac{\ddot{a}}{N_0^2} = - \frac{\kappa}{2} \left[ \frac{1}{2} \left( \frac{\dot{\phi}_0}{N_0} \right)^2 -V + \frac{1}{\kappa a^2}\left(1 + \frac{\dot{a}^2}{N_0^2} \right) \right] .
\end{align*}
The first equation corresponds to the formula \eqref{eq:pi-a} for the momentum $\pi_a$. The second equation can be simplified using the scalar constraint \eqref{eq:friedmann-h0}, leading to the acceleration equation,
\be
\frac{1}{a} \frac{\ddot{a}}{N_0^2} = -\frac{\kappa}{2} (\rho + 3P) \, .
\label{eq:acceleration-eq}
\ee
The Hamilton equations for the scalar field read:
\begin{align}
\dot{\phi}_0 &= \{\phi_0, H^{(0)}\} = N_0 \frac{\pi_{\phi_0}}{a^3V_0} \, , \nonumber \\
\dot{\pi}_{\phi_0} &= \{ \pi_{\phi_0}, H^{(0)}\} = - N_0 V_0 a^3 V' \nonumber \\
& \Rightarrow \frac{\ddot{\phi}_0}{N_0^2} + 3 H \frac{\dot{\phi}_0}{N_0} + V' = 0 \, .
\label{eq:klein-gordon}
\end{align}
The first equation corresponds to the formula \eqref{eq:pi-phi0} for the momentum $\pi_\phi$, and the second equation is the Klein-Gordon equation for a homogeneous field.

\subsection{Quadratic Hamiltonian for linear perturbations}

The ADM Hamiltonian is described in Eq.~\eqref{eq:adm-hamiltonian}. In order to determine the quadratic Hamiltonian that governs the dynamics of linear perturbations, we follow the steps of the analysis performed for the case of a flat universe in \cite{Agullo:2018bs}, adapted for a closed universe. Let us first sketch the main steps of the derivation, and then describe its detailed implementation. We will proceed as follows. The scalar and diffeomorphism constraints can be expanded as a series in the perturbations $\delta \gamma_{ij},\delta \tilde{\pi}^{ij},\delta \phi,\delta \tilde{\pi}_\phi$ in phase space:
\begin{align}
\mathcal{H} &= \mathcal{H}^{(0)} + \mathcal{H}^{(1)} + \mathcal{H}^{(2)} + \cdots\, , 
\label{eq:scalar-constraint-expansion}
\\
\mathcal{H}_i &= \mathcal{H}_i^{(1)} + \mathcal{H}_i^{(2)} + \cdots\, ,
\label{eq:diffeo-constraint-expansion}
\end{align}
where a superscript $^{(n)}$ indicates that the term is of $n$-th order in the perturbations. The zeroth-order, background diffeomorphism vector vanishes due to isotropy. Explicit expressions for the $n$-th order perturbations of the constraints can be obtained by direct substitution of the formulas \eqref{eq:first-order-expansion-vars} for the perturbed fields in the general expressions for the scalar and diffeomorphism constraints given in Eqs.~\eqref{eq:constraint-scalar} and \eqref{eq:constraint-diff}.

The zeroth-order scalar constraint corresponds to the Friedmann equation, as discussed in Section~\ref{sec:zeroth-adm}. In order to determine the perturbations of the constraints, we choose a special gauge in which they assume simple forms. As discussed in the Appendix \ref{sec:gauges}, perturbations of the spatial metric can be made to vanish, $\delta \gamma_{ij}=0$, with an adequate choice of coordinates. We say that this condition characterizes the spatially spherical gauge, in analogy with the spatially flat gauge employed on flat universes. The lapse function and shift vector are expressed in terms of a background part and a perturbation in Eq.~\eqref{eq:multipliers-pert}. The background parts of the shift vector and diffeomorphism constraint vanish due to isotropy. The quadratic Hamiltonian is then given by
\be
H^{(2)} = \int \dd^3 x \left( N_0 \mathcal{H}^{(2)} + \delta N \mathcal{H}^{(1)} + \delta N^i \mathcal{H}_i^{(1)} \right) \, ,
\label{eq:quadratic-adm}
\ee
where we have not included a term proportional to $\mathcal{H}^{(0)}$, which vanishes when the background equations are satisfied, and do not affect the determination of the equations of motion for the perturbations through the calculation of the Poisson brackets of the quadratic Hamiltonian and the linear perturbations.

Imposing the linearized constraints $\mathcal{H}^{(1)}$ and $\mathcal{H}_i^{(1)}$, we can fix the momentum perturbations $\delta \tilde{\pi}^{ij}$ in terms of the field perturbations $\delta \phi,\delta \tilde{\pi}_\phi$. Moreover, in order for the gauge condition $\delta \gamma_{ij}=0$ to be preserved under time-evolution, we must impose the consistency condition:
\be
\{H^{(2)}, \delta \gamma_{ij} \}=0 \, .
\ee
This fixes the perturbations $\delta N$ and $\delta N^i$ of the lapse function and shift vector in terms of the other perturbations. As the momentum perturbations $\delta \tilde{\pi}^{ij}$ are eliminated from the dynamics by the linearized constraints, the quadratic Hamiltonian can then be expressed exclusively in terms of the field perturbations $\delta \phi$ and $\delta \tilde{\pi}_\phi$. Computing then the Poisson brackets:
\begin{align}
\frac{d \delta \phi}{dt}&= \{\delta \phi, H^{(2)} \} \, , \\
\frac{d \delta \tilde{\pi}_\phi}{dt}&= \{\delta \tilde{\pi}_\phi, H^{(2)} \} \, ,
\end{align}
we obtain the equations of motion for the field perturbations. This will be done, in fact, for the amplitudes of the normal modes describing the perturbations of the field. We now describe in detail the steps sketched above leading to the determination of the equations of motion for the field perturbations in the spatially spherical gauge.

\subsubsection{Linearized constraints}

For generic linear perturbations of the form \eqref{eq:first-order-expansion-vars}, the first-order contribution to the scalar constraint \eqref{eq:constraint-scalar} assumes the form:
\begin{widetext}
\begin{align}
\mathcal{H}^{(1)} &=  \sqrt{\Omega}\left[ - \frac{\kappa}{72 V_0^2 a }\pi_a^2  - \frac{a}{2 \kappa} - \frac{1}{4 V_0^2 a^3}\pi_{\phi_0}^2 + \frac{a^3}{2}V(\phi_0) \right] \mathring{\gamma}^{ij} \delta \gamma_{ij} \notag \\
	&\qquad - \frac{\kappa}{3 V_0 a^2} \pi_a \mathring{\gamma}_{ij} \delta \tilde{\pi}^{ij} + \frac{1}{V_0 a^3} \pi_{\phi_0} \delta \tilde{\pi}_\phi + \sqrt{\Omega} a^3 \frac{\partial V}{\partial \phi} \delta \phi - \frac{\sqrt{\Omega}}{2 \kappa} a^3 (\mathring{D}^i \mathring{D}^j - \mathring{\gamma}^{ij} \mathring{D}_k \mathring{D}^k) \delta \gamma_{ij} \, ,
\label{eq:constraint-linear-m}
\end{align}
\end{widetext}
where $\mathring{D}^i = \mathring{\gamma}^{ij} \mathring{D}_j$, and we expressed the zeroth-order momenta in terms of background quantities using Eq.~\eqref{eq:momentum-densities-bg}. For scalar modes, the linear perturbations of the metric and conjugate momentum are given by Eq.~\eqref{eq:metric-perturbations-mode-expansion} with the sum over $\alpha$ restricted to $\alpha=1,2$. We will omit the subscripts $nlm$ in the amplitudes of the perturbations. Inserting the normal mode expansions~\eqref{eq:field-perturbations-mode-expansion} and \eqref{eq:metric-perturbations-mode-expansion} into Eq.~\eqref{eq:constraint-linear-m}, and writing the scalar modes in the explicit forms given in Eq.~\eqref{eq:metric-scalar-modes}, the scalar constraint decomposes into a set of decoupled normal modes,
\[
\mathcal{H}^{(1)} = \sum_{n \neq 1} \sum_{lm} \mathcal{H}^{(1)}_{n l m} \, ,
\]
given by
\begin{align*}
\frac{\mathcal{H}^{(1)}_{n l m}}{\sqrt{\Omega}} &= \left[ - \frac{\kappa \pi_a^2}{72 V_0^2 a^3 }  - \frac{\pi_{\phi_0}^2}{4 V_0^2 a^5} + \frac{a}{2}V(\phi_0) + \frac{2 n^2-5}{6 \kappa a}\right] \gamma^1 \\
	& \quad + \frac{n^2-4}{3 \kappa a} \gamma^2 -  \frac{\kappa \pi_a}{V_0} \pi^1 + \frac{\pi_{\phi_0}}{V_0 a^3} \pi^f + a^3 V'(\phi_0) f \, .
\end{align*}
In the spatially spherical gauge, $\gamma^1=\gamma^2=0$, and the normal modes of the scalar constraint reduce to the form:
\[
\mathcal{H}^{(1)}_{n l m} = \sqrt{\Omega} \left[ - \frac{\kappa \pi_a}{V_0} \pi^1 + \frac{\pi_{\phi_0}}{V_0 a^3} \pi^f + a^3 V'(\phi_0) f \right] \, .
\]
Imposing this constraint, we can fix the momentum perturbation $\pi^1$ in terms of the other linear perturbations,
\be
\pi^1 = \frac{1}{\kappa a^3} \frac{\pi_{\phi_0}}{\pi_a}\pi^f + \frac{V_0 a^3 V'(\phi_0)}{\kappa \pi_a} f \, .
\label{eq:fix-pi1}
\ee

Similarly, the first-order contribution to the diffeomorphism constraint \eqref{eq:constraint-diff} reads:
\begin{align*}
\mathcal{H}_i^{(1)} &= \frac{\sqrt{\Omega}}{V_0} \frac{\Omega^{jk}}{6a} \pi_a ( \mathring{D}_i \delta \gamma_{jk} - 2 \mathring{D}_j \delta \gamma_{ik} ) \\
	& \quad - 2 \mathring{\gamma}_{ki} \mathring{D}_j \delta \tilde{\pi}^{jk} + \frac{\sqrt{\Omega}}{V_0} \pi_{\phi_0} \mathring{D}_i \delta \phi \, ,
\end{align*}
where we have again expressed the zeroth-order momenta in terms of background quantities using the equations~\eqref{eq:momentum-densities-bg}. For scalar perturbations, the linearized constraint has the normal mode expansion:
\[
\mathcal{H}_i^{(1)} = \sum_{n \neq 1} \sum_{l m} \mathcal{H}^{(1)}_{i,n l m} \, ,
\]
with
\begin{align*}
\mathcal{H}^{(1)}_{i,n l m} &= \sqrt{\Omega} \left[ - 2 a^2 (\pi^1-\pi^2) \right.\\
& \quad + \left. \frac{\pi_a}{18 V_0 a} \left( \gamma^1 - 4 \frac{n^2-4}{n^2-1} \gamma^2 \right) + \frac{\pi_{\phi_0}}{V_0} f \right] \bar{D}_i Q_{n l m} .
\end{align*}
In the spatially spherical gauge,
\[
\mathcal{H}^{(1)}_{i,n l m} = \sqrt{\Omega} \left[ - 2 a^2 (\pi^1-\pi^2) + \frac{\pi_{\phi_0}}{V_0} f \right] \bar{D}_i Q_{n l m} \, .
\]
The expression within the brackets is a scalar constraint that implements the linearized diffeomorphism constraint $\mathcal{H}^{(1)}_{i,n l m}$. It can be used to fix the momentum $\pi^2$ in terms of the other linear perturbations. Together with the formula~\eqref{eq:fix-pi1} for $\pi^1$, it leads to:
\be
\pi^2 = \frac{\pi_{\phi_0}}{\kappa a^3 \pi_a} \pi^f + \left( \frac{V_0 a^3 V'(\phi_0)}{\kappa \pi_a} - \frac{\pi_{\phi_0}}{2 V_0 a^2} \right) f \, .
\label{eq:fix-pi2}
\ee

\subsubsection{Quadratic terms in the Hamiltonian constraint}

The quadratic terms in the scalar constraint \eqref{eq:constraint-scalar} are given in the spatially spherical gauge by:
\begin{align*}
\mathcal{H}^{(2)} &= \frac{2 \kappa a^3}{\sqrt{\Omega}} \left( \Omega_{ir} \Omega_{js} \delta \tilde{\pi}^{ij} \delta \tilde{\pi}^{rs} - \frac{1}{2} \Omega_{ij} \Omega_{rs} \delta \tilde{\pi}^{ij} \delta \tilde{\pi}^{rs} \right) \\ 
	& \quad + \frac{(\delta \tilde{\pi}_\phi)^2}{2 a^3 \sqrt{\Omega}} 
+ \frac{a^3 \sqrt{\Omega} V''(\phi_0)}{2} (\delta \phi)^2 \\
& \quad + \frac{a \sqrt{\Omega}}{2} \Omega^{ij} \partial_i \delta \phi \partial_j \delta \phi \, .
\end{align*}
Inserting the expansion of the perturbations in terms of hyperspherical harmonics and integrating on the $3$-sphere, we obtain:
\begin{multline*}
\int \dd^3 x \, \mathcal{H}^{(2)} = \sum_{n \neq 1} \sum_{l m} \left\{ -3 \kappa a \left[ (\pi^1)^2 - \frac{n^2 -1}{n^2-4} (\pi^2)^2  \right] \right. \\
	+ \left. \frac{1}{2 a^3} (\pi^f)^2 + \frac{a^3 V''(\phi_0)}{2} f^2 + \frac{a}{2} (n^2-1) f^2 \right\} \, .
\end{multline*}

\subsubsection{Quadratic Hamiltonian for the perturbations}

Substituting the formulas derived for the linearized constraints and for the quadratic scalar constraint into the quadratic Hamiltonian \eqref{eq:quadratic-adm}, it decomposes into a sum of decoupled normal modes,
\[
H^{(2)} = \sum_{n \neq 1} \sum_{l m} H^{(2)}_{n l m} \, ,
\]
expressed in terms of the amplitudes of the normal modes of the perturbations as
\begin{multline}
H^{(2)}_{n l m} = N_0 \left\{ -3 \kappa a \left[ (\pi^1)^2 - \frac{n^2 -1}{n^2-4} (\pi^2)^2  \right] \right. \\
	+ \left. \frac{1}{2 a^3} (\pi^f)^2 + \frac{a^3 V''(\phi_0)}{2} f^2 + \frac{a}{2} (n^2-1) f^2 \right\} \\
	+ g \left[ -\frac{\kappa \pi_a}{V_0} \pi^1 + \frac{\pi_{\phi_0}}{V_0 a^3} \pi^f + a^3 V '(\phi_0) f  \right] \\
	+ (n^2-1) k \left[ -2a^2 (\pi^1 - \pi^2) + \frac{\pi_{\phi_0}}{V_0} f\right] \, .
\label{eq:quadratic-adm-1}
\end{multline}
Imposing the consistency requirement that the gauge condition is preserved by the evolution under $H^{(2)}$:
\[
\{ \gamma^1,H^{(2)}\} = \{\gamma^2,H^{(2)}\} = 0 \, ,
\]
the amplitudes $g$ and $k$ of the perturbations of the lapse function and shift vector are fixed as
\begin{align}
k &= - \frac{3 \kappa N_0}{a(n^2-4)} \pi^2 \, , 
\label{eq:fix-k}
\\
g &= -6 N_0 V_0 \frac{a}{\pi_a} \left( \pi^1 - \frac{n^2-1}{n^2-4} \pi^2 \right) \, .
\label{eq:fix-g}
\end{align}
It can also be checked that $\{\gamma^1,H^{(1)}\} = \{\gamma^2,H^{(1)}\} = 0$.

Substituting the formulas \eqref{eq:fix-pi1}, \eqref{eq:fix-pi2}, \eqref{eq:fix-k} and \eqref{eq:fix-k} that express the perturbations $\pi^1, \pi^2,k,g$ in terms of amplitudes of field perturbations in Eq.~\eqref{eq:quadratic-adm-1}, we obtain an expression of the form
\[
H^{(2)}_{n l m} = \alpha f^2 + \beta f \pi^f + \gamma (\pi^f)^2 \, ,
\]
where the coefficients $\alpha, \beta,\gamma$ depend only on background quantities. This Hamiltonian can be brought into a diagonal form by the application of the canonical transformation
\[
\pi_T^f = \pi^f + \frac{\beta}{2 \gamma} f \, , \qquad f_T = f \, .
\]
After this transformation, the quadratic Hamiltonian is expressed in terms of the new canonical variables, which for simplicity we rename as $\pi_T^f \to \pi^f$ and $f_T \to f$, as:
\be
H^{(2)}_{n l m} =  \frac{c_1}{2} (\pi^f)^2 + \frac{c_2}{2} f^2  \, ,
\label{eq:quadratic-hamiltonian}
\ee
with
\begin{widetext}
\begin{align}
c_1 &= N_0 \left[ \frac{1}{a^3} + \frac{18 \pi_{\phi_0}^2}{(n^2-4)\kappa a^5 \pi_a^2} \right] \, ,
\label{eq:quadratic-hamiltonian-coeff1}
 \\
c_2 &= N_0 \left\{ a(n^2-1) + a^3 V''(\phi_0)-\frac{6 a^7 (V')^2 V_0^2}{\kappa \pi_a^2} + 6 \kappa a \frac{n^2-1}{n^2-4} \left( \frac{a^3 V' V_0}{\kappa \pi_a} - \frac{\pi_{\phi_0}}{2 a^2 V_0} \right)^2\right.
\nonumber \\
& \quad \left. - \left[\frac{3}{a^4} \frac{n^2-1}{n^2-4} \frac{\pi_{\phi_0}}{\pi_a} - \frac{18 a V_0^2 V'}{\kappa \pi_a^2 (n^2-4)} \right]^2 \left[ \frac{V_0^2}{a^3 \pi_{\phi_0}^2 } + \frac{18 V_0^2}{(n^2-4)\kappa a^5 \pi_a^2}\right]^{-1} \right\} \, .
\label{eq:quadratic-hamiltonian-coeff2}
\end{align}
\end{widetext}

The equations of motion are then simply given by:
\begin{align}
\dot{f} &= \{f,H^{(2)}_{n l m}\} = c_1 \pi^f \, , \\
\dot{\pi}^f &= \{\pi^f,H^{(2)}_{n l m}\} = - c_2 f \, ,
\end{align}
which can be combined in a second-order equation for the amplitude $f$:
\be
\ddot{f} - \frac{\dot{c}_1}{c_1} \dot{f} + c_1 c_2 f = 0 \, .
\label{eq:eom-amplitudes}
\ee

Our derivation of the quadratic Hamiltonian for scalar perturbations follows the steps employed in the flat case in \cite{Agullo:2018bs}, adapted to the case of a closed universe. In that work, the Hamiltonian is expanded up to third-order. We verified that the flat limit of the Hamiltonian \eqref{eq:quadratic-hamiltonian} corresponds to the Hamiltonian for scalar perturbations on a perturbed flat universe obtained following the methods of \cite{Agullo:2018bs}, restricting to an expansion up to second order, as discussed in Appendix \ref{sec:flat-limit}.

\section{Effective dynamics in LQC}
\label{sec:lqc-effective-dynamics}

Models of loop quantum cosmology are obtained through the application of loop techniques for the quantization of the gravitational field in symmetry-reduced cosmological models \cite{Ashtekar:2011ni,Agullo:2016tjh}. For instance, restricting to isotropic and homogeneous spacetimes, the spatial geometry is characterized by a single degree of freedom---the scale factor---, rather than by the infinite number of degrees of freedom of a generic configuration of the geometry. An additional finite number of degrees of freedom describes the matter content of the model, consisting of homogeneous distributions of distinct matter components. Under this severe simplification, the system formed by the gravitational field coupled to the matter components can be quantized, and in LQC this is done so that the quantization of the geometry mimics the main features of LQG \cite{Ashtekar:2011ni}. 

In the simplest models of LQC, one considers a universe filled by a single scalar field \cite{Ashtekar:2011ni,Agullo:2016tjh}. After quantization, the background is described by a wavefunction $\Psi(a,\phi_0)$, with a time-evolution described by a quantum Hamiltonian constraint. Considering a state sharply peaked at a configuration describing an expanding classical solution at late times, and evolving it towards the past, it was found that the mean geometry closely follows the classical evolution while far from the Planck scale. When the evolution approaches what would be the Big Bang in the classical dynamics, however, quantum corrections become relevant. The mean geometry then deviates from the classical trajectory, following a modified Friedmann equation. In this effective dynamics, the Big Bang singularity is resolved, being replaced by a Big Bounce. The resolution of the initial singularity is a robust prediction of LQC, obtained for a variety of cosmological models \cite{Ashtekar:2011ni,Agullo:2016tjh}.

One can also quantize linear cosmological perturbations together with the background quantities, as done, for instance, in the ``dressed metric approach'' \cite{Agullo:2012fc} based on the framework of quantum field theory on quantum cosmological spacetimes introduced in \cite{Ashtekar:2009mb} (alternative strategies are explored in LQC; for a review, see \cite{Agullo:2023rev,Agullo:2016tjh}). This approach provides an extension of cosmological perturbation theory to the Planck scale, and was first applied to the construction of a quantum gravity extension of an inflationary model in \cite{Agullo:2012sh,Agullo:2013ai}. In such an extension, quantum gravity effects were shown to introduce potentially observable corrections to the inflationary predictions for the primordial power spectra of cosmological perturbations. This allows the model to be tested with observations of the CMB. It was found that consistency with the observations requires the number of $e$-folds $N_B$ from the bounce until the onset of the observable part of inflation to be larger than a lower bound $N_{B,min} \sim 12$ \cite{Agullo:2015tca,Agullo:2018bs}. For large $N_B \gg N_{B,min}$, the quantum gravity corrections are redshifted to nonobservable scales. For $N_B \ll N_{B,min}$, they affect regions of the spectrum that are well described by the usual inflationary prediction, leading to inconsistencies with the observations. For $N_B \sim 12$, the corrections appear at scales that correspond to the CMB anomalies.

In the pioneering works \cite{Agullo:2012sh,Agullo:2013ai}, the case of a quadratic inflaton potential was considered. A Starobinsky potential, favored by the \planck{} observations, was later considered in \cite{Bonga:2015kaa,Bonga:2015xna}. Non-Gaussianities were analyzed for both potentials in \cite{Agullo:2018bs}. In these works, a flat universe was considered. Here, we are interested in the case of a closed universe and a Starobinsky potential. 

The starting point for the dressed metric approach is the usual theory of linear perturbations on a classical FLRW cosmological spacetime \cite{Mukhanov:1990me,Dodelson:2020,Weinberg:2008}. In a Hamiltonian approach, the system has a phase space $\Gamma = \Gamma_{hom} \times \Gamma_{pert}$, where $\Gamma_{hom}$ describes the background and $\Gamma_{pert}$ the linear perturbations. Expanding the ADM Hamiltonian up to second order in the perturbations, the zeroth-order terms describes the background dynamics, the first-order part can be used to eliminate redundant variables, and the quadratic terms generate the evolution of the physical linear perturbations, as described in the previous section for a closed universe. The case of a flat universe is discussed in detail in \cite{Agullo:2018bs}. Backreaction of the perturbations on the background is assumed to be negligible. The background can then be quantized as in the absence of the perturbations. 

In a flat universe, the linear perturbations are expanded in Fourier modes, and described by the amplitudes of an infinite set of time-dependent oscillators. The quadratic Hamiltonian that describes their evolution depends on the background variables $(a,\pi_a)$ and $(\phi_0, \pi_{\phi_0})$. As the perturbations now propagate on a quantum background, such quantities are represented by operators. Nevertheless, it turns out that the evolution of the perturbations on the quantum background is equivalent to that on a classical spacetime, but with a dressed metric $\tilde{g}$ computed from averages of background quantities and in this way incorporating quantum corrections. For sharply peaked states of the background, the dressed metric reduces to that described by the effective dynamics of the background. The end result is that the perturbations have the same equations of motion as in the classical theory, but with the background quantities $(a,\pi_a)$ and $(\phi_0, \pi_{\phi_0})$ computed in the effective Hamiltonian dynamics of the background.

For our analysis, we adopted a model that provides a natural extension to the case of a closed universe of the picture reached through the application of the dressed metric approach to the case of a flat universe. The effective dynamics of LQC on a closed universe was developed in \cite{Ashtekar:2006es}. In Sec.~\ref{sec:hamiltonian-formalism}, we derived the quadratic Hamiltonian and the equations of motion for scalar modes of the linear perturbations on a closed universe. In what follows, we analyze the evolution of such linear perturbations on the effective background described by LQC. Accordingly, the background quantities $(a,\pi_a)$ and $(\phi_0, \pi_{\phi_0})$ appearing in the equations of motion for the perturbations derived in Sec.~\ref{sec:hamiltonian-formalism} will be computed in the effective dynamics described in \cite{Ashtekar:2006es}, and not from the classical equations for the background. This gives a model that reproduces on a closed universe the main features of the dressed metric approach as applied for a flat universe.

In the next subsection, we briefly review the LQC effective dynamics of the homogeneous and isotropic sector of a closed universe. A detailed numerical analysis of the effective dynamics of the background is then presented in Sec.~\ref{sec:effective-background}, followed by an analysis of the evolution of the perturbations in Sec.~\ref{sec:evolution-perturbations}.

\subsection{Effective background dynamics}

The Hamiltonian formalism for the effective dynamics of the background geometry for a closed universe in loop quantum cosmology was developed in \cite{Ashtekar:2006es}, and is briefly reviewed in \cite{Gordon:2020gel}. An overview of the formalism is presented in this section. A canonical transformation is then implemented that leads to a description of the effective dynamics in terms of the scale factor and its conjugate momentum.

The effective dynamics of the background is described in \cite{Ashtekar:2006es,Gordon:2020gel} in terms of the geometrical variables
\[
v=|p|^{3/2}=2 \pi^2 a^3 \, , \qquad b=c |p|^{-1/2} = \frac{c}{V_0^{1/3} a} \, ,
\]
and the field variables $\phi_0,\pi_{\phi_0}$. The Poisson brackets read:
\[
\{b,v\} = 4 \pi G \gamma \, , \quad \{\phi_0,\pi_{\phi_0}\}=1 \, ,
\]
and the effective Hamiltonian is given by:
\begin{multline}
H_{\text{eff}} = -\frac{3 v}{8 \pi G \gamma^2\lambda^2}\left[ \sin^2(\lambda b-D) - \sin^2 D \right. \\
\left. + (1+\gamma^2) D^2 \right] + H_m \, ,
\label{eq:effective-hamiltonian-v}
\end{multline}
where
\[ 
D=\lambda \left( \frac{2\pi^2}{v} \right)^{1/3} \, ,
\]
and
\be
\lambda=\sqrt{\Delta}=2\sqrt{\sqrt{3} \pi \gamma}
\label{eq:lambda-def}
\ee
is the length scale associated with the area gap $\Delta=4\sqrt{3} \pi \gamma$ of loop quantum gravity, expressed in Planck units. The matter Hamiltonian for a scalar field is
\[
H_m = \frac{\pi_{\phi_0}^2}{2v}+vV(\phi_0) \, .
\]

The expansion of the universe under the effective Hamiltonian \eqref{eq:effective-hamiltonian-v} can be equivalently described in terms of the scale factor $a$, instead of in terms of the volume $v$. This is accomplished through the application of the canonical transformation:
\[
v=V_0 a^3 \, , \quad b= - \frac{4 \pi G \gamma}{3 V_0} \frac{\pi_a}{a^2} + \frac{1}{a} \, , 
\]
where $\pi_a$ is the canonical momentum associated with $a$,
\[
\{a, \pi_a\}=1 \, .
\]
The effective Hamiltonian is expressed in the new variables as:
\begin{multline*}
H_{\text{eff}} = -\frac{3}{8\pi G \gamma^2 \lambda^2} a^3 V_0 \left[ \sin^2\left( \frac{4 \pi G \gamma \lambda}{3 V_0} \frac{\pi_a}{a^2}\right) \right. \\
\left. - \sin^2\left( \frac{\lambda}{a}\right) + (1+\gamma^2) \frac{\lambda^2}{a^2} \right] + H_m \, ,
\end{multline*}
with
\[
H_m = \frac{\pi_{\phi_0}^2}{2 a^3 V_0} + a^3 V_0 V(\phi_0) \, .
\]

Computing the equations of motion for the configuration variables, we find
\begin{align}
\dot{a} 	&= - \frac{a}{2 \gamma \lambda} \sin\left( \frac{8 \pi G \gamma \lambda}{3 V_0} \frac{\pi_a}{a^2}\right) \, , 
\label{eq:eom-a-effective} \\
\dot{\phi_0} 	&= \frac{\pi_{\phi_0}}{a^3 V_0} \, .
\label{eq:eom-phi-effective} 
\end{align}
The matter Hamiltonian can then be expressed as:
\[
H_m= a^3 V_0 \rho \, , \qquad \rho = \frac{\dot{\phi_0}^2}{2} + V(\phi_0) \, .
\]

Introducing the critical density
\[
\rho_c=\frac{3}{8\pi G \gamma^2 \lambda^2} \, ,
\]
and
\[
\rho_{min} = \rho_c \left[ (1+\gamma^2) \frac{\lambda^2}{a^2} - \sin^2\left( \frac{\lambda}{a}\right) \right] \, ,
\]
the effective Hamiltonian assumes the form:
\[
H_{\text{eff}} = a^3 V_0 \left[ -\rho_c \sin^2\left( \frac{4 \pi G \gamma \lambda}{3 V_0} \frac{\pi_a}{a^2}\right) - \rho_{min} + \rho\right] \, .
\]
Imposing the Hamiltonian constraint, $H_{\text{eff}} =0$:
\be
\sin^2\left( \frac{4 \pi G \gamma \lambda}{3 V_0} \frac{\pi_a}{a^2}\right) = \frac{\rho - \rho_{min}}{\rho_c} \, .
\label{eq:hamiltonian-constraint-eff}
\ee
Combining the Eqs.~\eqref{eq:eom-a-effective} and \eqref{eq:hamiltonian-constraint-eff}, we obtain the effective Friedmann equation:
\be
H^2 = \frac{8\pi G}{3} (\rho - \rho_{min}) \left[ 1- \frac{\rho - \rho_{min}}{\rho_c} \right] \, .
\label{eq:modified-Friedmann}
\ee

Computing the equation of motion for the momentum associated with the scalar field, we find:
\[
\dot{\pi}_{\phi_0} = - a^3 V_0 V'(\phi_0) \, ,
\]
which, combined with Eq.~\eqref{eq:eom-phi-effective}, gives:
\[
\ddot{\phi}_0 + 3 H \dot{\phi}_0 + V'(\phi_0) = 0 \, .
\]
The Klein-Gordon equation implies the usual relation for the conservation of energy:
\[
\dot{\rho} = - 3H (\rho + P) \, .
\]

For the numerical implementation of the background dynamics, it will be convenient to use an acceleration equation for the scale factor, i.e., a second-order equation for $a(t)$ subjected to the constraint \eqref{eq:modified-Friedmann}. In order to determine such an equation, we can use the relation
\be
\frac{\ddot{a}}{a} = \dot{H} + H^2 \, .
\label{eq:dH2}
\ee
Introducing the quantity
\be
\zeta = \sin^2\left( \frac{\lambda}{a} \right) - \frac{\lambda}{a} \sin\left( \frac{\lambda}{a} \right) \cos\left( \frac{\lambda}{a} \right)\, ,
\ee
we can write:
\[
\dot{\rho}_{min} = -2H \rho_c \left( \zeta + \frac{\rho_{min}}{\rho_c} \right) \, .
\]
From:
\[
2 H \dot{H} = \frac{d}{dt} H^2 = \frac{8\pi G}{3} (\dot{\rho} - \dot{\rho}_{min}) \left[ 1- 2 \frac{(\rho - \rho_{min})}{\rho_c} \right] \, .
\]
it then follows that:
\begin{multline}
\dot{H} = \left[ - 4\pi G (\rho + P) + \frac{1}{\gamma^2 \lambda^2} \left( \zeta + \frac{\rho_{min}}{\rho_c} \right) \right] \\
\times \left[ 1- 2 \frac{(\rho - \rho_{min})}{\rho_c} \right] \, .
\label{eq:Hdot-effective}
\end{multline}
Inserting Eq.~\eqref{eq:Hdot-effective} and the modified Friedmann equation \eqref{eq:modified-Friedmann} into Eq.~\eqref{eq:dH2}, we obtain the modified acceleration equation:
\begin{multline}
\frac{\ddot{a}}{a} = \left[ - 4\pi G (\rho + P) + \frac{1}{\gamma^2 \lambda^2} \left( \zeta + \frac{\rho_{min}}{\rho_c} \right) \right] \\
	\times \left[ 1- 2 \frac{(\rho - \rho_{min})}{\rho_c} \right] 	\\
	+ \frac{8\pi G}{3} (\rho - \rho_{min}) \left[ 1- \frac{\rho - \rho_{min}}{\rho_c} \right] \, .
\label{eq:modified-acceleration-eq}
\end{multline}

The classical limit of the effective dynamics is obtained by letting the length scale associated with the area gap become negligibly small, $\lambda \to 0$. From Eq.~\eqref{eq:lambda-def}, the Immirzi parameter must then also vanish as a term of order $\mathcal{O}(\lambda^2)$. In this limit, the effective Hamiltonian reduces to:
\[
H_{\text{eff}} \simeq - \frac{2\pi G}{3 aV_0} \pi_a^2 - \frac{3}{8 \pi G}a V_0 + \frac{\pi_{\phi_0}^2}{2 a^3 V_0} + a^3 V_0 V(\phi_0) \, ,
\]
which exactly agrees with the classical background Hamiltonian $H^{(0)}$ given in Eq.~\eqref{eq:hamiltonian-constraint-zeroth}, with $N_0=1$.

\section{Inflationary and pre-inflationary evolution of the background}
\label{sec:effective-background}

\subsection{Slow-roll inflation}

The inflationary and pre-inflationary classical dynamics of the background geometry and scalar field are determined by the Friedmann or the acceleration equation, Eqs.~\eqref{eq:friedmann-h0} and \eqref{eq:acceleration-eq}, together with the Klein-Gordon equation, given in Eq.~\eqref{eq:klein-gordon}. Choosing the zeroth-order lapse function as $N_0=1$, the background equations read:
\begin{align}
&H^2 = \frac{8 \pi G}{3} \rho - \frac{1}{a^2} \, , 
\label{eq:friedmann-proper-time} \\
&\frac{\ddot{a}}{a} = - \frac{4 \pi G}{3} (\rho + 3P) \, , 
\label{eq:acceleration-proper-time}
\\
&\ddot{\phi}_0 + 3 \frac{\dot{a}}{a} \dot{\phi}_0 + V'(\phi_0) = 0 
\label{eq:kg-proper-time}\, ,
\end{align}
where $\rho$ and $P$ are the energy density and pressure of the background scalar field,
\begin{equation}
\rho = \frac{\dot{\phi}_0^2}{2} + V(\phi_0) \, , \quad P = \frac{\dot{\phi}_0^2}{2} - V(\phi_0) \, .
\label{eq:rho-P-scalar-field}
\end{equation}
We adopted the Starobinsky model to describe the inflationary and pre-inflationary dynamics. The Starobinsky potential is given by:
\begin{equation}
V(\phi_0) = \frac{3 m^2}{32 \pi G} \left[ 1- \exp\left( - \sqrt{\frac{16 \pi G}{3}} \phi_0 \right) \right]^2 \, .
\end{equation}

The regime of slow-roll inflation is characterized by the conditions:
\be
\epsilon \ll 1 \, , \qquad |\delta| \ll 1 \, ,
\label{eq:sr-conditions}
\ee
where the slow-roll parameters are defined by:
\be
\epsilon = -\frac{\dot{H}}{H^2} \, , \qquad  \delta =  \frac{\ddot{\phi}_0}{H \dot{\phi_0}} \, .
\label{eq:12sr-def}
\ee

In the presence of spatial curvature, the curvature term $1/a^2$ in the Friedmann equation \eqref{eq:friedmann-proper-time} becomes increasingly important towards the past, but it is negligible during the slow-roll inflationary regime. The equations of motion can then be approximated by:
\begin{align}
H^2 & \simeq \frac{8\pi G}{3}\rho \simeq \frac{8\pi G}{3} V(\phi_0) \, , 
\label{eq:friedmann-sr}
\\
\dot{H} &\simeq - 4 \pi G \dot{\phi}_0^2 \, , 
\label{eq:hdot-sr}\\
3 H \dot{\phi}_0 &\simeq - V'(\phi_0) \, . 
\label{eq:kg-sr}
\end{align}
Inserting the background equations \eqref{eq:friedmann-sr} and \eqref{eq:hdot-sr} in the definition \eqref{eq:12sr-def} of the first slow-roll potential and simplifying, we find:
\be
(\epsilon -3) \dot{\phi}_0^2 + 2 \epsilon V(\phi_0) = 0 \, .
\label{eq:phi-epsilon}
\ee
Slow-roll inflation lasts while the conditions \eqref{eq:sr-conditions} are satisfied. The onset and end of inflation are instants of time $t_{on}$ and $t_{end}$ at which $\epsilon=1$.

The number of $e$-folds elapsed from an initial time $t_i$ until some time $t_f$ is defined as
\begin{align*}
N_e(t) &= \log\left( \frac{a(t_f)}{a(t_i)} \right) \\
	&= \int_{t_i}^{t_f} H(t') dt' \, .
\end{align*}
Using the Klein-Gordon equation to change the integration variable from $dt$ to $d\phi_0$, and invoking the Friedmann equation, we find that:
\be
N_e = - 8 \pi G\int_{\phi_i}^{\phi_f} \frac{V}{V'} d\phi_0 \, .
\label{eq:duration-inflation}
\ee
For the Starobinsky potential:
\be
N_e = \sqrt{3 \pi G} (\phi_f - \phi_i) - \frac{3}{4} \left[ e^{\sqrt{16 \pi G/3}\, \phi_f} - e^{\sqrt{16 \pi G/3}\, \phi_i} \right] \, .
\label{eq:duration-inf-starobinsky}
\ee

During inflation, the first slow-roll parameter admits the approximation:
\be
\epsilon \simeq \epsilon_V = \frac{1}{16 \pi G} \frac{V^2_\phi}{V^2} = 
\frac{4}{3} \left( \frac{e^{-\sqrt{16 \pi G/3}\, \phi_0}}{1-e^{-\sqrt{16 \pi G/3}\, \phi_0}} \right)^2  \, ,
\label{eq:approx-epsilon}
\ee
By fixing the value of the field at the end of inflation through the condition $\epsilon_V = 1$,
we obtain the estimate:
\[
\phi_f = \log \left( 1+\frac{2}{\sqrt{3}} \right) \sqrt{\frac{3}{16 \pi G}} = 0.187 m_{pl} \, .
\]
Inserting this value in Eq.~\eqref{eq:duration-inf-starobinsky}, we find
\be
N_e \simeq -1.04 + \frac{3}{4} e^{\sqrt{16 \pi G/3}\, \phi_i} - \sqrt{3 \pi G} \phi_i \, ,
\label{eq:number-efolds}
\ee
which express the number of $e$-folds until the end of the inflationary regime in terms of the initial configuration of the background scalar field.

\subsection{Initial conditions from Planck data}
\label{sec:initial-conditions}

We are interested in backgrounds with a slow-roll inflationary regime compatible with the constraints on inflation set by the CMB observations by \planck{}~\cite{Planck:2018jri}. In order to construct such backgrounds, we first translate best estimates for inflationary parameters derived from the CMB observations into initial conditions for the background. In this section, we describe how this was done. In the next section, we numerically determine the evolution of the background, varying the spatial curvature and the initial conditions within uncertainties.

The base $\Lambda$CDM model has six parameters, and estimates for such cosmological parameters were derived by the \planck{} Collaboration \cite{Planck:2018vyg}. Two out of the six parameters characterize the primordial cosmological perturbations. More concretely, the primordial power spectrum of comoving curvature perturbations is assumed to have the power-law form:
\be
P_\mathcal{R}(k) = A_s \left( \frac{k}{k_*} \right)^{n_s-1} \, ,
\label{eq:tilted-spectrum}
\ee
with two free parameters: the scalar power spectrum amplitude $A_s$, and the scalar spectral index $n_s$. The variable $k$ is the comoving wavenumber, and $k_*$ is an arbitrary pivot scale, which in the \planck{} papers is set equal to $k_*^P=0.05 \text{ Mpc}^{-1}$, except in discussions related to inflation, where  $k_*=0.002 \text{ Mpc}^{-1}$ is adopted. The primordial parameters of the $\Lambda$CDM model are constrained by the \planck{} observations to the windows:
\begin{align}
A_s &= (2.10 \pm 0.03) \times 10^{-9} \, ,
\nonumber \\
n_s &= 0.9649 \pm 0.0042 \, ,
\label{eq:planck-estimate}
\end{align}
with respect to the pivot mode $k_*^{P}=0.05 \text{ Mpc}^{-1}$.

For a given choice of an inflaton potential, observational bounds can also be set on the duration of the inflationary regime, as measured by the number $N_*$ of $e$-folds from the horizon crossing for the inflationary pivot mode $k_*=0.002 \text{ Mpc}^{-1}$ until the end of inflation \cite{Planck:2018jri}. For the Starobinsky model, the duration of inflation is constrained to the range:
\be
49 < N_* < 59 \, ,
\ee
at a Confidence Level of $95\%$.

In the base $\Lambda$CDM model, the spatial curvature is set to zero. Allowing for a nonvanishing spatial curvature, one obtains a simple extension of the model with one extra parameter $\Omega_k$. The cases of $\Omega_k$ negative, positive and zero correspond to closed, open and flat universes, respectively. In a closed universe,
\be
\Omega_k = - \frac{c^2}{(r_0 H_0)^2} \, ,
\ee
where $r_0=a(t_0)$ describes the radius of the universe at the present. The value of $\Omega_k$ fixes the size of the universe, once $H_0$ is measured. The observational bound provided by \planck{}, obtained under the assumption of a primordial power spectrum given by the power-law spectrum \eqref{eq:tilted-spectrum} and considering the dynamics of the post-inflationary regime, is given by \cite{Planck:2018vyg}:
\be
\Omega_k = 0.001 \pm 0.002 \, .
\ee
Accordingly, we restrict to $|\Omega_k| \le 0.005$. This summarizes the observational input required for our analysis.

Let us now discuss the determination of initial conditions for the background. We fix them at the initial time $t_*$ when the inflationary pivot mode $k_*=0.002 \text{ Mpc}^{-1}$ crosses the Hubble horizon. We assume that this happens in a regime of slow-roll inflation with a negligible spatial curvature. We can then use the slow-roll approximations discussed in the previous section.

We first note that the pivot scale introduced in \eqref{eq:tilted-spectrum} is arbitrary:
If another pivot scale $k_*$ is adopted, the amplitude is simply redefined according to:
\be
A_{s*}= \left( \frac{k_*}{k_*^P} \right)^{n_s-1} A_s \, .
\ee

The inflationary predictions for the parameters $A_{s*}$ and $n_s$ in terms of background quantities are given by:
\begin{align}
A_{s*} &= \frac{G H_*^2}{\pi \epsilon} \, ,  
	\label{eq:inflationary-As} \\
n_s &= 1-4 \epsilon - 2 \delta \, .
	\label{eq:inflationary-ns}
\end{align}
where $H_*$ is the Hubble parameter at the time $t_*$ when the pivot mode $k_*$ leaves the horizon during inflation. The prediction for the number of $e$-folds $N_*$ is given by Eq.~\eqref{eq:number-efolds}, with $\phi_i=\phi_*$:
\be
N_* \simeq -1.04 + \frac{3}{4} e^{\sqrt{16 \pi G/3}\, \phi_*} - \sqrt{3 \pi G} \phi_* \, .
\label{eq:number-efolds-star}
\ee

We fix the parameters $A_s,n_s$ as equal to the best estimates \eqref{eq:planck-estimate} of \planck, and set $N_*=54$, at the center of the allowed range. This is sufficient for the determination of the initial conditions:
\[
a_*=a(t_*), \quad \phi_*=\phi_0(t_*), \quad \dot{\phi}_*=\dot{\phi}_0(t_*)\, .
\]
Let us first determine $a_*$. The dimensionless scale factor at the time when $k_*$ crosses the horizon,
\be
\tilde{a}_*=\frac{a_*}{r_0} \, ,
\label{eq:tilde-a}
\ee
is fixed by the horizon crossing condition:
\be
\tilde{a}_* H_*  = k_* \, .
\label{eq:horizon-crossing}
\ee
Combining Eqs.~\eqref{eq:inflationary-As}, \eqref{eq:inflationary-ns}, \eqref{eq:tilde-a} and \eqref{eq:horizon-crossing}, we obtain:
\be
a_* = 2 r_0 k_* \sqrt{\frac{G}{\pi A_{s*} (1-n_s)}} \, .
\label{eq:init-cond-1}
\ee
The initial condition $\phi_*$ is directly fixed by Eq.~\eqref{eq:number-efolds-star}. Now, for $\phi_0=\phi_*$ and $H=H_*(A_{s*},\epsilon)$ (using Eq.~\eqref{eq:inflationary-As}), the Eqs.~\eqref{eq:friedmann-sr}, \eqref{eq:kg-sr} and \eqref{eq:phi-epsilon} form a system of equations that can be numerically solved for $m$, $\dot{\phi}_0=\dot{\phi}_*$ and $\epsilon$. The slow-roll parameter $\delta$ is then determined by Eq.~\eqref{eq:inflationary-ns}. This completes the determination of the initial conditions $a_*$, $\phi_*$ and $\dot{\phi}_*$, together with the mass $m$ of the Starobinsky potential and the slow-roll parameters $\delta$, $\epsilon$.

The procedure described above can be implemented at any time during slow-roll inflation. Any mode $k_*$ that crosses the horizon during inflation can be chosen as a pivot mode, and it will be convenient for our purposes to consider nonstandard pivots. As the number of inflationary $e$-folds $N_*$ is measured from $t_*$ until the end of inflation, however, it is necessary to correct it under a change of the pivot mode from $k_*$ to $k_*'$. During slow-roll, the Hubble parameter is approximately constant. The horizon crossing condition \eqref{eq:horizon-crossing} then implies:
\be
\frac{\tilde{a}_*}{k_*} \simeq \frac{\tilde{a}'_*}{k'_*} \, ,
\ee
so that, for $k'_* < k_*$, there is an additional number of $e$-folds given by:
\be
\Delta N_e = \log\left( \frac{k_*}{k'_*} \right) \, .
\ee
The scalar amplitude must also be computed at the new pivot mode.

Our determination of initial conditions involves two main steps. We first choose a pivot mode $k^{(1)}_*< k_*$, and determine initial conditions for a flat background at the time $t_1$ when it crosses the horizon. We then translate the initial conditions to the time $t_*$ at which the mode $k_*=0.002 \text{ Mpc}^{-1}$ crosses the horizon, by integrating the flat background evolution from $t_1$ until $t_*$ and then reading the values of $a_*=a(t_*)$, $\phi_*=\phi(t_*)$ and $\dot{\phi}_*=\dot{\phi}(t_*)$ at the end of the numerical evolution.

The reason why we first determine initial conditions at $t_1$ and then translate them to $t_*$ is the following. The direct application of the described procedure to $t_*$ produces initial conditions that determine a background with approximately $54$ $e$-folds of slow-roll inflation for $t>t_*$. Integrating towards the past, however, only a few $e$-folds of slow-roll are obtained, producing a short inflationary regime. Longer inflationary regimes can be obtained by varying the observational input given by the parameters $A_s$, $n_s$ and $N_*$ within error bars, but not in a systematic way. For the analysis of the effects of spatial curvature, however, there is no reason to restrict to short inflationary regimes, and we wished to consider inflationary regimes of arbitrary durations. This was achieved by manipulating the choice of the auxiliary pivot mode $k^{(1)}_*$. Decreasing the value of $k^{(1)}_*$, initial conditions associated with longer inflationary regimes in the flat approximation are obtained, and this can be done in a controlled way. In the earlier time when this new pivot crosses the horizon, the flat approximation may not hold for the full window of values of $\Omega_k$ of interest, however. For this reason, after determining initial conditions at an earlier time $t_1$ under a flat approximation, we translate them back to the later time $t_*$ at which the flat approximation is valid for all values of $\Omega_k$ of interest.

\subsection{Numerical evolution of the effective background}
\label{sec:background-evolution}

The effective background evolution is determined numerically by integrating the system formed by the quantum acceleration equation, Eq.~\eqref{eq:modified-acceleration-eq}, and the Klein-Gordon equation, Eq.~\eqref{eq:kg-proper-time}. Initial conditions were always specified during slow-roll inflation at the time $t_*$. For $t>t_*$, all backgrounds considered have similar evolutions, satisfying the constraints on inflation set by \planck{}. This is necessary, since backgrounds with a distinct evolution for $t>t_*$ would violate known constraints on inflation, producing modifications in the primordial power spectrum for modes that cross the horizon after $t_*$, at scales where the observations are well described by the usual power-law spectrum. In the window of observable modes that cross the horizon before $t_*$, on the other hand, anomalies that are not accurately described by a power-law spectrum are observed, and modifications in the background for such earlier times cannot be ruled out on the same grounds. We thus allow for deviations from slow-roll inflation for $t<t_*$. The adopted strategy thus consists of starting the integration of the background from an era that is well understood, i.e., the slow-roll inflationary regime, and then determining the possibilities for the past evolution within the uncertainties of the inflationary parameters.

Spatial curvature is negligible for $t>t_*$, but its effects are amplified towards the past, and small differences in the value of $\Omega_k$ can produce very different pre-inflationary regimes. Small changes in the initial conditions at $t_*$ can also lead to markedly distinct behaviors for $t<t_*$. In a flat universe, the duration of inflation can be changed in a systematic way from changes in the initial conditions, as discussed in the previous section, being determined in our approach by the choice of $k^{(1)}_*$. In a closed universe, the parameter $k^{(1)}_*$ can still be expected to be related to the duration of inflation, but an interplay with the spatial curvature should also be expected. In our analysis, we allowed for variations of both $\Omega_k$ and $k^{(1)}_*$, and determined the possible types of  inflationary and pre-inflationary behavior of the background for $t<t_*$ in a closed universe. 

\begin{figure}
\hspace{-6pt}\includegraphics[scale=.6]{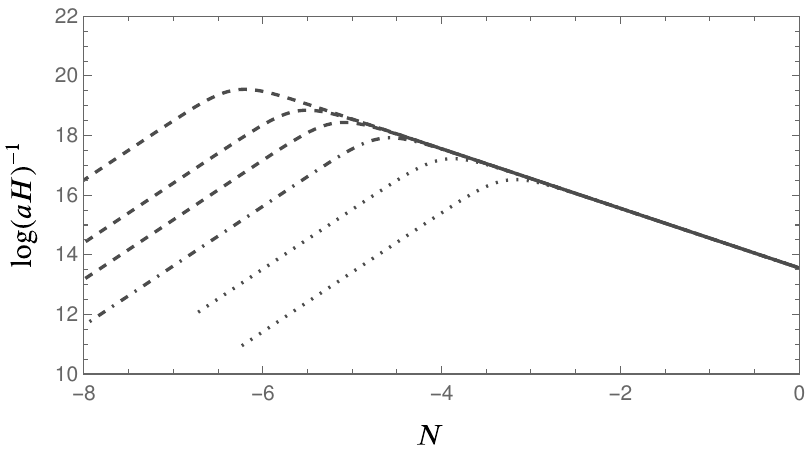} \\
\hspace{-6pt}\includegraphics[scale=.6]{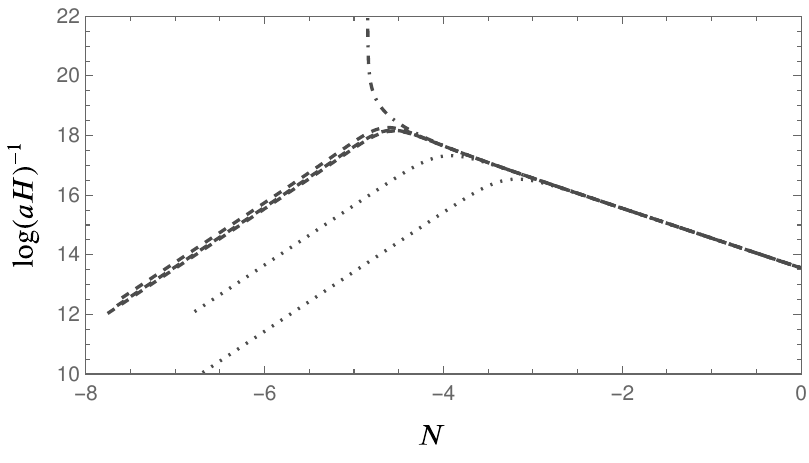}
\caption{Effective background dynamics for a flat universe (\emph{top}) and a closed universe with $\Omega_k=-0.005$ (\emph{bottom}), around the onset of inflation. The evolution of the Hubble horizon is represented in terms of the number $N$ of $e$-folds from $t_*$, for six distinct initial conditions specified at $t_*$.}
\label{fig:onset-varying-Omega}
\end{figure}

We started our analysis by determining how the change of the initial conditions affected the past evolution of the background for a fixed spatial curvature. For that purpose, we first determined a set of initial conditions leading to inflationary regimes of distinct durations for a flat universe, and then compared the background evolution in the presence and in the absence of spatial curvature. The results are displayed in Fig.~\ref{fig:onset-varying-Omega}, where we compare a flat universe and a closed universe with $\Omega_k = -0.005$. Background evolutions for six distinct initial conditions, obtained for $k^{(1)}_* \in (10^{-4} \text{ Mpc}^{-1}, 0.002 \text{ Mpc}^{-1})$, are represented. The slow-roll inflationary regime corresponds to the linearly decreasing behavior, reaching the initial time $t_*$ at $N=0$. All displayed evolutions agree in the inflationary regime. With respect to the onset of inflation, three distinct behaviors are observed, which are represented by dotted, dot-dashed and dashed lines, and will be called type I, II and III, respectively.

The two evolutions represented in dotted lines (type~I) are those with the smallest durations of the inflationary regime. For these evolutions, the presence of spatial curvature does not affect the background evolution. The integrated evolution is approximately the same for the flat and closed universes, and the duration of inflation is solely determined by the choice of the initial conditions, with the spatial curvature playing no significant role.

The evolutions represented in dashed lines (type III) are those with the longer inflationary regimes in the flat evolution. For these backgrounds, the onset of inflation occurs approximately at the same time in the closed universe dynamics for all initial conditions, regardless of how the duration of inflation differed for them in a flat universe. This type of background clearly displays strong effects of the spatial curvature, which determines the duration of inflation.

A background of type II is represented by the dot-dashed line in Fig.~\ref{fig:onset-varying-Omega}. In a flat universe, the initial conditions produce a background that has an inflationary regime with an intermediate duration with respect to the two previous cases. In a closed universe, a markedly distinct behavior is met. Towards the past, the Hubble horizon diverges before the inflationary regime, in contrast with the other backgrounds, in which it decreases towards the past. The Hubble parameter vanishes before the onset of inflation, which indicates the presence of a bounce. That a bounce is met is not surprising, as a bounce is generically observed in the effective dynamics of LQC; in this case, however, the bounce is not driven by quantum effects, but by the spatial curvature.

\begin{figure}
\includegraphics[scale=.56]{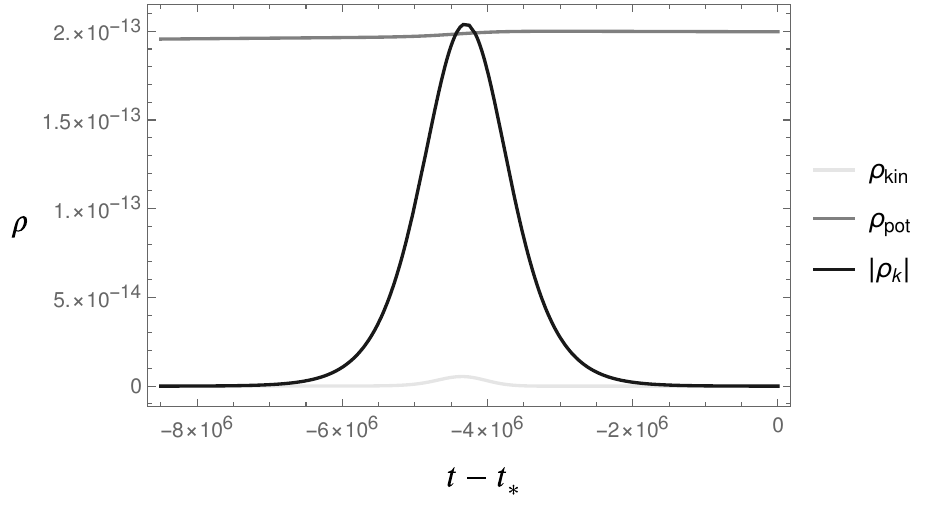}
\caption{Classical bounce: energy balance. Contributions from kinetic energy ($\dot{\phi}_0^2/2$), potential energy ($V(\phi_0)$) and spatial curvature ($\rho_k$) to the Friedmann equation, around the bounce, in Planck units. Background dynamics determined for initial conditions obtained with $k^{(1)}_*=0.0005 \text{ Mpc}^{-1}$ and $\Omega_k=-0.005$.}
\label{fig:energy-balance-2}
\end{figure}

This can be seen as follows. First, we checked that the effective LQC dynamics and the classical dynamics for the initial conditions corresponding to the dot-dashed line in Fig.~\ref{fig:onset-varying-Omega} agree at all times up to the bounce and before that. Hence, quantum effects in the background dynamics are negligible, and the evolution is well described by the classical background equations. In addition, an analysis of the contributions of the distinct energy terms to the Friedmann equation is presented in Fig.~\ref{fig:energy-balance-2}. The contribution of the spatial curvature is described by:
\[
\rho_k = - \frac{3}{8 \pi G} \frac{1}{a(t)^2} \, ,
\]
while the kinetic and potential energies are described by $\rho_{kin}=\dot{\phi}_0^2/2$ and $\rho_{pot}=V(\phi_0)$, respectively. At the initial time $t_*$, during inflation, the evolution is dominated by the potential energy. For smaller times $t$, the contribution $\rho_k$ due to the spatial curvature becomes increasingly important, until it reaches approximately the same value as the potential energy, at which time inflation reaches an endpoint, and the universe undergoes a bounce, around $t\sim -4.2 \times 10^6 \, t_p$. Before that, the contribution of spatial curvature decreases towards the past and becomes negligible again. The kinetic energy is negligible along the whole evolution. Before the bounce, another slow-roll regime is found in the contracting era. For this background, the spatial curvature clearly plays an important role, driving the occurrence of the classical bounce, but quantum gravity effects are irrelevant.

\begin{figure}
\includegraphics[scale=.6]{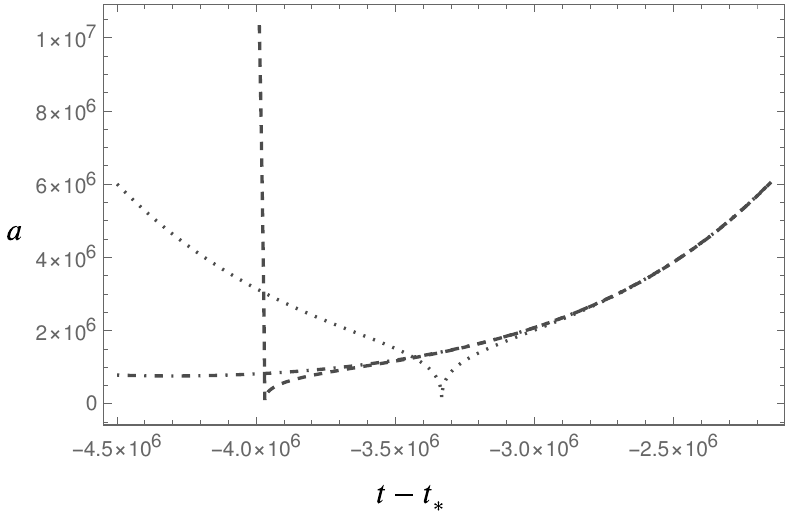}
\caption{Background evolution in the LQC effective dynamics for a closed universe with $\Omega_k=-0.005$ around the bounce. The scale factor is plotted against the proper time, in Planck units. Backgrounds of the types I, II and III are represented in dotted, dot-dashed and dashed lines, respectively.}
\label{fig:three-bounces}
\end{figure}

For backgrounds of types I and III, a bounce dominated by quantum effects in the effective dynamic occurs. The pre-bounce evolution is quite distinct in the two cases. In Fig.~\ref{fig:three-bounces}, we plot the evolution of the scale factor for an example of each kind of background, again representing the types I, II and III as dotted, dot-dashed and dashed lines, respectively. For backgrounds such that the spatial curvature is negligible at all times (type I), we found that the bounce is symmetric: around the bounce, the pre-bounce evolution mirrors the post-bounce evolution. For initial conditions such that the duration of inflation is determined by the spatial curvature (type III), the bounce is highly asymmetric, with a faster contraction in the pre-bounce era as compared to the expansion in the post-bounce era. The bounce of type I occurs closer to the inflationary regime than the bounce of type III. The classical bounce (type II) is smoother and takes place at a larger scale factor in comparison with the quantum bounces. It occurs farther in the past, but the number of $e$-folds from the bounce until the initial time $t_*$ is smaller.

\begin{figure}[t]
\includegraphics[scale=.54]{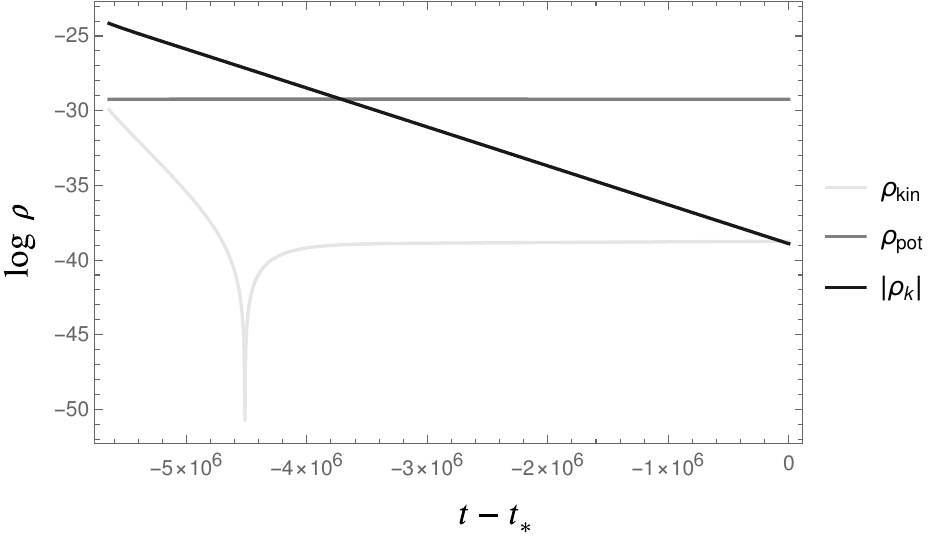}
\includegraphics[scale=.54]{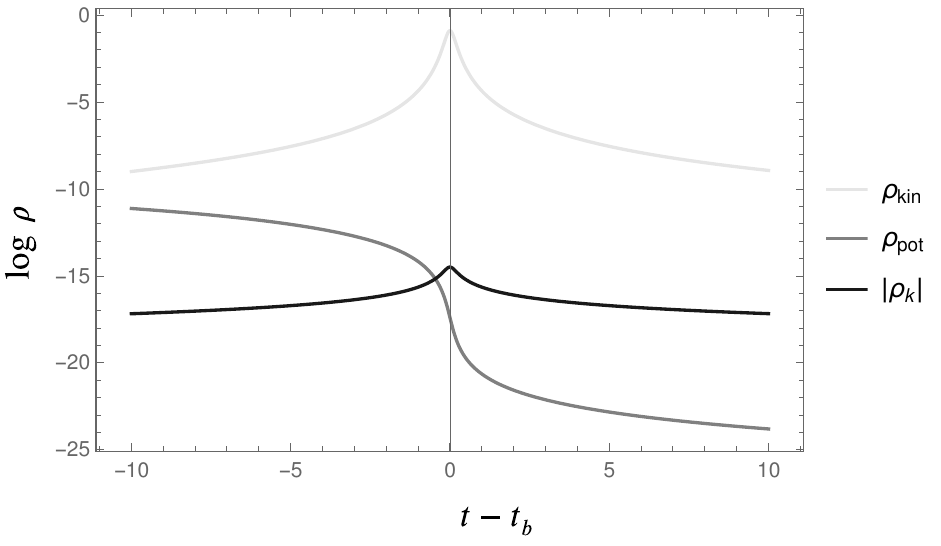}
\caption{Quantum bounce of type III: energy balance. Evolution of the kinetic energy $\rho_{kin}$, potential energy $\rho_{pot}$ and spatial curvature energy density $\rho_k$, in Planck units. Background dynamics determined for initial conditions obtained with $k^{(1)}_*=10^{-6} \text{ Mpc}^{-1}$ and $\Omega_k=-10^{-6}$. In the top panel, the evolution of the energy terms is represented from the onset of inflation at $t_{on}-t_*=-5.64 \times 10^{-6}$ until $t_*$. The bottom panel depicts their evolution around the bounce at $t_b$.}
\label{fig:bounce-II-energy-balance}
\end{figure}

For the analysis of the evolution of scalar perturbations, we will consider backgrounds of the type III, in which the spatial curvature is not negligible and there is a quantum bounce. We now discuss in more detail the properties of their effective dynamics and how they are affected by variations of the spatial curvature parameter $\Omega_k$. For this analysis, we fixed the choice of $k^{(1)}_*=10^{-6} \text{ Mpc}^{-1}$. This gives the following initial conditions at $t_*$:
\begin{align}
a_*&= 9.67 \times 10^7 \, l_p \, , \notag \\
\phi_*&= 1.06 \, m_p \, , \notag \\
\dot{\phi}_* &= -4.88 \times 10^{-9} \, m_p^2 \, ,
\label{eq:bg-initial-conditions}
\end{align}
and an inflaton mass of
\begin{align}
m = 2.636 \times 10^{-6} \, m_p\, .
\label{eq:bg-inflaton-mass}
\end{align}
For a flat universe, these initial conditions determine an evolution with $15.9$ $e$-folds from the bounce until $t_*$, long enough so that imprints of the bounce on observable modes of the CMB are negligible for LQC in a flat universe \cite{Agullo:2013ai,Agullo:2015tca}.

The energy balance of a typical solution in a closed universe is displayed in Fig.~\ref{fig:bounce-II-energy-balance}, which describes a background with $\Omega_k=-10^{-6}$. Near $t_*$, during the slow-roll regime, the evolution is dominated by potential energy, with negligible contributions from spatial curvature and kinetic energy. Moving towards the past, the contribution from spatial curvature becomes increasingly important, until it surpasses that of potential energy. The onset of inflation is dominated by spatial curvature. The kinetic energy is negligible at the onset of inflation, but becomes dominant around the quantum bounce.

\begin{figure}
\includegraphics[scale=.47]{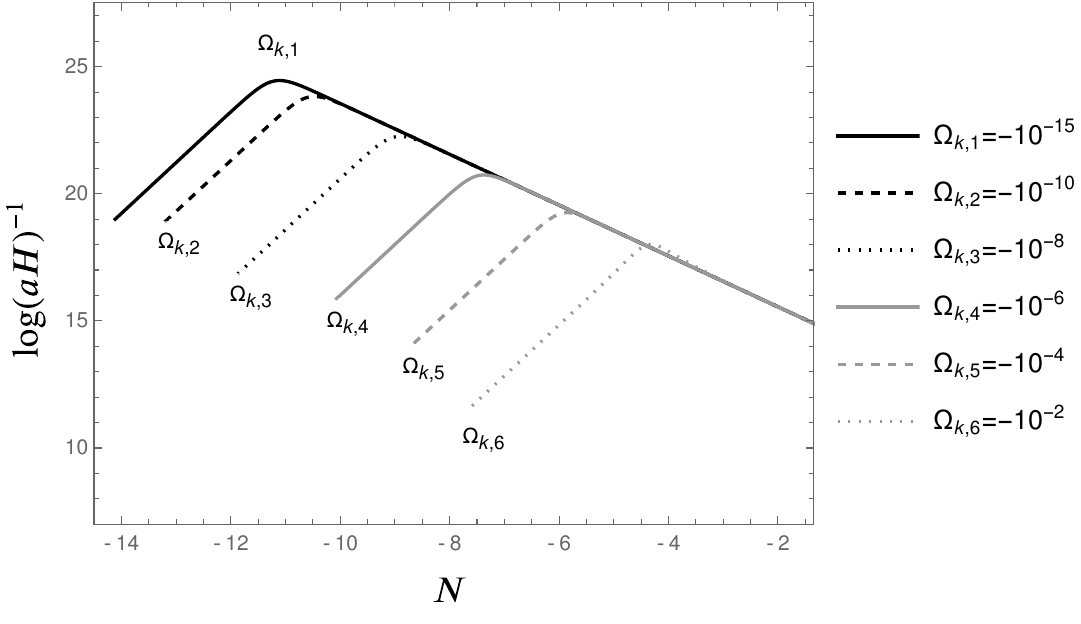}
\caption{Dependence of the onset of inflation on the spatial curvature parameter $\Omega_k$. The evolution of the Hubble horizon in Planck units is represented in terms of the number $N$ of $e$-folds from $t_*$, for six values of $\Omega_k$. The case $\Omega_{k,1}$ is indistinguishable from the flat evolution. The onset of inflation occurs at later times for larger spatial curvatures.}
\label{fig:vary-Omega-k}
\end{figure}

\begin{figure*}
\hspace{-14pt}\includegraphics[scale=.5]{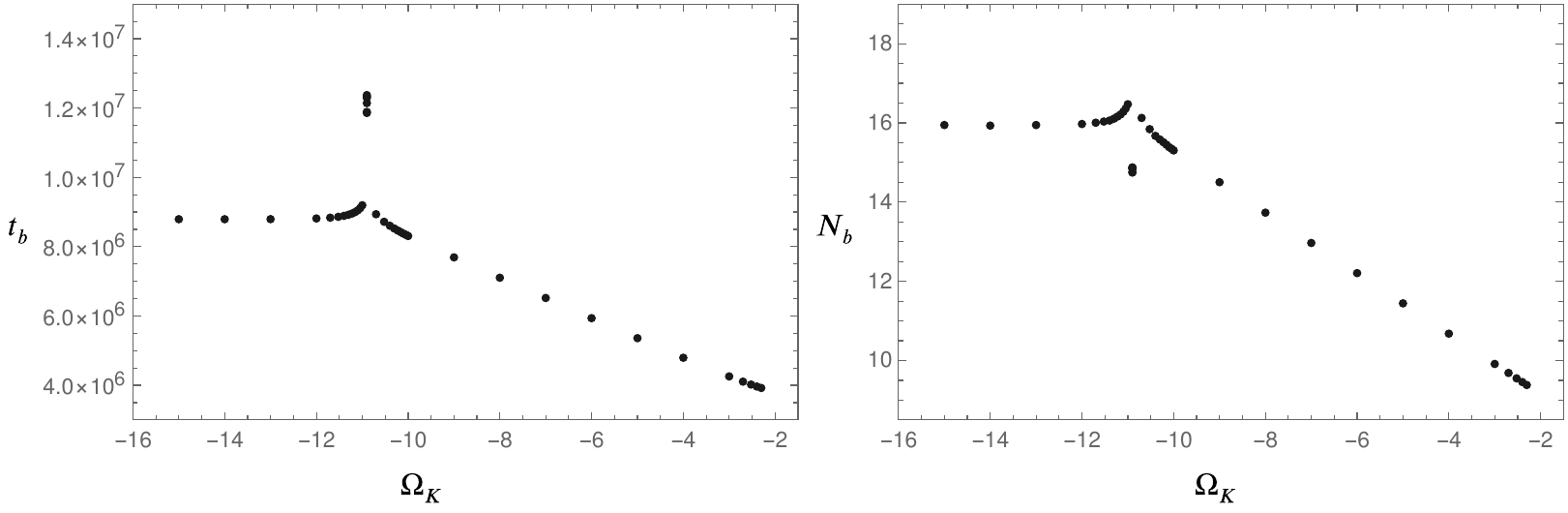}
\includegraphics[scale=.5]{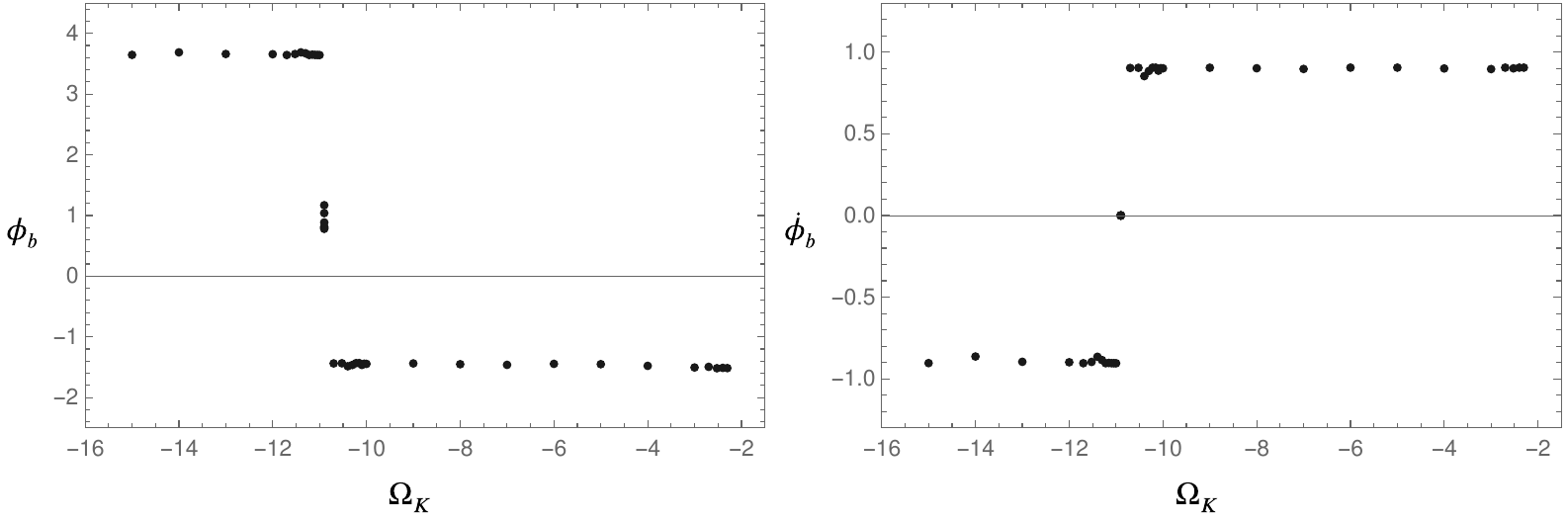}
\caption{Properties of the bounce for distinct spatial curvatures varying in the interval $\Omega_k \in (-10^{-2},-10^{-15})$. All quantities are represented in Planck units. \emph{Top left}: Time $t_b$ elapsed from the bounce until the initial time $t_*$ at which the mode $k_*=0.002 \text{ Mpc}^{-1}$ crosses the Hubble horizon. \emph{Top right}: Number of $e$-folds from the bounce until $t_*$. \emph{Bottom left}: Background scalar field $\phi_b$ at the bounce. \emph{Bottom right}: Derivative of the background scalar field $\dot{\phi}_b$ at the bounce.}
\label{fig:bounce-properties}
\end{figure*}

The time at which the onset of inflation takes place depends on the spatial curvature. In Fig.~\ref{fig:vary-Omega-k}, the Hubble horizon is plotted as a function of the number of $e$-folds from the time $t_*$. The background with $\Omega_{k,1}=-10^{-15}$ is of type I, corresponding to the flat limit. As $|\Omega_k|$ is increased, the onset of inflation moves to the right in the figure, occurring at later times. The evolutions of the backgrounds agree for $t>t_*$, so that the end of inflation happens at the same time in all cases. The duration of inflation is thus smaller for the larger spatial curvatures. As discussed before, the onset of inflation is dominated by spatial curvature. For the backgrounds with larger $|\Omega_k|$, such a dominance of spatial curvature takes place at a later time, decreasing the duration of inflation.

In Fig.~\ref{fig:bounce-properties}, properties related to the bounce are represented for $\Omega_k \in (-10^{-2},-10^{-15})$. The three kinds of backgrounds can be clearly identified as distinct regimes in the plots. Backgrounds of type I, in which the curvature is negligible along the whole evolution, are observed on the left part of the plots. As the spatial curvature does not play a significant role for these backgrounds, they form plateaus in the plots. Backgrounds of type III are observed on the right part of the plots, for spatial curvatures roughly of order $|\Omega_k| \gtrsim 10^{-10}$. The duration of the pre-inflationary regime, as described by the time from the bounce until $t_*$, decreases as the curvature is increased. The number of $e$-folds from the bounce until $t_*$ also progressively decreases for larger spatial curvatures. The values of the background scalar field and its derivative, however, are not strongly affected by the variation of spatial curvature. A background of type II describes a classical bounce, and corresponds to a transition between the two previous cases, taking place approximately at $\Omega_k \sim - 10^{-11}$. The duration of the pre-inflationary regime is larger for a background of type II, as the approach to the bounce is smoother for a classical bounce in comparison with quantum bounces. As the classical bounce occurs at a larger scale factor, however, the number of $e$-folds is smaller than in the previous cases. We observe that the classical bounce takes place with $\dot{\phi}_b$ close to zero. The value of the background scalar field $\phi_b$ at the bounce takes an intermediate values as compared to its value in backgrounds of types I and III.

\section{Primordial power spectrum for scalar perturbations}
\label{sec:evolution-perturbations}

The amplitudes $f_{nlm}$ of the normal modes of perturbations $\delta \phi$ of the scalar field, which describe the scalar cosmological perturbations in the spatially spherical gauge, evolve according to the equation of motion given in Eq.~\eqref{eq:eom-amplitudes}. Including the dependence on the indices $nlm$, the equation has the form:
\be
\ddot{f}_{nlm} + \alpha_n \dot{f}_{nlm}+\beta_n f_{nlm} =0 \, ,
\ee
where
\be
\alpha_n=-\frac{\dot{c}_1}{c_1} \, , \quad \beta_n= c_1 c_2 \, ,
\ee
with $c_1,c_2$ given in Eqs.~\eqref{eq:quadratic-hamiltonian-coeff1} and \eqref{eq:quadratic-hamiltonian-coeff2}. As the indices $lm$ do not enter the coefficients of the equation, the evolution of $f_{nlm}$ will depend only on the index $n$. Under the transformation:
\[
f_{nlm} = \frac{h_n}{a^{3/2}} \, ,
\]
the equation of motion transforms into:
\be
\ddot{h}_n + (\alpha_n-3H) \dot{h}_n + \Omega_n^2 h_n = 0 \, ,
\label{eq:mode-h-eom}
\ee
with
\[
\Omega_n^2 = -\frac{3}{2} \frac{\ddot{a}}{a}+ \frac{15}{4} \frac{\dot{a}^2}{a^2}-\frac{3}{2} \alpha_n H + \beta_n \,. 
\]

We studied the evolution of the normal modes from an initial time $t_i<t_b$, before the bounce, until the end of inflation at the time $t_{end}$, for backgrounds of type III with spatial curvatures $|\Omega_k|>10^{-10}$. For concreteness, we considered spatial curvatures $\Omega_k = 10^{-p}$, for $p<10$, in addition to the maximal curvature $\Omega_k=-0.005$, for the analysis of the perturbations. Initial conditions for the background are specified in Eq.~\eqref{eq:bg-initial-conditions}. The perturbations were set at an initial adiabatic vacuum state, specified by a choice of normal modes that will be explicitly described later in this section. Numerically evolving the normal modes until the end of inflation, the primordial power spectrum for the comoving curvature perturbations $\mathcal{R}$ can be computed at the end of inflation in terms of the amplitudes of the evolved normal modes in the same manner as in a spatially flat gauge in flat space, as follows.

We closely follow the strategy for the analysis of scalar perturbations in LQC employed in \cite{Agullo:2018bs} for a flat universe. We first determine the two-point function for the perturbation of the scalar field $\delta \phi$ at the end of inflation:
\be
\mathcal{P}_{\delta \phi} = \left| f_{nlm}(t_{end}) \right|^2 \, .
\label{eq:phi-phi}
\ee
As the coefficients of the equation of the modes do not depend on $l$ or $m$  and we choose initial conditions that only depend on $n$, the power spectrum is a function $\mathcal{P}_{\delta \phi}(n)$ of $n$. We compute this quantity in the spatially spherical gauge. For any choice of gauge, however, invariant quantities can be constructed out of the gauge-dependent variables. In particular, invariant pertubations in a closed universe were constructed in \cite{Langlois:1994ec} and discussed in more detail in \cite{Bonga:2016iuf}. In the spatially spherical gauge, the invariant comoving curvature perturbation $\mathcal{R}$ can be expressed in terms of the perturbation of the scalar field through the same relation as in flat space:
\be
\mathcal{R} = - \frac{\dot{a}}{a \dot{\phi}_0} \delta \phi \, .
\label{eq:R-phi}
\ee
From Eqs. \eqref{eq:phi-phi} and \eqref{eq:R-phi}, the power spectrum for the comoving curvature perturbations then reads:
\be
\overline{\mathcal{P}}_{\mathcal{R}}(n) = \left| f_{nlm}(t_{end}) \right|^2 \left[\frac{\dot{a}(t_{end})}{a(t_{end}) \dot{\phi}_0(t_{end})}\right]^2 \, .
\ee

In order to compare this result with the power spectrum in a flat universe, we can invoke the flat limit of the normal modes on the $3$-sphere (see \cite{Abbott:1986ct,Bonga:2016iuf,Bonga:2016cje} and Appendix \ref{sec:flat-limit}). In a small region of the $3$-sphere and for short wavelengths, i.e. for $\chi \ll \pi$ and large $n\gg 1$, the hyperspherical harmonics $Q_{nlm}$ cannot be distinguished from harmonics $u_{klm}$ of a spatially flat universe, as spatial curvature is negligible in this regime. The flat harmonics are given in spherical coordinates by:
\be
u_{klm}(r,\theta,\varphi) =  \sqrt{\frac{2}{\pi}} k j_l(kr) Y_{l m}(\theta,\varphi) \, ,
\ee
where $j_l$ is a spherical Bessel function and $Y_{l m}$ are spherical harmonics. The flat limit of the hyperspherical harmonics is given by \cite{Abbott:1986ct}:
\be
Q_{nlm}(\chi,\theta,\varphi) \simeq \sqrt{\frac{2 n^2}{\pi}} j_l(k_1 \chi) Y_{lm}(\theta, \varphi)\, ,
\ee
with $k_1^2=n^2-1$. Considering a $3$-sphere with radius $r_0$, the distance $r$ along the radius is given by $r_0 \chi$. Introducing the quantity:
\be
k^2 = \frac{n^2-1}{r_0^2} \, ,
\label{eq:n-to-k}
\ee
we can then write, for large $n$:
\begin{align}
Q_{nlm}(\chi,\theta,\varphi) &\simeq r_0 \sqrt{\frac{2}{\pi}} k j_l(k r) Y_{lm}(\theta, \varphi) \\
&\simeq r_0 u_{klm} \,.
\label{eq:n-to-k-modes}
\end{align}
Therefore, in the flat limit, discrete modes labeled by $nlm$ correspond to flat modes $klm$ under the relation \eqref{eq:n-to-k}, up to a factor of $r_0$, which can be absorbed in the mode amplitude,
\be
f_{klm} = r_0 f_{nlm} \, .
\ee

In addition, as the power spectrum is a density in the $k$-space, it must be multiplied by the density of states in the translation from the discrete $n$-space to the $k$-space. According to Eq.~\eqref{eq:n-to-k}, for large $n$, the density of states simply consists of a factor of $r_0$. We thus obtain the following identification for the power spectrum in the flat limit:
\be
\overline{\mathcal{P}}_{\mathcal{R}}(k) = r_0^3 \overline{\mathcal{P}}_{\mathcal{R}}(n) \, .
\ee
The adimensional power spectrum is defined as
\be
\mathcal{P}_{\mathcal{R}}(k) \equiv \frac{k^3}{2 \pi^3} \overline{\mathcal{P}}_{\mathcal{R}}(k) \, ,
\ee
from which we obtain the following formula in the flat limit for a closed universe:
\be
\mathcal{P}_{\mathcal{R}}(k) \simeq \frac{n^3}{2 \pi^2} \left| f_{nlm}(t_{end}) \right|^2 \left[\frac{\dot{a}(t_{end})}{a(t_{end}) \dot{\phi}_0(t_{end})}\right]^2 \, .
\label{eq:pk}
\ee
At the end of inflation, spatial curvature is indeed negligible. Moreover, for the spatial curvatures that we will consider, the modes observable in the CMB are such that $n \gg 1$. We will use Eq.~\eqref{eq:pk} to express the results of the numerical calculations in a manner convenient for the comparison with the usual flat universe results.

Let us first discuss the behavior of the coefficients of the equation of motion, before setting the initial conditions. In particular, we wish to show that a convenient approximation can be used. Next, we discuss the results for the primordial power spectrum.

\begin{figure}
\includegraphics[scale=.55]{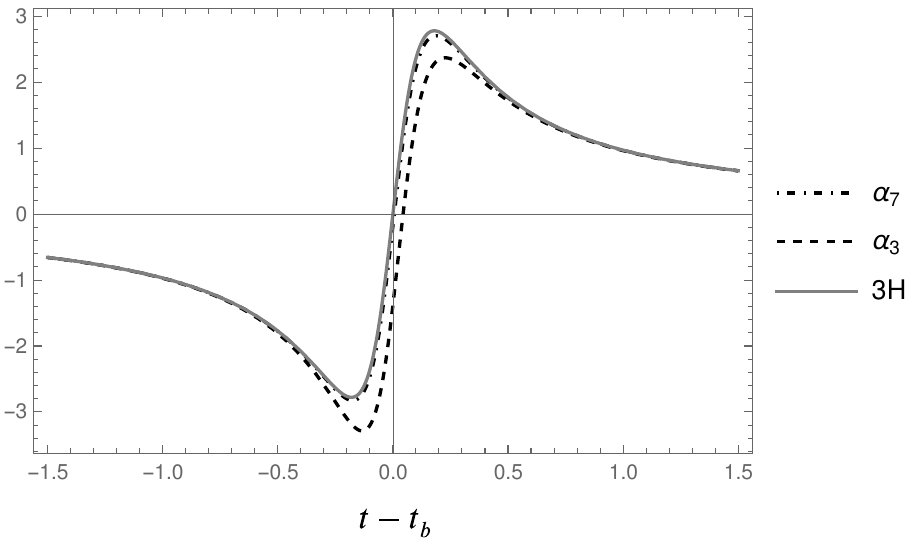}
\caption{Evolution of the coefficients $\alpha_n$ near the bounce at $t_b$, for $\Omega_k=-0.005$, in Planck units. The cases of $n=3$ and $n=7$ are compared with the function $3H$. The coefficient $\alpha_3$ slightly deviates from $3H$. The coefficient $\alpha_7$ is well approximated by $3H$. For larger values of $n$, the coefficients $\alpha_n$ become indistinguishable from $3H$ at the scale of the plot.}
\label{fig:alpha-approx}
\end{figure}

Consider the coefficient $\alpha_n-3H$ of the equation of motion as given in Eq.~\eqref{eq:mode-h-eom}. When this coefficient can be neglected, the equation reduces to that of a time-dependent oscillator. This turns out to be the case for the scalar modes that can affect predictions for the spectrum of temperature anisotropies of the CMB. From \cite{Planck:2018jri}, such observable modes correspond roughly to the window $10^{-4} \text{ Mpc}^{-1} \lesssim k \lesssim 10^{-1} \text{ Mpc}^{-1}$. This can be translated into an interval of modes $n$ through \eqref{eq:n-to-k}. For $\Omega_k=-0.005$, for instance, we obtain the observable window $6 \lesssim n \lesssim 6294$. For $\Omega_k=-10^{-6}$, the observable window corresponds to $445 \lesssim n \lesssim 445 \times 10^3$. Consider the case of the maximal allowed spatial curvature $\Omega_k=-0.005$. In Fig.~\ref{fig:alpha-approx}, we compare the evolution of the coefficients $\alpha_n$ and the function $3H$ near the bounce. For $n=3$, the coefficient slightly deviates from $3H$, but for $n=7$, the evolutions already agree well. Away from the bounce, all $\alpha_n$ are well approximated by $3H$ for all times. For larger $n$, the difference between $\alpha_n$ and $3H$ rapidly decreases. The same behavior is observed for smaller spatial curvatures. This allows us to adopt the approximation $\alpha_n \simeq 3H$, which reduces the equation of motion to that of a time-dependent oscillator with frequency $\Omega_n$.

The variation of the frequency $\Omega_n$ in time leads to the production of excitations of the scalar modes. When the rate of variation is small, the production of excitations can be neglected, and the mode evolves adiabatically. In this case, an initial vacuum state evolves to another vacuum state at a later time \cite{Parker:2009uva,Birrell:1982ix}. We set the perturbations initially in a $4$-th order adiabatic vacuum, as usually done in the dressed metric approach to LQC \cite{Agullo:2012fc,Agullo:2013ai}. This means that around the initial time $t_i$, the mode function can be expressed in the form:
\be
h_n \simeq \frac{1}{\sqrt{2 W^{(4)}_n}} \exp\left[- i \int W^{(4)}_n(t') \dd t'\right] \, ,
\ee
with
\[
W_n^{(4)} = \Omega_n \sqrt{1+\epsilon_2(n)} \sqrt{1+\epsilon_4(n)} \, ,
\]
where
\begin{align*}
\epsilon_2(n) &= \frac{3}{4\Omega_n^4} \left( \frac{d\Omega_n}{dt} \right)^2 - \frac{1}{2 \Omega_n^3}\frac{d^2\Omega_n}{dt^2}\, , \\
\epsilon_4(n) &= \frac{3}{4 \Omega_{n,1}^4 \Omega_n^2} \left( \frac{d\Omega_{n,1}}{dt} \right)^2 + \frac{1}{2 \Omega_{n,1}^3 \Omega_n^3} \frac{d\Omega_n}{dt} \frac{d\Omega_{n,1}}{dt} \\
	& \quad - \frac{1}{2 \Omega_{n,1}^3 \Omega_n^2} \frac{d^2\Omega_{n,1}}{dt^2} \, ,
\end{align*}
and $\Omega_{n,1}=\sqrt{1+\epsilon_2(n)}$.

\begin{figure*}
\hspace{-18pt}
\includegraphics[scale=.6]{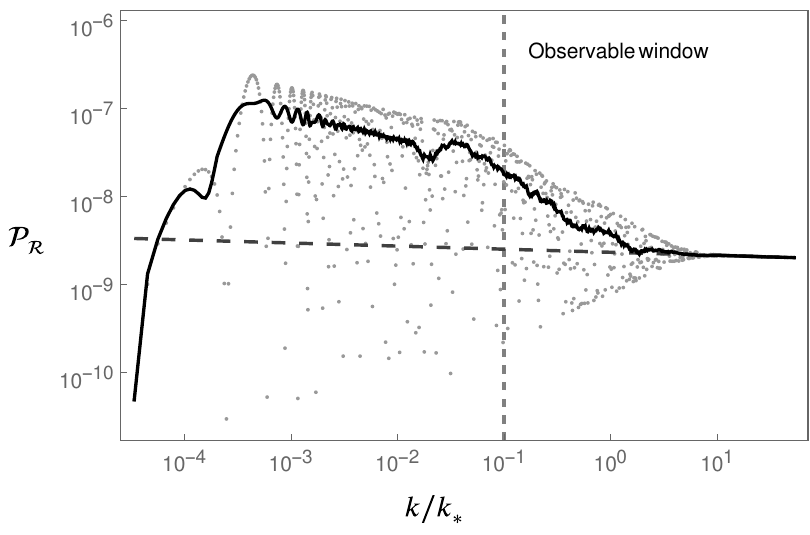}
\hspace{18pt}
\includegraphics[scale=.6]{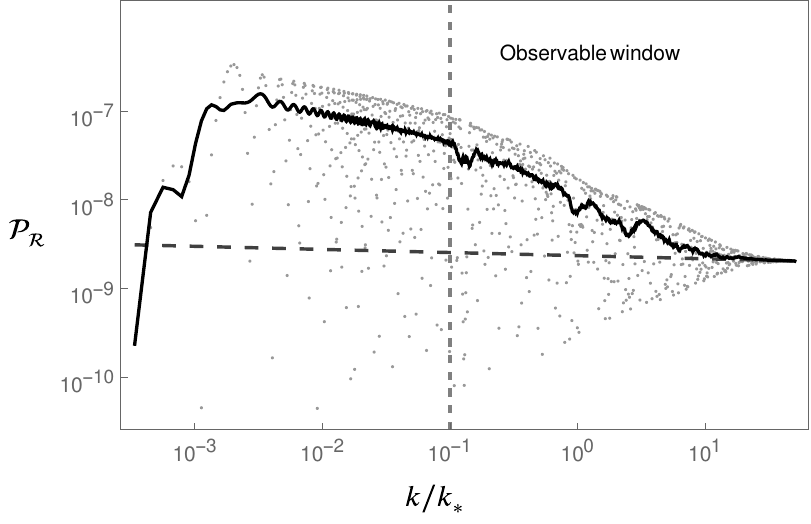}
\caption{Primordial power spectrum for comoving curvature perturbations on a closed universe with $\Omega_k=-10^{-8}$ (\textit{left panel}) and $\Omega_k=-10^{-6}$ (\textit{right panel}). The grey points represent the results of numerical calculations. The solid line describes their local average. The dashed curve corresponds to a power law spectrum with amplitude $A_s$ and spectral index $n_s$ set equal to their best estimates by the \planck{} Collaboration \cite{Planck:2018jri}. The scale of the largest observable modes, at $k/k_* \sim 0.1$, is represented as a vertical line that bounds the window of observable modes.}
\label{fig:power-spectra}
\end{figure*}

Before the bounce, we can move sufficiently towards the past until the rate of variation of the frequency $\Omega_n$ is negligible. We checked that particle production for the chosen initial $4$-th order adiabatic vacua is negligible for the observable modes near the initial time by comparing the absolute value of the evolution of the normal mode defining the initial adiabatic vacuum with those of normal modes defining adiabatic vacua at later times, for the largest observable modes. The initial adiabatic vacuum indeed evolves into new adiabatic vacua up to the vicinity of the bounce, where excitations are then produced. The evolution of the excitations produced around the bounce ultimately affect the predictions of the model for the primordial power spectrum at the end of inflation, and constitute the contribution from quantum gravity to the modifications of the inflationary predictions.

The primordial power spectra $\mathcal{P}_\mathcal{R}(k)$ obtained for spatial curvatures $\Omega_k=-10^{-8}$ and $\Omega_k=-10^{-6}$ are shown in Fig.~\ref{fig:power-spectra}. The case of $\Omega_k=-10^{-8}$ represents a scale at which corrections to the inflationary predictions affect the smallest wavenumbers in the observable part of the primordial power spectrum. The power law spectrum is recovered for large momenta, but for $k/k_* \lesssim 1$, significant modifications are observed. Fast oscillations develop around the power law spectrum, which are asymmetric around the usual inflationary predictions. On average, they describe an enhancement of power, more pronounced for smaller $k$. The enhancement of power is close to one order of magnitude at the boundary of the observable window. Increasing the spatial curvature, the part of the spectrum displaying deviations from the power law spectrum extends to include progressively larger wavenumbers $k$, moving to the right in the plot. For $\Omega_k=-10^{-6}$, the corrections are clearly visible for $k/k_*$ around $\sim 30$, i.e., for wavenumbers far larger than the smallest observable wavenumbers, and affect a considerable share of the observable window. For a spatial curvature $|\Omega_k|\sim 10^{-9}$, which we do not represent in the figure, the observable part of the spectrum does not display significant corrections, except for a few oscillations at the scale of the largest observable modes. In the flat limit, the considered initial conditions determine a background with $\sim 16$ $e$-folds from the bounce until $t_*$, with negligible corrections to the primordial power spectrum in the observable window.

The power spectra have the same general features as in LQC on a flat universe (compare with \cite{Agullo:2018bs}, for instance). In contrast with the case of a flat universe, however, the region of the spectrum displaying deviations from the power law spectrum cannot be arbitrarily chosen by changing initial conditions for the background at $t_*$, thereby changing the duration of inflation. In a closed universe, under the assumption of the existence of an inflationary regime such that modes $k>k_*$ cross the Hubble horizon during a slow-roll regime compatible with the observations of the CMB, and that both spatial curvature and quantum gravity effects are non-negligible, i.e., for a background of type III, the duration of inflation is determined by the spatial curvature, as discussed in Sec.~\ref{sec:background-evolution}. Accordingly, the interval of wavenumbers for which corrections to the power law spectrum are relevant, and how far it extends into the observable window of the spectrum, is determined by the spatial curvature parameter $\Omega_k$. This is a main difference between the cases of flat and closed universes in LQC.

Fast oscillations in the primordial power spectrum tend to average out in predictions for the CMB spectra. Observable signals in the CMB anisotropies due to modifications in the primordial power spectrum can then be expected when the local average of the primordial power spectrum deviates from the power law spectrum in the observable window. In order to roughly evaluate the significance of any deviations, the averaged local spectrum (shown as a solid line in Fig.~\ref{fig:power-spectra}) can be compared with nonparametric reconstructions of the primordial power spectrum from observations of the anisotropies of the CMB \cite{Hunt:2013bha,Planck:2018jri}. The results represented in Fig.~\ref{fig:power-spectra} then indicate the following.

For $\Omega_k =-10^{-8}$, the local average of $\mathcal{P}_\mathcal{R}(k)$ significantly deviates from the power law spectrum roughly for $k/k_* \lesssim 1$, i.e., for the largest observable modes. This part of the spectrum is affected by a large cosmic variance and by the presence of anomalies. Accordingly, reconstructions of the primordial power spectrum display two characteristic features around the lowest observable $k$'s: a power suppression and large uncertainties. In addition, oscillations are found in the same region \cite{Hunt:2013bha,Planck:2018jri}. Although modifications in the spectrum were found in the predictions of our model at the scale of the anomalies for $\Omega_k =-10^{-8}$, they cannot explain or alleviate the anomalies, as an enhancement of power is predicted, instead of the observed suppression. In addition, the amplification of power is larger than allowed by the uncertainties in the reconstructed primordial power spectrum, putting the model in tension with the observations at such scales.

For $\Omega_k = -10^{-6}$, the averaged power spectrum displays visible deviations from the power law spectrum for $k/k_*$ close to $10$. But this part of the observable region of the spectrum is associated with modes of the CMB that are well described by the usual inflationary predictions. In addition, for a large window of modes with smaller $k$'s, the power is enhanced by about an order of magnitude. The results obtained for $\Omega_k=-10^{-6}$ are clearly inconsistent with the observations of the CMB, and the same is true for larger spatial curvatures, for which the corrections move to still larger wavenumbers. On the other hand, for $\Omega_k = -10^{-9}$, we did not find significant modifications in the observable part of the spectrum, only corrections consisting of mild oscillations for the largest observable modes, and the spectrum does not display any evident conflict with the spectrum reconstructed from the observations of the CMB. As the spectrum reconstructed from the observations include oscillations not present in a power-law spectrum, it might be possible that for some curvatures around $\Omega_k = -10^{-9}$, the predictions of the model provide a better fit of the observations.

\section{Summary and outlook}
\label{ref:conclusion}

We analyzed the inflationary and pre-inflationary evolution of scalar modes of linear cosmological perturbations on a closed FLRW universe described by loop quantum cosmology. We adopted a model that reproduces in the context of a closed universe the picture resulting from the application of the dressed metric approach of LQC to the case of a flat universe \cite{Agullo:2012sh,Agullo:2013ai,Agullo:2015tca}. For this purpose, we first derived the Hamilton equations for the linear scalar perturbations in a suitable gauge, analogous to the spatially flat gauge. The coefficients of the equations depend on the background canonical variables $(a,\phi_a)$ and $(\phi_0, \pi_{\phi_0})$. In our model, these are evaluated in the LQC effective dynamics of the background, given by a modified Friedmann equation obtained in \cite{Ashtekar:2006es}. In the effective dynamics, the Big Bang singularity is replaced by a bounce. We numerically evolved the perturbations from before the bounce until the end of inflation and determined the primordial power spectrum for comoving curvature perturbations for a sample of backgrounds consistent with observational constraints on inflation \cite{Planck:2018jri}.

The Hamiltonian formalism for the linear scalar perturbations was derived in full detail from the ADM action in Sec.~\ref{sec:hamiltonian-formalism}. We chose to work on a specific gauge, instead of using gauge invariant perturbations. Specifically, we adopted a spatially spherical gauge, in which the spatial three-dimensional metric is not perturbed (only the lapse and shift), and the scalar perturbations are encoded in the perturbations of the scalar field. In addition to the relative simplicity of this gauge, the main reason for not directly using equations for the comoving curvature perturbations $\mathcal{R}$, in terms of which the primordial power spectrum is written, is that they become ill-behaved whenever $\dot{\phi}_0=0$. This is not an issue during inflation, but does happen in the pre-inflationary regime. In contrast, the scalar perturbations $\delta \phi$ are well-behaved in the pre-inflationary era. The situation is similar to that met in a flat universe, discussed in \cite{Agullo:2018bs}. Our result for the quadratic Hamiltonian in the spatially spherical gauge is given in Eqs.~\eqref{eq:quadratic-hamiltonian}--\eqref{eq:quadratic-hamiltonian-coeff2}, for normal modes of $\delta \phi$ labelled by indices $nlm$ associated with hyperspherical harmonics on the $3$-sphere. This new explicit representation for the dynamics of scalar perturbations on closed universes can also be employed in other models of the pre-inflationary universe, as far as a universe filled by a single scalar field is considered, as it only relies on classical gravity as described by general relativity.

The dynamics of the quantum-corrected background was determined from the numerical integration of the effective dynamics of LQC. For numerical reasons, it was convenient to use the modified acceleration equation \eqref{eq:modified-acceleration-eq}, instead of the modified Friedmann equation \eqref{eq:modified-Friedmann}, coupled to the Klein-Gordon equation. This makes the integration well behaved across the bounce. Initial conditions were set during inflation, at the time $t_*$ when the inflationary pivot mode $k_*=0.002 \text{ Mpc}^{-1}$ crosses the Hubble horizon. For the considered range of values of $\Omega_k$, the spatial curvature is negligible at this time. As quantum gravity effects are also irrelevant during inflation, the usual picture of slow-roll inflation in flat space is valid at $t_*$, and was used to translate the observational estimates on $A_s$, $n_s$ and $N_*$ provided by \planck{} into initial conditions for the background, as discussed in Sec.~\ref{sec:initial-conditions}. In the integration towards the past of $t_*$, spatial curvature effects are amplified, and a short era dominated by spatial curvature can occur. Farther in the past, quantum-gravity effects in general manifest.

Three types of pre-inflationary backgrounds were obtained, depending on the value of $\Omega_k$. We restricted to spatial curvatures within the bound $|\Omega_k| < 0.005$ derived from observations of the CMB considering the effects of spatial curvature only in the post-inflationary $\Lambda$CDM era \cite{Planck:2018vyg}. In the first class (type I), obtained for the smallest considered curvatures, the spatial curvature is negligible throughout the whole evolution. This gives the flat limit of the model. A symmetric quantum bounce then takes place in the far past, as in a flat universe. For backgrounds of type II, a classical bounce dominated by spatial curvature occurs right before inflation, connecting an era of deflation in the contracting era with the inflationary regime. This corresponds to the picture analyzed in \cite{Renevey:2020zdj,Renevey:2022xzh}. In this case, quantum-gravity effects are negligible throughout the whole evolution. Backgrounds of type III were obtained for the larger values of $\Omega_k$ in the range considered. The spatial curvature also becomes dominant near the onset of inflation, but instead of producing a classical curvature bounce, it is taken over by the kinetic energy of the inflaton (towards the past), and a kinetically dominated quantum bounce takes place in the far past. This bounce is highly asymmetric. Since only for backgrounds of type III both spatial curvature and quantum-gravity effects are relevant, we restricted our numerical analysis of the perturbations to backgrounds of this type. The results for the first class must reduce to those in a flat universe, and perturbations on backgrounds of the second type were analyzed in \cite{Renevey:2020zdj,Renevey:2022xzh}.

In the determination of the initial conditions for the background, the observational input given by the parameters $A_s$, $n_s$ and $N_*$ can be varied within error bars. In a flat universe, this produces inflationary regimes with distinct durations. A common late behavior is obtained for $t>t_*$, consistently with the fact that slow-roll inflation is an attractor for the dynamics, but the past behavior for $t<t_*$ is strongly affected by variations of the initial conditions at $t_*$, as shown in Fig.~\ref{fig:onset-varying-Omega}. In a closed universe, a quite distinct picture was found. In backgrounds of type III, for a given $\Omega_k$, the duration of inflation is roughly insensitive to variations of the initial conditions, being determined by the spatial curvature. In addition, if $\Omega_k$ is increased, then the onset of inflation occurs at a later time in the past of $t_*$, so that the inflationary regime is shorter. Decreasing $\Omega_k$, it happens progressively farther in the past, producing longer inflationary regimes, until a critical value for which a classical curvature bounce then takes place (type II). For even smaller $\Omega_k$, a background of the type I is obtained, with negligible spatial curvature along the whole evolution. The critical value of $\Omega_k$ that determines which class of background is obtained depends on the initial conditions, but as long as a background of the third type is considered, the duration of inflation is determined by the spatial curvature, as discussed in Sec.~\ref{sec:background-evolution}.

For the numerical evolution of the scalar perturbations, we chose backgrounds of type III and $|\Omega_k| \gtrsim 10^{-11}$, with non-negligible spatial curvature effects and a quantum bounce. We set the perturbations in a fourth-order adiabatic vacuum state at a time $t_i<t_b$ before the bounce, around which particle production was checked to be negligible. Excitations are produced as the perturbations cross the bounce and evolve along the pre-inflationary regime, including the era dominated by curvature near the onset of inflation. As a result, the perturbations are not in a vacuum state at the onset of inflation. This leads to significant modifications in the primordial power spectrum for comoving curvature perturbations at the end of inflation. For curvatures slightly larger than $|\Omega_k| \sim 10^{-11}$, the number of $e$-folds between the bounce and the end of inflation is sufficiently large to move these modifications to nonobservable scales, so that the observable part of the spectrum is well described by the usual power-law spectrum. This is similar to what happens in a flat universe, where LQC corrections to the primordial power spectrum are redshifted to nonobservable scales for sufficiently long inflationary regimes. Increasing the spatial curvature in our model, the modifications in the primordial power spectrum affect modes of progressively larger $k$ and reach observable scales.

For $|\Omega_k| \sim 10^{-9}$, we found corrections to the primordial power spectrum only for the very largest observable modes, roughly within the uncertainties in the observational constraints on the primordial power spectrum. For $|\Omega_k| \sim 10^{-8}$, modes at the largest observable scales, for $k/k_* \lesssim 1$, start to be significantly affected, as shown in Fig.~\ref{fig:power-spectra}. At such scales, the spectrum develops fast oscillations and, on the average, there is an enhancement of power. For $k/k_* \gtrsim 1$, a power-law spectrum is recovered, leading to the usual inflationary predictions. The modifications to the power-law spectrum are restricted to a small part of the observable spectrum, where anomalies are observed. It is not possible to explain the anomalies from such modifications, however, as an enhancement of power is predicted, instead of the observed power suppression. Moreover, the enhancement of power is close to an order of magnitude for the very largest observable modes, too large to be accounted for even by the large uncertainties in this region of the spectrum. The results for $|\Omega_k| \sim 10^{-6}$ are also shown in Fig.~\ref{fig:power-spectra}. The deviations from the power-law spectrum in this case extend into a wider range of modes, reaching scales for which the usual inflationary predictions provide an accurate description of the observations, putting the predictions of the model in conflict with the observations. From these results, we find that consistency with the observations of the CMB requires the spatial curvatures to be rather small, not much larger than $|\Omega_k| \sim 10^{-9}$.

The analysis was restricted to backgrounds of type III, and larger spatial curvatures remain viable for a different type of background. In the scenario of a classical curvature bounce (background of type II), analyzed in \cite{Renevey:2020zdj,Renevey:2022xzh}, even a curvature of $\Omega_k=-0.044$ remains consistent with the observations, as discussed in \cite{Renevey:2022xzh}. On the other hand, for a given $\Omega_k$, a background of type I (flat limit) requires initial conditions determining an inflationary regime shorter than for a type III background. For $|\Omega_k| \gg 10^{-9}$, the number of $e$-folds from the bounce until the end of inflation would then be too small for this scenario to remain consistent with the observations.

The window of modes in the primordial power spectrum affected by the pre-inflationary dynamics strongly depends on the number of $e$-folds from the bounce until the end of inflation. As this duration depends on the inflaton potential, it would be interesting to analyze alternative potentials. Moreover, distinct initial conditions have been explored for the perturbations in LQC, and could also be compared with our choice of adiabatic vacua. In particular, the initial conditions adopted in \cite{Ashtekar:2016wpi,Ashtekar:2016pqn}, which lead to a power suppression in flat universes, could be considered. In general, stricter bounds on the spatial curvature can be expected to be obtained when its effects in the pre-inflationary regime are taken into account, in comparison with bounds obtained from models where only the post-inflationary era is considered. For small spatial curvatures $|\Omega_k| \sim 10^{-9}$ consistent with the observations, one could also analyze the production of non-Gaussianities. As stressed in \cite{Agullo:2018bs}, large non-Gaussianities can be produced at a quantum bounce. A similar analysis on closed universes would require going to a higher order in the perturbation theory, including terms of third-order in the expansion of the ADM Hamiltonian.

\appendix
\section{Hyperspherical harmonics on the three-sphere}
\label{sec:harmonics}

The line element in the unit $3$-sphere reads:
\be
d\Omega^2 = d\chi^2 + \sin^2 \chi (d\theta^2 + \sin^2 \theta \, d\varphi^2) \, ,
\label{eq:s3-metric}
\ee
where $\chi, \theta \in [0,\pi]$ and $\varphi \in [0,2\pi]$. Denote the metric tensor by $\Omega_{ij}$, and let $\bar{D}$ be the torsion-free covariant derivative compatible with $\Omega_{ij}$. The determinant of the metric is $\Omega=\sin^4 \chi \sin^2 \theta$. Scalar, vector and tensor harmonics on the $3$-sphere are discussed in \cite{Halliwell:1984eu,Abbott:1986ct,Gerlach:1978gy}. Here we gather some basic definitions and results used in the main text of this paper.

Hyperspherical harmonics on $S^3$ are defined as
\be
Q_{nl m}(\chi,\theta,\varphi) = \Phi_{nl}(\chi) Y_{l m}(\theta,\varphi) \, ,
\ee
where $n=1,2,\dots$, $l=0,1,2,\dots$ and $m=-l, -l+1, \dots, l$. The normalized radial Fock harmonics can be expressed in terms of associated Legendre functions as
\[
\Phi_{nl}(\chi) = \sqrt{\frac{M_{nl}}{\sin \chi}} P^{-1/2-l}_{-1/2 +n}(\cos \chi) \, , \quad M_{nl} = \prod_{r=0}^l (n^2-r^2) \, .
\]
The functions $Y_{lm}(\theta,\varphi)$ are the usual spherical harmonics on the $2$-sphere:
\[
Y_{lm}(\theta, \varphi) = (-1)^m \sqrt{\frac{(2l+1)}{4\pi} \frac{(l-m)!}{(l+m)!}} P_l^m(\cos \theta) e^{i m \varphi} \, ,
\]
which satisfy
\[
\Delta_{S^2} Y_{lm} = - l(l+ 1) Y_{lm} \, .
\]

The hyperspherical harmonics
\be
Q_{nlm} = \Phi_{nl}(\chi) Y_{lm}(\theta,\varphi)
\ee
are eigenstates of the Laplacian,
\[
\Delta_{S^3} Q_{nlm} = - (n^2 - 1) Q_{nlm} \, ,
\]
and form a complete orthonormal set in the space $L^2(S^3)$ of square-integrable functions on the 3-sphere,
\be
\int \dd \Omega \, Q_{nlm} Q_{n'l'm'} = \delta_{nn'} \delta_{ll'} \delta_{mm'} \, ,
\label{eq:scalar-Q-orthonormality}
\ee
where $\dd \Omega = \dd^3x \sqrt{\Omega}$. Any scalar function $f$ on $S^3$ can be expanded as
\[
f(\chi,\theta,\varphi) = \sum_{n=1}^\infty \sum_{l=0}^{n-1} \sum_{m=-l}^l f_{nlm} Q_{nlm}(\chi,\theta,\varphi) \, .
\]
The expansion coefficients can be determined from the resolution of the identity \eqref{eq:scalar-Q-orthonormality},
\[
f_{nlm} = \int \dd \Omega \, Q_{nlm}(x) f(\vec x) \, .
\]

Distinct conventions are used for the hyperspherical harmonics in the literature. In particular, we use normalized harmonics, as expressed in Eq. \eqref{eq:scalar-Q-orthonormality}. The hyperspherical harmonics $Q_{\beta lm}$ studied in \cite{Abbott:1986ct}, in contrast, are not normalized, and must be multiplied by a factor of $\sqrt{2n^2/\pi}$ so as to match our convention, with $\beta \to n$.

The flat limit of the hyperspherical harmonics is described in \cite{Abbott:1986ct}. At small distances and short wavelengths, effects of spatial curvature can be neglected, as space is locally flat. In this limit, given by $\chi \ll \pi$ and $n \gg 1$, the harmonics $Q_{nlm}$ indeed reduce to harmonics of a flat universe. In our notation, we have:
\be
Q_{nlm}(\chi,\theta,\varphi) \simeq \sqrt{\frac{2 n^2}{\pi}} j_l(k_1 \chi) Y_{lm}(\theta, \varphi)\, ,
\ee
with $k_1^2=n^2-1$. Considering a $3$-sphere with radius $r_0$, the distance along the radius is given by $r_0 \chi$. It is convenient to introduce the quantity:
\be
k^2 = \frac{n^2-1}{r_0^2} \, ,
\label{eq:n-to-k-app}
\ee
in terms of which we can write, for large $n$:
\begin{align}
Q_{nlm}(\chi,\theta,\varphi) &\simeq r_0 \sqrt{\frac{2}{\pi}} k j_l(k r) Y_{lm}(\theta, \varphi) \\
&\simeq r_0 u_{klm} \,,
\end{align}
where $u_{klm} =  \sqrt{2/\pi} k j_l(k r) Y_{lm}(\theta, \varphi)$ is a normalized mode of a flat universe in spherical coordinates.

\section{Spatially spherical gauge}
\label{sec:gauges}

Generic perturbed metrics over a homogeneous and isotropic closed universe can be represented in the form:
\begin{align}
g_{00}(t, \vec x) &= -1 + h_{00}(t, \vec x) \, , \nonumber \\
g_{0i}(t, \vec x) &= a(t) h_{0i}(t, \vec x) \, , \nonumber \\
g_{ij}(t, \vec x) &= a(t)^2 [ \Omega_{ij}(\vec x) + h_{ij}(t,\vec x) ] \, ,
\label{eq:metric-perturbation}
\end{align}
choosing a background lapse function $N_0(t)=1$. The components of the perturbation can be decomposed into scalar, vector and tensor contributions:
\begin{align}
h_{00} &= -2A \, , \nonumber \\
h_{0i} &= - \bar{D}_i B - B_i \, , \nonumber \\
h_{ij} &= 2 D \Omega_{ij} - \bar{D}_i \bar{D}_j E + \bar{D}_j V_i + \bar{D}_i V_j + h_{ij}^{TT} \, ,
\label{eq:svt}
\end{align}
where
\begin{align*}
\bar{D}_i B^i = \bar{D}_i V^i = 0 \, , \quad \bar{D}^i h_{ij}^{TT} = 0 \, , \quad \Omega^{ij} h_{ij}^{TT}=0 \, .
\end{align*}
We raise and lower indices of $3$-vectors with the metric $\Omega_{ij}$ and its inverse. Indices of $4$d tensors are raised and lowered with the full metric $g_{\mu\nu}$ and its inverse, as usual. The unperturbed geometry is represented by $\mathring{g}_{\mu \nu}$. Similarly, a circle over any given quantity means that its unperturbed version is being considered.

The explicit form of the perturbations depends on the choice of a coordinate system. A gauge transformation is defined as an infinitesimal change of coordinates
\[
x^\mu \mapsto \hat{x}^\mu \simeq x^\mu + \lambda \xi^\mu \, ,
\]
where we are considering a first-order expansion in the parameter $\lambda$ of the one-parameter group of diffeomorphisms generated by the vector field $\xi^\mu$. We can make $\lambda=1$ with a redefinition of the vector $\xi^\mu$. The variation of any tensor under such a passive transformation is given to first-order by
\[
\delta T \simeq - \mathcal{L}_\xi T \, .
\]
In particular, the metric tensor transforms as:
\be
g_{\mu \nu} \mapsto \hat{g}_{\mu \nu} \simeq g_{\mu \nu} - \nabla_\mu \xi_\nu - \nabla_\nu \xi_\mu \, .
	\label{eq:delta-g}
\ee
Assuming that the variation of the components of the metric is comparable to the magnitude of the perturbations, the transformed metric is a new perturbed FLRW metric, and can again be written as in Eqs.~\eqref{eq:metric-perturbation} and \eqref{eq:svt}.

It is convenient to introduce the Helmholtz decomposition of the spatial part of the generator $\xi^\mu$ of the transformation,
\begin{align}
\xi^0 &= \zeta(t,\vec x) \, , \nonumber \\
 \xi^i &= \Omega^{ij} \partial_j \xi(t,\vec x) + \alpha^i(t,\vec x) \, .
\label{eq:gauge-transformation-modes}
\end{align}
To first-order, the indices of $\xi^\mu$ can be lowered with the unperturbed metric,
\begin{align}
\xi_0 &= -\zeta \, , \nonumber \\
\xi_i &= a^2 (\partial_i \xi + \alpha_i) \, .
\label{eq:xi-lowered}
\end{align}
The covariant derivative of the vector field $\xi_\mu$ can be computed to first-order with the Christoffel symbols of the background geometry,
\begin{align}
\nabla_0 \xi_0 	& \simeq - \frac{\zeta'}{a} \, ,\\ 
\nabla_0 \xi_i 	&\simeq a (\partial_i \xi' + \alpha'_i) + a'(\partial_i \xi + \alpha_i)  \, , \\
\nabla_i \xi_0 	&\simeq - \partial_i \zeta - a' (\partial_i \xi + \alpha_i) \, , \\
\nabla_i \xi_j &\simeq a^2 \bar{D}_i(\partial_j \xi + \alpha_j) + H a^2 \zeta \Omega_{ij} \, .
\end{align}

Inserting the formulas above in Eq.~\eqref{eq:delta-g}, we obtain:
\begin{align*}
\hat{g}_{00} 	&= -1 - 2\left( A - \frac{\zeta'}{a}  \right) \, , \\
\hat{g}_{0i} 	&= -a \bar{D}_i\left( B+\xi'-\frac{\zeta}{a}\right) - a (B_i + \alpha'_i) \, , \\
\hat{g}_{ij}	&= a^2 \left[ (1+2D -2H\zeta) \Omega_{ij} -2 \bar{D}_i \bar{D}_j (E + \xi) \right. \\
				&\qquad \left. + \bar{D}_i(V_j-\alpha_j) + \bar{D}_i (V_i - \alpha_i) + h^{TT}_{ij}\right] \, .
\end{align*}
Comparing with \eqref{eq:svt}, we find that the transformed scalar, vector and tensor perturbations read:
\begin{align*}
\hat{A} &= A - \frac{\zeta'}{a} \, , \\
\hat{B} &= B+ \xi'- \frac{\zeta}{a} \, , \\
\hat{D} &= D - H\zeta \, , \\
\hat{E} &= E + \xi \, , \\
\hat{B}_i &= B_i + \alpha'_i \, , \\
\hat{V}_i &= V_i - \alpha_i \, , \\
\hat{h}_{ij}^{TT}{}' &= h^{TT}_{ij} \, .
\end{align*}

In the absence of tensor perturbations, $h^{TT}_{ij}=0$, perturbations of the spatial part of the metric can be made to vanish by an adequate choice of coordinates. This is obtained through a gauge transformations satisfying
\begin{align*}
\hat{V}^i = 0 &\quad \Rightarrow \quad \alpha^i = V^i \, , \\
\hat{D} = 0 &\quad \Rightarrow \quad \zeta=D/H \, , \\
\hat{E}=0 &\quad \Rightarrow \quad \xi = - E \, .
\end{align*}
The metric in the spherical slicing gauge has the form
\[
ds^2 = -(1+2A) dt^2 -2 a (\bar{D}_i B + B_i )dx^i dt + a^2 \Omega_{ij} \, .
\]

\section{Flat limit of the quadratic Hamiltonian}
\label{sec:flat-limit}

The quadratic Hamiltonian for scalar perturbations in a closed universe in the spherical gauge is given in Eqs.~\eqref{eq:quadratic-hamiltonian}--\eqref{eq:quadratic-hamiltonian-coeff2}. Let us show that its flat limit corresponds to the quadratic Hamiltonian for scalar perturbations in a flat universe obtained in a spatially flat gauge \cite{Agullo:2018bs}.

The flat limit is obtained by locally considering modes with $n \gg 1$, as discussed in Appendix \ref{sec:harmonics}. At any given time, for a sufficiently large $n$, the physical wavelength of the mode is much smaller than the radius of curvature. As a result, locally, the mode behaves as in a flat space. 
In the large $n$ limit, the closed universe Hamiltonian \eqref{eq:quadratic-hamiltonian} becomes:
\begin{multline*}
H^{(2)}_{nlm} =  \frac{N_0}{2} \left\{ \frac{1}{a^3} (\pi^f)^2 \right.\\
+ \left[ a(n^2-1) + a^3 V''(\phi_0)-6a^2 V' \frac{\pi_\phi}{\pi_a} \right. \\
\left. \left. + \frac{3 \kappa}{2 a^3}  \left( \frac{\pi_{\phi_0}}{V_0} \right)^2
- \frac{9}{a^5} \frac{\pi_{\phi_0}^4}{\pi_a^2 V_0^2}\right] f^2 \right\} \, .
\end{multline*}
This formula can be compared with the quadratic Hamiltonian obtained following the methods of \cite{Agullo:2018bs}. Our conventions for the background momenta $\pi_{\phi_0},\pi_a$ differ from the background momenta $p_\phi, \pi_a$ employed in \cite{Agullo:2018bs} by a factor of $1/V_0$, as can be seen from the distinct normalizations of the Poisson brackets, so that we must identify:
\begin{align}
\pi_{\phi_0} \to V_0 p_\phi \, ,\nonumber \\
\pi_a \to V_0 \pi_a \, .
\label{eq:redef-momenta}
\end{align}
We reproduced the steps of the derivation of the perturbed Hamiltonian in a flat universe in the spatially flat gauge, restricting to terms up to quadratic order, and found an agreement with the flat limit of our Hamiltonian under the redefinitions \eqref{eq:redef-momenta} and the map \eqref{eq:n-to-k-app}.

\acknowledgments

N.Y. acknowledges financial support from the Conselho Nacional de Desenvolvimento Cient\'ifico e Tecnol\'ogico (CNPq) under the Grant No.~306744/2018-0. N.Y. thanks Ivan Agullo for a discussion during the formulation of the project which was invaluable for shaping the strategy for the analysis of the perturbations. The authors thank Teresa Seabra Antunes and Henrique de Paiva Costa for extended discussions on closed universes and cosmological perturbations in our Group for Fundamental Theory (GFT) at the Federal University of Minas Gerais (UFMG).


\begin{thebibliography}{49}%
\makeatletter
\providecommand \@ifxundefined [1]{%
 \@ifx{#1\undefined}
}%
\providecommand \@ifnum [1]{%
 \ifnum #1\expandafter \@firstoftwo
 \else \expandafter \@secondoftwo
 \fi
}%
\providecommand \@ifx [1]{%
 \ifx #1\expandafter \@firstoftwo
 \else \expandafter \@secondoftwo
 \fi
}%
\providecommand \natexlab [1]{#1}%
\providecommand \enquote  [1]{``#1''}%
\providecommand \bibnamefont  [1]{#1}%
\providecommand \bibfnamefont [1]{#1}%
\providecommand \citenamefont [1]{#1}%
\providecommand \href@noop [0]{\@secondoftwo}%
\providecommand \href [0]{\begingroup \@sanitize@url \@href}%
\providecommand \@href[1]{\@@startlink{#1}\@@href}%
\providecommand \@@href[1]{\endgroup#1\@@endlink}%
\providecommand \@sanitize@url [0]{\catcode `\\12\catcode `\$12\catcode
  `\&12\catcode `\#12\catcode `\^12\catcode `\_12\catcode `\%12\relax}%
\providecommand \@@startlink[1]{}%
\providecommand \@@endlink[0]{}%
\providecommand \url  [0]{\begingroup\@sanitize@url \@url }%
\providecommand \@url [1]{\endgroup\@href {#1}{\urlprefix }}%
\providecommand \urlprefix  [0]{URL }%
\providecommand \Eprint [0]{\href }%
\providecommand \doibase [0]{https://doi.org/}%
\providecommand \selectlanguage [0]{\@gobble}%
\providecommand \bibinfo  [0]{\@secondoftwo}%
\providecommand \bibfield  [0]{\@secondoftwo}%
\providecommand \translation [1]{[#1]}%
\providecommand \BibitemOpen [0]{}%
\providecommand \bibitemStop [0]{}%
\providecommand \bibitemNoStop [0]{.\EOS\space}%
\providecommand \EOS [0]{\spacefactor3000\relax}%
\providecommand \BibitemShut  [1]{\csname bibitem#1\endcsname}%
\let\auto@bib@innerbib\@empty
\bibitem [{\citenamefont {Dodelson}\ and\ \citenamefont
  {Schmidt}(2020)}]{Dodelson:2020}%
  \BibitemOpen
  \bibfield  {author} {\bibinfo {author} {\bibfnamefont {S.}~\bibnamefont
  {Dodelson}}\ and\ \bibinfo {author} {\bibfnamefont {F.}~\bibnamefont
  {Schmidt}},\ }\href {https://books.google.com.br/books?id=GGjfywEACAAJ}
  {\emph {\bibinfo {title} {Modern Cosmology}}}\ (\bibinfo  {publisher}
  {Elsevier},\ \bibinfo {year} {2020})\BibitemShut {NoStop}%
\bibitem [{\citenamefont {Weinberg}(2008)}]{Weinberg:2008}%
  \BibitemOpen
  \bibfield  {author} {\bibinfo {author} {\bibfnamefont {S.}~\bibnamefont
  {Weinberg}},\ }\href {https://books.google.com.br/books?id=nqQZdg020fsC}
  {\emph {\bibinfo {title} {Cosmology}}} (\bibinfo  {publisher}
  {Oxford University Pres},\ \bibinfo {year} {2008})\BibitemShut {NoStop}%
\bibitem [{\citenamefont {Schwarz}\ \emph {et~al.}(2016)\citenamefont
  {Schwarz}, \citenamefont {Copi}, \citenamefont {Huterer},\ and\ \citenamefont
  {Starkman}}]{Schwarz:2015cma}%
  \BibitemOpen
  \bibfield  {author} {\bibinfo {author} {\bibfnamefont {D.~J.}\ \bibnamefont
  {Schwarz}}, \bibinfo {author} {\bibfnamefont {C.~J.}\ \bibnamefont {Copi}},
  \bibinfo {author} {\bibfnamefont {D.}~\bibnamefont {Huterer}},\ and\ \bibinfo
  {author} {\bibfnamefont {G.~D.}\ \bibnamefont {Starkman}},\ }\bibfield
  {title} {\bibinfo {title} {{CMB Anomalies after Planck}},\ }\href
  {https://doi.org/10.1088/0264-9381/33/18/184001} {\bibfield  {journal}
  {\bibinfo  {journal} {Class. Quant. Grav.}\ }\textbf {\bibinfo {volume}
  {33}},\ \bibinfo {pages} {184001} (\bibinfo {year} {2016})}
  \BibitemShut {NoStop}%
\bibitem [{\citenamefont {Bonga}\ \emph {et~al.}(2016)\citenamefont {Bonga},
  \citenamefont {Gupt},\ and\ \citenamefont {Yokomizo}}]{Bonga:2016iuf}%
  \BibitemOpen
  \bibfield  {author} {\bibinfo {author} {\bibfnamefont {B.}~\bibnamefont
  {Bonga}}, \bibinfo {author} {\bibfnamefont {B.}~\bibnamefont {Gupt}},\ and\
  \bibinfo {author} {\bibfnamefont {N.}~\bibnamefont {Yokomizo}},\ }\bibfield
  {title} {\bibinfo {title} {{Inflation in the closed FLRW model and the
  CMB}},\ }\href {https://doi.org/10.1088/1475-7516/2016/10/031} {\bibfield
  {journal} {\bibinfo  {journal} {JCAP}\ }\textbf {\bibinfo {volume} {10}},\
  \bibinfo {pages} {031} (\bibinfo {year} {2016})}
  \BibitemShut {NoStop}%
\bibitem [{\citenamefont {Bonga}\ \emph {et~al.}(2017)\citenamefont {Bonga},
  \citenamefont {Gupt},\ and\ \citenamefont {Yokomizo}}]{Bonga:2016cje}%
  \BibitemOpen
  \bibfield  {author} {\bibinfo {author} {\bibfnamefont {B.}~\bibnamefont
  {Bonga}}, \bibinfo {author} {\bibfnamefont {B.}~\bibnamefont {Gupt}},\ and\
  \bibinfo {author} {\bibfnamefont {N.}~\bibnamefont {Yokomizo}},\ }\bibfield
  {title} {\bibinfo {title} {{Tensor perturbations during inflation in a
  spatially closed Universe}},\ }\href
  {https://doi.org/10.1088/1475-7516/2017/05/021} {\bibfield  {journal}
  {\bibinfo  {journal} {JCAP}\ }\textbf {\bibinfo {volume} {05}},\ \bibinfo
  {pages} {021} (\bibinfo {year} {2017})} \BibitemShut {NoStop}%
\bibitem [{\citenamefont {Handley}(2019)}]{Handley:2019anl}%
  \BibitemOpen
  \bibfield  {author} {\bibinfo {author} {\bibfnamefont {W.}~\bibnamefont
  {Handley}},\ }\bibfield  {title} {\bibinfo {title} {{Primordial power spectra
  for curved inflating universes}},\ }\href
  {https://doi.org/10.1103/PhysRevD.100.123517} {\bibfield  {journal} {\bibinfo
   {journal} {Phys. Rev. D}\ }\textbf {\bibinfo {volume} {100}},\ \bibinfo
  {pages} {123517} (\bibinfo {year} {2019})}
  \BibitemShut {NoStop}%
\bibitem [{\citenamefont {Renevey}\ \emph {et~al.}(2021)\citenamefont
  {Renevey}, \citenamefont {Barrau}, \citenamefont {Martineau},\ and\
  \citenamefont {Touati}}]{Renevey:2020zdj}%
  \BibitemOpen
  \bibfield  {author} {\bibinfo {author} {\bibfnamefont {C.}~\bibnamefont
  {Renevey}}, \bibinfo {author} {\bibfnamefont {A.}~\bibnamefont {Barrau}},
  \bibinfo {author} {\bibfnamefont {K.}~\bibnamefont {Martineau}},\ and\
  \bibinfo {author} {\bibfnamefont {S.}~\bibnamefont {Touati}},\ }\bibfield
  {title} {\bibinfo {title} {{Curvature bounce in general relativity:
  background and primordial spectrum}},\ }\href
  {https://doi.org/10.1088/1475-7516/2021/01/018} {\bibfield  {journal}
  {\bibinfo  {journal} {JCAP}\ }\textbf {\bibinfo {volume} {01}},\ \bibinfo
  {pages} {018} (\bibinfo {year} {2021})} \BibitemShut {NoStop}%
\bibitem [{\citenamefont {Renevey}\ \emph {et~al.}(2022)\citenamefont
  {Renevey}, \citenamefont {Barrau},\ and\ \citenamefont
  {Martineau}}]{Renevey:2022xzh}%
  \BibitemOpen
  \bibfield  {author} {\bibinfo {author} {\bibfnamefont {C.}~\bibnamefont
  {Renevey}}, \bibinfo {author} {\bibfnamefont {A.}~\bibnamefont {Barrau}},\
  and\ \bibinfo {author} {\bibfnamefont {K.}~\bibnamefont {Martineau}},\
  }\bibfield  {title} {\bibinfo {title} {{Detailed analysis of the curvature
  bounce: background dynamics and imprints in the CMB}},\ }\href
  {https://doi.org/10.1140/epjc/s10052-022-10745-8} {\bibfield  {journal}
  {\bibinfo  {journal} {Eur. Phys. J. C}\ }\textbf {\bibinfo {volume} {82}},\
  \bibinfo {pages} {775} (\bibinfo {year} {2022})} \BibitemShut
  {NoStop}%
\bibitem [{\citenamefont {Hergt}\ \emph {et~al.}(2022)\citenamefont {Hergt},
  \citenamefont {Agocs}, \citenamefont {Handley}, \citenamefont {Hobson},\ and\
  \citenamefont {Lasenby}}]{Hergt:2022fxk}%
  \BibitemOpen
  \bibfield  {author} {\bibinfo {author} {\bibfnamefont {L.~T.}\ \bibnamefont
  {Hergt}}, \bibinfo {author} {\bibfnamefont {F.~J.}\ \bibnamefont {Agocs}},
  \bibinfo {author} {\bibfnamefont {W.~J.}\ \bibnamefont {Handley}}, \bibinfo
  {author} {\bibfnamefont {M.~P.}\ \bibnamefont {Hobson}},\ and\ \bibinfo
  {author} {\bibfnamefont {A.~N.}\ \bibnamefont {Lasenby}},\ }\bibfield
  {title} {\bibinfo {title} {{Finite inflation in curved space}},\ }\href
  {https://doi.org/10.1103/PhysRevD.106.063529} {\bibfield  {journal} {\bibinfo
   {journal} {Phys. Rev. D}\ }\textbf {\bibinfo {volume} {106}},\ \bibinfo
  {pages} {063529} (\bibinfo {year} {2022})}
  \BibitemShut {NoStop}%
\bibitem [{\citenamefont {Kiefer}\ and\ \citenamefont
  {Vardanyan}(2022)}]{Kiefer:2021iko}%
  \BibitemOpen
  \bibfield  {author} {\bibinfo {author} {\bibfnamefont {C.}~\bibnamefont
  {Kiefer}}\ and\ \bibinfo {author} {\bibfnamefont {T.}~\bibnamefont
  {Vardanyan}},\ }\bibfield  {title} {\bibinfo {title} {{Power spectrum for
  perturbations in an inflationary model for a closed universe}},\ }\href
  {https://doi.org/10.1007/s10714-022-02918-3} {\bibfield  {journal} {\bibinfo
  {journal} {Gen. Rel. Grav.}\ }\textbf {\bibinfo {volume} {54}},\ \bibinfo
  {pages} {30} (\bibinfo {year} {2022})} \BibitemShut
  {NoStop}%
\bibitem [{\citenamefont {Contaldi}\ \emph {et~al.}(2003)\citenamefont
  {Contaldi}, \citenamefont {Peloso}, \citenamefont {Kofman},\ and\
  \citenamefont {Linde}}]{Contaldi:2003zv}%
  \BibitemOpen
  \bibfield  {author} {\bibinfo {author} {\bibfnamefont {C.~R.}\ \bibnamefont
  {Contaldi}}, \bibinfo {author} {\bibfnamefont {M.}~\bibnamefont {Peloso}},
  \bibinfo {author} {\bibfnamefont {L.}~\bibnamefont {Kofman}},\ and\ \bibinfo
  {author} {\bibfnamefont {A.~D.}\ \bibnamefont {Linde}},\ }\bibfield  {title}
  {\bibinfo {title} {{Suppressing the lower multipoles in the CMB
  anisotropies}},\ }\href {https://doi.org/10.1088/1475-7516/2003/07/002}
  {\bibfield  {journal} {\bibinfo  {journal} {JCAP}\ }\textbf {\bibinfo
  {volume} {07}},\ \bibinfo {pages} {002} (\bibinfo {year} {2003})}
  \BibitemShut {NoStop}%
\bibitem [{\citenamefont {Boyanovsky}\ \emph {et~al.}(2006)\citenamefont
  {Boyanovsky}, \citenamefont {de~Vega},\ and\ \citenamefont
  {Sanchez}}]{Boyanovsky:2006pm}%
  \BibitemOpen
  \bibfield  {author} {\bibinfo {author} {\bibfnamefont {D.}~\bibnamefont
  {Boyanovsky}}, \bibinfo {author} {\bibfnamefont {H.~J.}\ \bibnamefont
  {de~Vega}},\ and\ \bibinfo {author} {\bibfnamefont {N.~G.}\ \bibnamefont
  {Sanchez}},\ }\bibfield  {title} {\bibinfo {title} {{CMB quadrupole
  suppression. 2. The early fast roll stage}},\ }\href
  {https://doi.org/10.1103/PhysRevD.74.123007} {\bibfield  {journal} {\bibinfo
  {journal} {Phys. Rev. D}\ }\textbf {\bibinfo {volume} {74}},\ \bibinfo
  {pages} {123007} (\bibinfo {year} {2006})}
  \BibitemShut {NoStop}%
\bibitem [{\citenamefont {Ramirez}\ and\ \citenamefont
  {Schwarz}(2012)}]{Ramirez:2011kk}%
  \BibitemOpen
  \bibfield  {author} {\bibinfo {author} {\bibfnamefont {E.}~\bibnamefont
  {Ramirez}}\ and\ \bibinfo {author} {\bibfnamefont {D.~J.}\ \bibnamefont
  {Schwarz}},\ }\bibfield  {title} {\bibinfo {title} {{Predictions of
  just-enough inflation}},\ }\href {https://doi.org/10.1103/PhysRevD.85.103516}
  {\bibfield  {journal} {\bibinfo  {journal} {Phys. Rev. D}\ }\textbf {\bibinfo
  {volume} {85}},\ \bibinfo {pages} {103516} (\bibinfo {year} {2012})}
  \BibitemShut {NoStop}%
\bibitem [{\citenamefont {Handley}\ \emph {et~al.}(2014)\citenamefont
  {Handley}, \citenamefont {Brechet}, \citenamefont {Lasenby},\ and\
  \citenamefont {Hobson}}]{Handley:2014bqa}%
  \BibitemOpen
  \bibfield  {author} {\bibinfo {author} {\bibfnamefont {W.~J.}\ \bibnamefont
  {Handley}}, \bibinfo {author} {\bibfnamefont {S.~D.}\ \bibnamefont
  {Brechet}}, \bibinfo {author} {\bibfnamefont {A.~N.}\ \bibnamefont
  {Lasenby}},\ and\ \bibinfo {author} {\bibfnamefont {M.~P.}\ \bibnamefont
  {Hobson}},\ }\bibfield  {title} {\bibinfo {title} {{Kinetic Initial
  Conditions for Inflation}},\ }\href
  {https://doi.org/10.1103/PhysRevD.89.063505} {\bibfield  {journal} {\bibinfo
  {journal} {Phys. Rev. D}\ }\textbf {\bibinfo {volume} {89}},\ \bibinfo
  {pages} {063505} (\bibinfo {year} {2014})}
  \BibitemShut {NoStop}%
\bibitem [{\citenamefont {Cicoli}\ \emph {et~al.}(2014)\citenamefont {Cicoli},
  \citenamefont {Downes}, \citenamefont {Dutta}, \citenamefont {Pedro},\ and\
  \citenamefont {Westphal}}]{Cicoli:2014bja}%
  \BibitemOpen
  \bibfield  {author} {\bibinfo {author} {\bibfnamefont {M.}~\bibnamefont
  {Cicoli}}, \bibinfo {author} {\bibfnamefont {S.}~\bibnamefont {Downes}},
  \bibinfo {author} {\bibfnamefont {B.}~\bibnamefont {Dutta}}, \bibinfo
  {author} {\bibfnamefont {F.~G.}\ \bibnamefont {Pedro}},\ and\ \bibinfo
  {author} {\bibfnamefont {A.}~\bibnamefont {Westphal}},\ }\bibfield  {title}
  {\bibinfo {title} {{Just enough inflation: power spectrum modifications at
  large scales}},\ }\href {https://doi.org/10.1088/1475-7516/2014/12/030}
  {\bibfield  {journal} {\bibinfo  {journal} {JCAP}\ }\textbf {\bibinfo
  {volume} {12}},\ \bibinfo {pages} {030} (\bibinfo {year} {2014})} \BibitemShut
  {NoStop}%
\bibitem [{\citenamefont {Ashtekar}\ and\ \citenamefont
  {Barrau}(2015)}]{Ashtekar:2015dja}%
  \BibitemOpen
  \bibfield  {author} {\bibinfo {author} {\bibfnamefont {A.}~\bibnamefont
  {Ashtekar}}\ and\ \bibinfo {author} {\bibfnamefont {A.}~\bibnamefont
  {Barrau}},\ }\bibfield  {title} {\bibinfo {title} {{Loop quantum cosmology:
  From pre-inflationary dynamics to observations}},\ }\href
  {https://doi.org/10.1088/0264-9381/32/23/234001} {\bibfield  {journal}
  {\bibinfo  {journal} {Class. Quant. Grav.}\ }\textbf {\bibinfo {volume}
  {32}},\ \bibinfo {pages} {234001} (\bibinfo {year} {2015})} \BibitemShut
  {NoStop}%
\bibitem [{\citenamefont {Agullo}\ \emph {et~al.}(2023)\citenamefont {Agullo},
  \citenamefont {Wang},\ and\ \citenamefont {Wilson-Ewing}}]{Agullo:2023rev}%
  \BibitemOpen
  \bibfield  {author} {\bibinfo {author} {\bibfnamefont {I.}~\bibnamefont
  {Agullo}}, \bibinfo {author} {\bibfnamefont {A.}~\bibnamefont {Wang}},\ and\
  \bibinfo {author} {\bibfnamefont {E.}~\bibnamefont {Wilson-Ewing}},\
  }\bibfield  {title} {\bibinfo {title} {Loop quantum cosmology: relation
  between theory and observations},\ }in\ \href
  {https://doi.org/10.1007/978-981-19-3079-9_103-1} {\emph {\bibinfo
  {booktitle} {Handbook of Quantum Gravity}}},\ \bibinfo {editor} {edited by\
  \bibinfo {editor} {\bibfnamefont {C.}~\bibnamefont {Bambi}}, \bibinfo
  {editor} {\bibfnamefont {L.}~\bibnamefont {Modesto}},\ and\ \bibinfo {editor}
  {\bibfnamefont {I.}~\bibnamefont {Shapiro}}}\ (\bibinfo  {publisher}
  {Springer},\ \bibinfo {year} {2023}) \BibitemShut
  {NoStop}%
\bibitem [{\citenamefont {Chataignier}\ \emph {et~al.}(2023)\citenamefont
  {Chataignier}, \citenamefont {Kiefer},\ and\ \citenamefont
  {Moniz}}]{Chataignier:2023rkq}%
  \BibitemOpen
  \bibfield  {author} {\bibinfo {author} {\bibfnamefont {L.}~\bibnamefont
  {Chataignier}}, \bibinfo {author} {\bibfnamefont {C.}~\bibnamefont
  {Kiefer}},\ and\ \bibinfo {author} {\bibfnamefont {P.}~\bibnamefont
  {Moniz}},\ }\bibfield  {title} {\bibinfo {title} {{Observations in quantum
  cosmology}},\ }\href {https://doi.org/10.1088/1361-6382/acfa5b} {\bibfield
  {journal} {\bibinfo  {journal} {Class. Quant. Grav.}\ }\textbf {\bibinfo
  {volume} {40}},\ \bibinfo {pages} {223001} (\bibinfo {year} {2023})}
  \BibitemShut {NoStop}%
\bibitem [{\citenamefont {Martin}\ and\ \citenamefont
  {Brandenberger}(2001)}]{Martin:2000xs}%
  \BibitemOpen
  \bibfield  {author} {\bibinfo {author} {\bibfnamefont {J.}~\bibnamefont
  {Martin}}\ and\ \bibinfo {author} {\bibfnamefont {R.~H.}\ \bibnamefont
  {Brandenberger}},\ }\bibfield  {title} {\bibinfo {title} {{The
  Trans-Planckian problem of inflationary cosmology}},\ }\href
  {https://doi.org/10.1103/PhysRevD.63.123501} {\bibfield  {journal} {\bibinfo
  {journal} {Phys. Rev. D}\ }\textbf {\bibinfo {volume} {63}},\ \bibinfo
  {pages} {123501} (\bibinfo {year} {2001})} \BibitemShut
  {NoStop}%
\bibitem [{\citenamefont {Aghanim}\ \emph {et~al.}(2020)\citenamefont {Aghanim}
  \emph {et~al.}}]{Planck:2018vyg}%
  \BibitemOpen
  \bibfield  {author} {\bibinfo {author} {\bibfnamefont {N.}~\bibnamefont
  {Aghanim}} \emph {et~al.} (\bibinfo {collaboration} {Planck}),\ }\bibfield
  {title} {\bibinfo {title} {{Planck 2018 results. VI. Cosmological
  parameters}},\ }\href {https://doi.org/10.1051/0004-6361/201833910}
  {\bibfield  {journal} {\bibinfo  {journal} {Astron. Astrophys.}\ }\textbf
  {\bibinfo {volume} {641}},\ \bibinfo {pages} {A6} (\bibinfo {year} {2020})},\
  \bibinfo {note} {[Erratum: Astron.Astrophys. 652, C4 (2021)]}
  \BibitemShut {NoStop}%
\bibitem [{\citenamefont {Handley}(2021)}]{Handley:2019tkm}%
  \BibitemOpen
  \bibfield  {author} {\bibinfo {author} {\bibfnamefont {W.}~\bibnamefont
  {Handley}},\ }\bibfield  {title} {\bibinfo {title} {{Curvature tension:
  evidence for a closed universe}},\ }\href
  {https://doi.org/10.1103/PhysRevD.103.L041301} {\bibfield  {journal}
  {\bibinfo  {journal} {Phys. Rev. D}\ }\textbf {\bibinfo {volume} {103}},\
  \bibinfo {pages} {L041301} (\bibinfo {year} {2021})}
  \BibitemShut {NoStop}%
\bibitem [{\citenamefont {Di~Valentino}\ \emph {et~al.}(2019)\citenamefont
  {Di~Valentino}, \citenamefont {Melchiorri},\ and\ \citenamefont
  {Silk}}]{DiValentino:2019qzk}%
  \BibitemOpen
  \bibfield  {author} {\bibinfo {author} {\bibfnamefont {E.}~\bibnamefont
  {Di~Valentino}}, \bibinfo {author} {\bibfnamefont {A.}~\bibnamefont
  {Melchiorri}},\ and\ \bibinfo {author} {\bibfnamefont {J.}~\bibnamefont
  {Silk}},\ }\bibfield  {title} {\bibinfo {title} {{Planck evidence for a
  closed Universe and a possible crisis for cosmology}},\ }\href
  {https://doi.org/10.1038/s41550-019-0906-9} {\bibfield  {journal} {\bibinfo
  {journal} {Nature Astron.}\ }\textbf {\bibinfo {volume} {4}},\ \bibinfo
  {pages} {196} (\bibinfo {year} {2019})}
  \BibitemShut {NoStop}%
  \bibitem [{\citenamefont {Vagnozzi}\ \emph
  {et~al.}(2021{\natexlab{a}})\citenamefont {Vagnozzi}, \citenamefont
  {Di~Valentino}, \citenamefont {Gariazzo}, \citenamefont {Melchiorri},
  \citenamefont {Mena},\ and\ \citenamefont {Silk}}]{Vagnozzi:2020rcz}%
  \BibitemOpen
  \bibfield  {author} {\bibinfo {author} {\bibfnamefont {S.}~\bibnamefont
  {Vagnozzi}}, \bibinfo {author} {\bibfnamefont {E.}~\bibnamefont
  {Di~Valentino}}, \bibinfo {author} {\bibfnamefont {S.}~\bibnamefont
  {Gariazzo}}, \bibinfo {author} {\bibfnamefont {A.}~\bibnamefont
  {Melchiorri}}, \bibinfo {author} {\bibfnamefont {O.}~\bibnamefont {Mena}},\
  and\ \bibinfo {author} {\bibfnamefont {J.}~\bibnamefont {Silk}},\ }\bibfield
  {title} {\bibinfo {title} {{The galaxy power spectrum take on spatial
  curvature and cosmic concordance}},\ }\href
  {https://doi.org/10.1016/j.dark.2021.100851} {\bibfield  {journal} {\bibinfo
  {journal} {Phys. Dark Univ.}\ }\textbf {\bibinfo {volume} {33}},\ \bibinfo
  {pages} {100851} (\bibinfo {year} {2021}{\natexlab{a}})}
  \BibitemShut {NoStop}%
\bibitem [{\citenamefont {Vagnozzi}\ \emph
  {et~al.}(2021{\natexlab{b}})\citenamefont {Vagnozzi}, \citenamefont {Loeb},\
  and\ \citenamefont {Moresco}}]{Vagnozzi:2020dfn}%
  \BibitemOpen
  \bibfield  {author} {\bibinfo {author} {\bibfnamefont {S.}~\bibnamefont
  {Vagnozzi}}, \bibinfo {author} {\bibfnamefont {A.}~\bibnamefont {Loeb}},\
  and\ \bibinfo {author} {\bibfnamefont {M.}~\bibnamefont {Moresco}},\
  }\bibfield  {title} {\bibinfo {title} {{Eppur \`e piatto? The Cosmic
  Chronometers Take on Spatial Curvature and Cosmic Concordance}},\ }\href
  {https://doi.org/10.3847/1538-4357/abd4df} {\bibfield  {journal} {\bibinfo
  {journal} {Astrophys. J.}\ }\textbf {\bibinfo {volume} {908}},\ \bibinfo
  {pages} {84} (\bibinfo {year} {2021}{\natexlab{b}})}
  \BibitemShut {NoStop}%
\bibitem [{\citenamefont {Dhawan}\ \emph {et~al.}(2021)\citenamefont {Dhawan},
  \citenamefont {Alsing},\ and\ \citenamefont {Vagnozzi}}]{Dhawan:2021mel}%
  \BibitemOpen
  \bibfield  {author} {\bibinfo {author} {\bibfnamefont {S.}~\bibnamefont
  {Dhawan}}, \bibinfo {author} {\bibfnamefont {J.}~\bibnamefont {Alsing}},\
  and\ \bibinfo {author} {\bibfnamefont {S.}~\bibnamefont {Vagnozzi}},\
  }\bibfield  {title} {\bibinfo {title} {{Non-parametric spatial curvature
  inference using late-Universe cosmological probes}},\ }\href
  {https://doi.org/10.1093/mnrasl/slab058} {\bibfield  {journal} {\bibinfo
  {journal} {Mon. Not. Roy. Astron. Soc.}\ }\textbf {\bibinfo {volume} {506}},\
  \bibinfo {pages} {L1} (\bibinfo {year} {2021})}
  \BibitemShut {NoStop}%
\bibitem [{\citenamefont {Borde}\ and\ \citenamefont
  {Vilenkin}(1996)}]{Borde:1996pt}%
  \BibitemOpen
  \bibfield  {author} {\bibinfo {author} {\bibfnamefont {A.}~\bibnamefont
  {Borde}}\ and\ \bibinfo {author} {\bibfnamefont {A.}~\bibnamefont
  {Vilenkin}},\ }\bibfield  {title} {\bibinfo {title} {{Singularities in
  inflationary cosmology: A Review}},\ }\href
  {https://doi.org/10.1142/S0218271896000497} {\bibfield  {journal} {\bibinfo
  {journal} {Int. J. Mod. Phys. D}\ }\textbf {\bibinfo {volume} {5}},\ \bibinfo
  {pages} {813} (\bibinfo {year} {1996})} \BibitemShut
  {NoStop}%
\bibitem [{\citenamefont {Ashtekar}\ \emph {et~al.}(2009)\citenamefont
  {Ashtekar}, \citenamefont {Kaminski},\ and\ \citenamefont
  {Lewandowski}}]{Ashtekar:2009mb}%
  \BibitemOpen
  \bibfield  {author} {\bibinfo {author} {\bibfnamefont {A.}~\bibnamefont
  {Ashtekar}}, \bibinfo {author} {\bibfnamefont {W.}~\bibnamefont {Kaminski}},\
  and\ \bibinfo {author} {\bibfnamefont {J.}~\bibnamefont {Lewandowski}},\
  }\bibfield  {title} {\bibinfo {title} {{Quantum field theory on a
  cosmological, quantum space-time}},\ }\href
  {https://doi.org/10.1103/PhysRevD.79.064030} {\bibfield  {journal} {\bibinfo
  {journal} {Phys. Rev. D}\ }\textbf {\bibinfo {volume} {79}},\ \bibinfo
  {pages} {064030} (\bibinfo {year} {2009})} \BibitemShut
  {NoStop}%
\bibitem [{\citenamefont {Agullo}\ \emph {et~al.}(2012)\citenamefont {Agullo},
  \citenamefont {Ashtekar},\ and\ \citenamefont {Nelson}}]{Agullo:2012sh}%
  \BibitemOpen
  \bibfield  {author} {\bibinfo {author} {\bibfnamefont {I.}~\bibnamefont
  {Agullo}}, \bibinfo {author} {\bibfnamefont {A.}~\bibnamefont {Ashtekar}},\
  and\ \bibinfo {author} {\bibfnamefont {W.}~\bibnamefont {Nelson}},\
  }\bibfield  {title} {\bibinfo {title} {{A Quantum Gravity Extension of the
  Inflationary Scenario}},\ }\href
  {https://doi.org/10.1103/PhysRevLett.109.251301} {\bibfield  {journal}
  {\bibinfo  {journal} {Phys. Rev. Lett.}\ }\textbf {\bibinfo {volume} {109}},\
  \bibinfo {pages} {251301} (\bibinfo {year} {2012})} \BibitemShut
  {NoStop}%
\bibitem [{\citenamefont {Agullo}\ \emph
  {et~al.}(2013{\natexlab{a}})\citenamefont {Agullo}, \citenamefont
  {Ashtekar},\ and\ \citenamefont {Nelson}}]{Agullo:2013ai}%
  \BibitemOpen
  \bibfield  {author} {\bibinfo {author} {\bibfnamefont {I.}~\bibnamefont
  {Agullo}}, \bibinfo {author} {\bibfnamefont {A.}~\bibnamefont {Ashtekar}},\
  and\ \bibinfo {author} {\bibfnamefont {W.}~\bibnamefont {Nelson}},\
  }\bibfield  {title} {\bibinfo {title} {{The pre-inflationary dynamics of loop
  quantum cosmology: Confronting quantum gravity with observations}},\ }\href
  {https://doi.org/10.1088/0264-9381/30/8/085014} {\bibfield  {journal}
  {\bibinfo  {journal} {Class. Quant. Grav.}\ }\textbf {\bibinfo {volume}
  {30}},\ \bibinfo {pages} {085014} (\bibinfo {year} {2013}{\natexlab{a}})}
  \BibitemShut {NoStop}%
\bibitem [{\citenamefont {Agullo}\ and\ \citenamefont
  {Morris}(2015)}]{Agullo:2015tca}%
  \BibitemOpen
  \bibfield  {author} {\bibinfo {author} {\bibfnamefont {I.}~\bibnamefont
  {Agullo}}\ and\ \bibinfo {author} {\bibfnamefont {N.~A.}\ \bibnamefont
  {Morris}},\ }\bibfield  {title} {\bibinfo {title} {{Detailed analysis of the
  predictions of loop quantum cosmology for the primordial power spectra}},\
  }\href {https://doi.org/10.1103/PhysRevD.92.124040} {\bibfield  {journal}
  {\bibinfo  {journal} {Phys. Rev. D}\ }\textbf {\bibinfo {volume} {92}},\
  \bibinfo {pages} {124040} (\bibinfo {year} {2015})} \BibitemShut
  {NoStop}%
\bibitem [{\citenamefont {Bonga}\ and\ \citenamefont
  {Gupt}(2016{\natexlab{a}})}]{Bonga:2015kaa}%
  \BibitemOpen
  \bibfield  {author} {\bibinfo {author} {\bibfnamefont {B.}~\bibnamefont
  {Bonga}}\ and\ \bibinfo {author} {\bibfnamefont {B.}~\bibnamefont {Gupt}},\
  }\bibfield  {title} {\bibinfo {title} {{Inflation with the Starobinsky
  potential in Loop Quantum Cosmology}},\ }\href
  {https://doi.org/10.1007/s10714-016-2071-0} {\bibfield  {journal} {\bibinfo
  {journal} {Gen. Rel. Grav.}\ }\textbf {\bibinfo {volume} {48}},\ \bibinfo
  {pages} {71} (\bibinfo {year} {2016}{\natexlab{a}})} \BibitemShut
  {NoStop}%
\bibitem [{\citenamefont {Bonga}\ and\ \citenamefont
  {Gupt}(2016{\natexlab{b}})}]{Bonga:2015xna}%
  \BibitemOpen
  \bibfield  {author} {\bibinfo {author} {\bibfnamefont {B.}~\bibnamefont
  {Bonga}}\ and\ \bibinfo {author} {\bibfnamefont {B.}~\bibnamefont {Gupt}},\
  }\bibfield  {title} {\bibinfo {title} {{Phenomenological investigation of a
  quantum gravity extension of inflation with the Starobinsky potential}},\
  }\href {https://doi.org/10.1103/PhysRevD.93.063513} {\bibfield  {journal}
  {\bibinfo  {journal} {Phys. Rev. D}\ }\textbf {\bibinfo {volume} {93}},\
  \bibinfo {pages} {063513} (\bibinfo {year} {2016}{\natexlab{b}})} \BibitemShut
  {NoStop}%
\bibitem [{\citenamefont {Agullo}\ \emph {et~al.}(2018)\citenamefont {Agullo},
  \citenamefont {Bolliet},\ and\ \citenamefont {Sreenath}}]{Agullo:2018bs}%
  \BibitemOpen
  \bibfield  {author} {\bibinfo {author} {\bibfnamefont {I.}~\bibnamefont
  {Agullo}}, \bibinfo {author} {\bibfnamefont {B.}~\bibnamefont {Bolliet}},\
  and\ \bibinfo {author} {\bibfnamefont {V.}~\bibnamefont {Sreenath}},\
  }\bibfield  {title} {\bibinfo {title} {Non-gaussianity in loop quantum
  cosmology},\ }\href {https://doi.org/10.1103/PhysRevD.97.066021} {\bibfield
  {journal} {\bibinfo  {journal} {Phys. Rev. D}\ }\textbf {\bibinfo {volume}
  {97}},\ \bibinfo {pages} {066021} (\bibinfo {year} {2018})}\BibitemShut
  {NoStop}%
\bibitem [{\citenamefont {Ashtekar}\ \emph {et~al.}(2007)\citenamefont
  {Ashtekar}, \citenamefont {Pawlowski}, \citenamefont {Singh},\ and\
  \citenamefont {Vandersloot}}]{Ashtekar:2006es}%
  \BibitemOpen
  \bibfield  {author} {\bibinfo {author} {\bibfnamefont {A.}~\bibnamefont
  {Ashtekar}}, \bibinfo {author} {\bibfnamefont {T.}~\bibnamefont {Pawlowski}},
  \bibinfo {author} {\bibfnamefont {P.}~\bibnamefont {Singh}},\ and\ \bibinfo
  {author} {\bibfnamefont {K.}~\bibnamefont {Vandersloot}},\ }\bibfield
  {title} {\bibinfo {title} {{Loop quantum cosmology of k=1 FRW models}},\
  }\href {https://doi.org/10.1103/PhysRevD.75.024035} {\bibfield  {journal}
  {\bibinfo  {journal} {Phys. Rev. D}\ }\textbf {\bibinfo {volume} {75}},\
  \bibinfo {pages} {024035} (\bibinfo {year} {2007})} \BibitemShut
  {NoStop}%
\bibitem [{\citenamefont {Halliwell}\ and\ \citenamefont
  {Hawking}(1985)}]{Halliwell:1984eu}%
  \BibitemOpen
  \bibfield  {author} {\bibinfo {author} {\bibfnamefont {J.~J.}\ \bibnamefont
  {Halliwell}}\ and\ \bibinfo {author} {\bibfnamefont {S.~W.}\ \bibnamefont
  {Hawking}},\ }\bibfield  {title} {\bibinfo {title} {{The Origin of Structure
  in the Universe}},\ }\href {https://doi.org/10.1103/PhysRevD.31.1777}
  {\bibfield  {journal} {\bibinfo  {journal} {Phys. Rev. D}\ }\textbf {\bibinfo
  {volume} {31}},\ \bibinfo {pages} {1777} (\bibinfo {year}
  {1985})}\BibitemShut {NoStop}%
\bibitem [{\citenamefont {Ashtekar}\ and\ \citenamefont
  {Singh}(2011)}]{Ashtekar:2011ni}%
  \BibitemOpen
  \bibfield  {author} {\bibinfo {author} {\bibfnamefont {A.}~\bibnamefont
  {Ashtekar}}\ and\ \bibinfo {author} {\bibfnamefont {P.}~\bibnamefont
  {Singh}},\ }\bibfield  {title} {\bibinfo {title} {{Loop Quantum Cosmology: A
  Status Report}},\ }\href {https://doi.org/10.1088/0264-9381/28/21/213001}
  {\bibfield  {journal} {\bibinfo  {journal} {Class. Quant. Grav.}\ }\textbf
  {\bibinfo {volume} {28}},\ \bibinfo {pages} {213001} (\bibinfo {year}
  {2011})} \BibitemShut {NoStop}%
\bibitem [{\citenamefont {Agullo}\ and\ \citenamefont
  {Singh}(2017)}]{Agullo:2016tjh}%
  \BibitemOpen
  \bibfield  {author} {\bibinfo {author} {\bibfnamefont {I.}~\bibnamefont
  {Agullo}}\ and\ \bibinfo {author} {\bibfnamefont {P.}~\bibnamefont {Singh}},\
  }\bibinfo {title} {{Loop Quantum Cosmology}},\ in\ \href
  {https://doi.org/10.1142/9789813220003_0007} {\emph {\bibinfo {booktitle}
  {{Loop Quantum Gravity}: {The First 30 Years}}}},\ \bibinfo {editor} {edited
  by\ \bibinfo {editor} {\bibfnamefont {A.}~\bibnamefont {Ashtekar}}\ and\
  \bibinfo {editor} {\bibfnamefont {J.}~\bibnamefont {Pullin}}}\ (\bibinfo
  {publisher} {WSP},\ \bibinfo {year} {2017})\ pp.\ \bibinfo {pages}
  {183--240} \BibitemShut {NoStop}%
\bibitem [{\citenamefont {Agullo}\ \emph
  {et~al.}(2013{\natexlab{b}})\citenamefont {Agullo}, \citenamefont
  {Ashtekar},\ and\ \citenamefont {Nelson}}]{Agullo:2012fc}%
  \BibitemOpen
  \bibfield  {author} {\bibinfo {author} {\bibfnamefont {I.}~\bibnamefont
  {Agullo}}, \bibinfo {author} {\bibfnamefont {A.}~\bibnamefont {Ashtekar}},\
  and\ \bibinfo {author} {\bibfnamefont {W.}~\bibnamefont {Nelson}},\
  }\bibfield  {title} {\bibinfo {title} {{Extension of the quantum theory of
  cosmological perturbations to the Planck era}},\ }\href
  {https://doi.org/10.1103/PhysRevD.87.043507} {\bibfield  {journal} {\bibinfo
  {journal} {Phys. Rev. D}\ }\textbf {\bibinfo {volume} {87}},\ \bibinfo
  {pages} {043507} (\bibinfo {year} {2013}{\natexlab{b}})} \BibitemShut
  {NoStop}%
\bibitem [{\citenamefont {Mukhanov}\ \emph {et~al.}(1992)\citenamefont
  {Mukhanov}, \citenamefont {Feldman},\ and\ \citenamefont
  {Brandenberger}}]{Mukhanov:1990me}%
  \BibitemOpen
  \bibfield  {author} {\bibinfo {author} {\bibfnamefont {V.~F.}\ \bibnamefont
  {Mukhanov}}, \bibinfo {author} {\bibfnamefont {H.~A.}\ \bibnamefont
  {Feldman}},\ and\ \bibinfo {author} {\bibfnamefont {R.~H.}\ \bibnamefont
  {Brandenberger}},\ }\bibfield  {title} {\bibinfo {title} {{Theory of
  cosmological perturbations. Part 1. Classical perturbations. Part 2. Quantum
  theory of perturbations. Part 3. Extensions}},\ }\href
  {https://doi.org/10.1016/0370-1573(92)90044-Z} {\bibfield  {journal}
  {\bibinfo  {journal} {Phys. Rept.}\ }\textbf {\bibinfo {volume} {215}},\
  \bibinfo {pages} {203} (\bibinfo {year} {1992})}\BibitemShut {NoStop}%
\bibitem [{\citenamefont {Gordon}\ \emph {et~al.}(2021)\citenamefont {Gordon},
  \citenamefont {Li},\ and\ \citenamefont {Singh}}]{Gordon:2020gel}%
  \BibitemOpen
  \bibfield  {author} {\bibinfo {author} {\bibfnamefont {L.}~\bibnamefont
  {Gordon}}, \bibinfo {author} {\bibfnamefont {B.-F.}\ \bibnamefont {Li}},\
  and\ \bibinfo {author} {\bibfnamefont {P.}~\bibnamefont {Singh}},\ }\bibfield
   {title} {\bibinfo {title} {{Quantum gravitational onset of Starobinsky
  inflation in a closed universe}},\ }\href
  {https://doi.org/10.1103/PhysRevD.103.046016} {\bibfield  {journal} {\bibinfo
   {journal} {Phys. Rev. D}\ }\textbf {\bibinfo {volume} {103}},\ \bibinfo
  {pages} {046016} (\bibinfo {year} {2021})} \BibitemShut
  {NoStop}%
\bibitem [{\citenamefont {Akrami}\ \emph {et~al.}(2020)\citenamefont {Akrami}
  \emph {et~al.}}]{Planck:2018jri}%
  \BibitemOpen
  \bibfield  {author} {\bibinfo {author} {\bibfnamefont {Y.}~\bibnamefont
  {Akrami}} \emph {et~al.} (\bibinfo {collaboration} {Planck}),\ }\bibfield
  {title} {\bibinfo {title} {{Planck 2018 results. X. Constraints on
  inflation}},\ }\href {https://doi.org/10.1051/0004-6361/201833887} {\bibfield
   {journal} {\bibinfo  {journal} {Astron. Astrophys.}\ }\textbf {\bibinfo
  {volume} {641}},\ \bibinfo {pages} {A10} (\bibinfo {year} {2020})}
  \BibitemShut {NoStop}%
  \bibitem [{\citenamefont {Langlois}(1994)}]{Langlois:1994ec}%
  \BibitemOpen
  \bibfield  {author} {\bibinfo {author} {\bibfnamefont {D.}~\bibnamefont
  {Langlois}},\ }\bibfield  {title} {\bibinfo {title} {{Hamiltonian formalism
  and gauge invariance for linear perturbations in inflation}},\ }\href
  {https://doi.org/10.1088/0264-9381/11/2/011} {\bibfield  {journal} {\bibinfo
  {journal} {Class. Quant. Grav.}\ }\textbf {\bibinfo {volume} {11}},\ \bibinfo
  {pages} {389} (\bibinfo {year} {1994})}\BibitemShut {NoStop}%
\bibitem [{\citenamefont {Abbott}\ and\ \citenamefont
  {Schaefer}(1986)}]{Abbott:1986ct}%
  \BibitemOpen
  \bibfield  {author} {\bibinfo {author} {\bibfnamefont {L.~F.}\ \bibnamefont
  {Abbott}}\ and\ \bibinfo {author} {\bibfnamefont {R.~K.}\ \bibnamefont
  {Schaefer}},\ }\bibfield  {title} {\bibinfo {title} {{A General, Gauge
  Invariant Analysis of the Cosmic Microwave Anisotropy}},\ }\href
  {https://doi.org/10.1086/164525} {\bibfield  {journal} {\bibinfo  {journal}
  {Astrophys. J.}\ }\textbf {\bibinfo {volume} {308}},\ \bibinfo {pages} {546}
  (\bibinfo {year} {1986})}\BibitemShut {NoStop}%
\bibitem [{\citenamefont {Parker}\ and\ \citenamefont
  {Toms}(2009)}]{Parker:2009uva}%
  \BibitemOpen
  \bibfield  {author} {\bibinfo {author} {\bibfnamefont {L.~E.}\ \bibnamefont
  {Parker}}\ and\ \bibinfo {author} {\bibfnamefont {D.}~\bibnamefont {Toms}},\
  }\href {https://doi.org/10.1017/CBO9780511813924} {\emph {\bibinfo {title}
  {{Quantum Field Theory in Curved Spacetime}: {Quantized Field and
  Gravity}}}}\ (\bibinfo
  {publisher} {Cambridge University Press},\ \bibinfo {year}
  {2009})\BibitemShut {NoStop}%
\bibitem [{\citenamefont {Birrell}\ and\ \citenamefont
  {Davies}(1984)}]{Birrell:1982ix}%
  \BibitemOpen
  \bibfield  {author} {\bibinfo {author} {\bibfnamefont {N.~D.}\ \bibnamefont
  {Birrell}}\ and\ \bibinfo {author} {\bibfnamefont {P.~C.~W.}\ \bibnamefont
  {Davies}},\ }\href {https://doi.org/10.1017/CBO9780511622632} {\emph
  {\bibinfo {title} {{Quantum Fields in Curved Space}}}} \ (\bibinfo  {publisher} {Cambridge University Press},\ \bibinfo {year} {1984})\BibitemShut
  {NoStop}%
\bibitem [{\citenamefont {Hunt}\ and\ \citenamefont
  {Sarkar}(2014)}]{Hunt:2013bha}%
  \BibitemOpen
  \bibfield  {author} {\bibinfo {author} {\bibfnamefont {P.}~\bibnamefont
  {Hunt}}\ and\ \bibinfo {author} {\bibfnamefont {S.}~\bibnamefont {Sarkar}},\
  }\bibfield  {title} {\bibinfo {title} {{Reconstruction of the primordial
  power spectrum of curvature perturbations using multiple data sets}},\ }\href
  {https://doi.org/10.1088/1475-7516/2014/01/025} {\bibfield  {journal}
  {\bibinfo  {journal} {JCAP}\ }\textbf {\bibinfo {volume} {01}},\ \bibinfo
  {pages} {025} (\bibinfo {year} {2014})} \BibitemShut {NoStop}%
\bibitem [{\citenamefont {Ashtekar}\ and\ \citenamefont
  {Gupt}(2017{\natexlab{a}})}]{Ashtekar:2016wpi}%
  \BibitemOpen
  \bibfield  {author} {\bibinfo {author} {\bibfnamefont {A.}~\bibnamefont
  {Ashtekar}}\ and\ \bibinfo {author} {\bibfnamefont {B.}~\bibnamefont
  {Gupt}},\ }\bibfield  {title} {\bibinfo {title} {{Quantum Gravity in the Sky:
  Interplay between fundamental theory and observations}},\ }\href
  {https://doi.org/10.1088/1361-6382/34/1/014002} {\bibfield  {journal}
  {\bibinfo  {journal} {Class. Quant. Grav.}\ }\textbf {\bibinfo {volume}
  {34}},\ \bibinfo {pages} {014002} (\bibinfo {year} {2017}{\natexlab{a}})}
  \BibitemShut {NoStop}%
\bibitem [{\citenamefont {Ashtekar}\ and\ \citenamefont
  {Gupt}(2017{\natexlab{b}})}]{Ashtekar:2016pqn}%
  \BibitemOpen
  \bibfield  {author} {\bibinfo {author} {\bibfnamefont {A.}~\bibnamefont
  {Ashtekar}}\ and\ \bibinfo {author} {\bibfnamefont {B.}~\bibnamefont
  {Gupt}},\ }\bibfield  {title} {\bibinfo {title} {{Initial conditions for
  cosmological perturbations}},\ }\href
  {https://doi.org/10.1088/1361-6382/aa52d4} {\bibfield  {journal} {\bibinfo
  {journal} {Class. Quant. Grav.}\ }\textbf {\bibinfo {volume} {34}},\ \bibinfo
  {pages} {035004} (\bibinfo {year} {2017}{\natexlab{b}})} \BibitemShut
  {NoStop}%
\bibitem [{\citenamefont {Gerlach}\ and\ \citenamefont
  {Sengupta}(1978)}]{Gerlach:1978gy}%
  \BibitemOpen
  \bibfield  {author} {\bibinfo {author} {\bibfnamefont {U.~H.}\ \bibnamefont
  {Gerlach}}\ and\ \bibinfo {author} {\bibfnamefont {U.~K.}\ \bibnamefont
  {Sengupta}},\ }\bibfield  {title} {\bibinfo {title} {{Homogeneous Collapsing
  Star: Tensor and Vector Harmonics for Matter and Field Asymmetries}},\ }\href
  {https://doi.org/10.1103/PhysRevD.18.1773} {\bibfield  {journal} {\bibinfo
  {journal} {Phys. Rev. D}\ }\textbf {\bibinfo {volume} {18}},\ \bibinfo
  {pages} {1773} (\bibinfo {year} {1978})}\BibitemShut {NoStop}%
\end{thebibliography}
\end{document}